\documentclass[12pt]{iopart}
\usepackage{iopams}
\usepackage{amsmath2}
\usepackage{graphicx}
\usepackage{color}
\usepackage{amssymb}
\usepackage{graphics}

\begin{document}

\title{The Anderson-Josephson quantum dot--A theory perspective}

\author{V Meden}

\address{Institut f\"ur Theorie der Statistischen Physik, RWTH Aachen University 
and JARA--Fundamentals of Future Information Technology, 52056 Aachen, Germany}

\ead{meden@physik.rwth-aachen.de}

\begin{abstract}
Recent progress in nanoscale manufacturing allowed to experimentally 
investigate quantum dots coupled to two superconducting leads in 
controlled and tunable setups. The equilibrium Josephson current was 
measured in on-chip SQUID devices and subgap states were investigated
using weakly coupled metallic leads for spectroscopy. This put back
two ``classic'' problems 
also on the agenda of theoretical condensed matter physics: the Josephson effect 
and quantum spins in superconductors. The relevance of the former is obvious 
as the barrier separating the two superconductors in a standard
Josephson junction is merely replaced 
by the quantum dot with well separated energy levels. For odd filling of the
dot it acts as a quantum mechanical spin-1/2 and the relevance of 
the latter becomes apparent as well. 
For normal conducting leads
and at odd dot filling the Kondo effect strongly modifies the transport 
properties as can, e.g., be studied within the Anderson model. One can expect
that also for superconducting leads and in certain parameter regimes remnants of
Kondo physics, i.e. strong electronic correlations, will affect the
Josephson current.   

In this topical review we discuss the status of the theoretical understanding 
of the Anderson-Josephson quantum dot in equilibrium mainly focusing on the 
Josephson current. We introduce a minimal model consisting of a dot which 
can only host one spin-up and one spin-down electron repelling each other
by a local Coulomb interaction. The dot is tunnel-coupled to two
superconducting leads described by the 
BCS Hamiltonian. This model was investigated using a variety of 
methods, some capturing aspects of Kondo physics others failing in this 
respect. We briefly review this. The model shows a first order 
level-crossing quantum phase transition when varying any  
parameter provided the others are within appropriate ranges. At vanishing
temperature it leads to a jump of the Josephson current. When being interested in the 
qualitative behavior of the phase diagram or the Josephson current several 
of the methods can be used. However, for a quantitative description elaborate 
quantum many-body methods must be employed.

We show that a quantitative agreement 
between accurate results obtained for the simple model and 
measurements of the current can be reached. This confirms that the experiments
reveal the finite temperature signatures of the zero temperature transition.

In addition, we consider two examples of more complex dot geometries which might be 
experimentally realized in the near future. The first is characterized by the interplay 
of the above level-crossing physics and the Fano effect, the second by the interplay 
of superconductivity and almost degenerate singlet and triplet two-body states.

\end{abstract}


\maketitle

\section{Introduction}
\label{sec:introduction}

Today nanostructuring techniques allow to routinely manufacture small quantum systems 
coupled to metallic leads. Many insights were gained by studying their linear response
transport properties. The small system might be a semiconductor 
heterostructure or a molecule, e.g., a carbon nanotube. Often the parameters can 
be tuned such that the typical level spacing of the system becomes the 
largest energy scale; for setups placed in a cryostat even larger than the energy 
associated to the temperature. For all practical purposes one 
can then focus on a single level, which might be degenerate due to spin and 
orbital symmetries. Such systems are commonly referred to as single-level quantum 
dots. Here we are mainly interested in the situation in which the level is (doubly) 
spin-degenerate in the absence of a Zeeman field.
In experiments the level energy (position) can be varied by tuning the voltage 
applied to a properly designed gate.  

Due to the strong spatial 
confinement the energy scale $U$ characterizing the electron-electron repulsion 
on the dot is sizeable, while the two-particle interaction in the leads 
is generically small and only leads to a slight modification of the parameters 
such as, e.g., the effective electron mass (Fermi liquid theory \cite{Pines61}); 
for the leads the 
independent electron approximation holds. The local on-dot interaction 
can be expected to alter the transport properties by e.g. Coulomb blockade\cite{Grabert92} but 
also by the more intriguing many-body Kondo effect.\cite{Hewson97} In this the 
single electron occupying the dot for an appropriate level energy acts as a localized 
spin which is screened by an effective exchange interaction with the spin of the itinerant 
lead electrons this way affecting the transport properties. 

Prior to the age of mesoscopic electron transport the Kondo effect was observed 
experimentally as the resistance minimum as a function of temperature 
of metals containing a small concentration of 
magnetic impurities. In 1964 Kondo provided a
satisfying explanation of the  
minimum by showing that the resistance increases logarithmically 
for intermediate temperatures down to the emergent low temperature Kondo 
scale $T_{\rm K}$.\cite{Kondo64} 
Effective models such as the Kondo model (s-d-model) and the single impurity 
Anderson model (SIAM)\cite{Hewson97} were suggested to describe Kondo physics.
In the former the charge fluctuations of the impurity level are suppressed from the 
outset and only spin fluctuations are kept while in the latter the charge fluctuations
are considered as well. 
However, even these simplified models resisted a fully satisfying theoretical 
treatment at energy scales below $T_{\rm K}$ until elaborate methods such as the 
numerical renormalization group (NRG)\cite{Hewson97,Bulla08} and the Bethe ansatz solutions\cite{Tsvelick83} 
were developed. Other standard approaches of quantum many-body physics (perturbation theory, 
the unrestricted mean-field approximation, equation-of-motion approaches,\ldots) can only capture 
certain aspects of Kondo physics but suffer from shortcomings of varying severity; some 
of these are mentioned below. Studying the Kondo effect in quantum dot setups instead of bulk
systems has the clear advantage of control and tunability of the parameters.

When superconducting materials are used for the two leads, transport through the quantum 
dot can be expected to show another interesting phenomenon first described two years
prior to Kondo's seminal work: the Josephson effect.\cite{Josephson62} If two 
superconductors are coupled via a structureless tunnel barrier an equilibrium (Josephson) supercurrent 
may flow across the barrier. Forgetting about the internal  degrees of freedom and 
thus also the local Coulomb interaction for a moment a dot level tuned away from 
resonance acts as a tunnel barrier and depending on the difference in the phase $\phi$ 
of the superconducting order parameter of the two leads one expects the appearance of a  
Josephson current. In the present review we exclusively consider conventional s-wave
superconducting lead materials which can accurately be described by the BCS model.\cite{Tinkham04} 

We here review the theoretical understanding of the physics of a single-level quantum dot
coupled to two BCS superconducting leads taking the internal dot degree of 
freedom and the local on-dot Coulomb repulsion into account. This system is modeled 
in terms of the SIAM with BCS leads. We discuss how the Josephson
current is modified by the dot degrees of freedom and the local two-particle 
interaction. As the most dramatic effect at temperature $T=0$ a jump of 
the supercurrent occurs if one of the system parameters is varied provided the others are taken 
from appropriate ranges. It results from an underlying level crossing quantum phase 
transition at which the ground state changes from being a singlet to being a doublet. 
This, e.g., implies that the well known sinusoidal current-phase relation (CPR) of an 
ordinary Josephson junction (in the large barrier limit) is strongly altered. In particular,
the current becomes negative in the doublet phase even for $\phi \in [0,\pi]$. This
so-called $\pi$-junction behavior must be contrasted to the noninteracting $0$-junction
behavior with a positive current in this interval of the phase difference. One therefore
also speaks off a $0$-to-$\pi$-transition. Signatures of the transition can be
observed  for $T>0$, at which the jump in the supercurrent is smeared  out. We show that this main effect 
although being rooted in the local Coulomb interaction is 
not directly related to Kondo correlations. It occurs and can be understood in 
various limiting cases in which the Kondo effect does obviously not play any role, e.g. 
the one of a large superconducting gap $\Delta$. In accordance with this the 
$\pi$-junction behavior of the Josephson current can be described at least 
qualitatively by several approximate many-body methods which are of limited 
usefulness when it comes to the description of the Kondo effect in systems 
with metallic leads. 
However, we argue that even for a qualitative understanding of the Josephson 
current only methods should be employed which do obey 
fundamental physical principles, such as, e.g., spin symmetry (in the absence of a
symmetry-breaking external Zeeman field). 
Although Kondo correlations are not the driving  force behind the quantum phase 
transition the most interesting parameter regime is still the one in which 
the dot is in the Kondo regime for suppressed superconductivity and Kondo correlations
and superconductivity compete.

The theoretical description becomes more 
involved if the dot parameters, that is the level position, the level-lead couplings, 
the on-dot interaction, and the Zeeman splitting are taken from the Kondo regime
and the superconducting gap is comparable in size to the normal state Kondo temperature 
$T_{\rm K}$. In this case one expects remnants of Kondo physics to affect the
supercurrent even though the Kondo effect cannot fully develop due to the lead superconductivity. 
We argue that for a quantitative description one then has to use an advanced many-body method to 
compute the current which proved to be able to capture the Kondo effect for metallic 
leads such as NRG\cite{Hewson97,Bulla08} or quantum Monte Carlo (QMC)\cite{Gubernatis16} approaches.
For the largest part of the parameter space, however, the Kondo effect is by far less important 
for the physics of the Anderson-Josephson quantum dot
than frequently suggested implicitly or explicitely in the literature. Accordingly typical
Kondo-related concepts such as, e.g., universality are of minor importance. This will become
more explicit further down.         

After unraveling the theoretical description we compare the parameter dependence of 
the Josephson current including the CPR computed by a QMC approach to recent 
measurements of the current through dots tuned to the regime of the interplay 
of the Josephson and the Kondo effect. We show that in this case to even properly 
estimate the level-lead coupling in the metallic state one needs to employ the above 
mentioned advanced methods and compare to the measured normal state linear conductance. 
As the Zeeman energy used to suppress the superconductivity must be of the order of 
$\Delta$ and thus of the order of (the normal state) $T_{\rm K}$ the Zeeman field must 
be taken into account in the normal-state calculation. Proceeding this way we achieve 
a satisfying agreement between the measurements of the Josephson current and the results 
based on the model calculations. This indicates that the experiments show the finite
temperature remnants of the $T=0$ quantum phase transition and can indeed be 
understood and quantitatively be described in terms of the simple SIAM with BCS leads 
in which many system specific details are ignored. 
      
We finally use one of the methods (only) capturing certain aspects of Kondo physics, namely
the functional renormalization group (FRG),\cite{Metzner12} to study two more complex dot 
setups with superconducting leads. In the first the 
Anderson-Josephson quantum dot is embedded in a Aharonov-Bohm-like geometry in which 
the two superconductors  are directly tunnel coupled besides being linked to the dot. 
In this case the Fano effect is of relevance and leads to an interesting reentrance 
behavior. The second illustrates the interplay of superconductivity and almost degenerate
singlet and triplet two-body states in a multi-level quantum dot. Both these dot setups 
might be experimentally realized in the near future. The examples indicate that interesting 
quantum many-body physics can be expected in other more complex dot setups as well. 
      
The remainder of this review is structured as follows. Next, in Subsect. \ref{subsec:josephson} 
we give a brief account of the conventional Josephson effect 
of two superconductors coupled via a structureless barrier and in 
Subsect. \ref{subsec:spinssup} we review the effect of a localized 
magnetic moment in a metal and a superconductor.
While in the former case spin and charge fluctuations are considered (SIAM with BCS leads) in the latter 
we focus on spin fluctuations (Kondo model) only. In 
Sect. \ref{sec:minimalmodel} we present the full minimal model and its physics. 
The model is introduced in Subsect.~\ref{subsec:model} and the Josephson current
in the noninteracting limit is discussed in Subsect.~\ref{subsec:U0}.
We investigate the $U >0$ level crossing physics and the supercurrent in 
the large $\Delta$ limit in Subsect. \ref{subsec:infinitegap}. By presenting 
``nearly'' exact results obtained by NRG and a QMC approach in 
Subsect. \ref{subsec:nearlyexact} we show that the main characteristics of the large $\Delta$ 
limit survive at finite $\Delta$ and indicate the effect of $T>0$. 
The in-gap bound states are briefly discussed.
We then review 
alternative approaches which are also capable of capturing the underlying physics
qualitatively 
in Subsect. \ref{subsec:alternative} but argue that others should be avoided as they 
spoil fundamental principles. A particular focus is put in Subsect. \ref{subsec:funrg} 
on the approximate FRG approach which does not suffer from such artifacts, is rather flexible, 
and will be used to study the more complex dot setups discussed  in  
Sect. \ref{sec:complexdots}. In Sect. \ref{sec:experiments} we 
directly compare theoretical and experimental results on the Josephson current.       
Section \ref{sec:summary} summarizes our considerations.  

We reemphasize that we exclusively consider setups with two superconducting leads 
in equilibrium. Systems involving in addition to superconducting leads 
metallic ones show interesting physics as well but are beyond the scope of 
the present review. Similarly, applying a bias voltage across the dot beyond 
the linear response regime leads to interesting effects. Both extensions 
are reviewed in, e.g., Ref.~\cite{Martin11}. 

\subsection{The Josephson effect}
\label{subsec:josephson}

To describe the essence of the Josephson effect we consider the simple model 
of two BCS superconducting leads labeled by $s=L,R$ with Hamiltonian 
\begin{eqnarray}
\label{eq:Hleads} 
H_{\rm lead}^{s} =  \sum_{k,\sigma} \epsilon_{k} c_{s,k,\sigma}^\dag 
c_{s,k,\sigma}^{\phantom{\dagger}} - \Delta \left[ e^{i \phi_s} \sum_k 
c^\dagger_{s,k,\uparrow}c^\dagger_{s,-k,\downarrow} + {\rm H.c.} \right] .  
\end{eqnarray}
Here the $c_{s,k,\sigma}^\dag$ are fermionic creation operators in lead $s$, of momentum $k$, 
and spin orientation $\sigma = \uparrow,\downarrow$.
For simplicity but without loss of generality we assume that the leads
are one-dimensional. They are characterized by 
their single-particle dispersion $\epsilon_k$, their superconducting gap $\Delta$, 
and their superconducting phase $\phi_s$.
We made the reasonable 
assumption that the dispersion and the gap of the two leads are identical; they are made from 
the same material.
The two leads are coupled by a term 
\begin{eqnarray}
\label{eq:Hdirect}
H_{\rm direct} = - t_{\rm d} \sum_{\sigma} c_{L,\sigma}^\dag c_{R,\sigma}^{\phantom{\dag}} + \mbox{\rm H.c.} 
\end{eqnarray}
with direct hopping amplitude $t_{\rm d} \geq 0 $ and $c_{s,\sigma} = \sum_k  c_{s,k,\sigma} / \sqrt{N}$, 
where $N$ is the number of lead sites. Later we will consider the limit $N \to \infty$. 

As it is often the case when dealing with superconductivity it turns out to be advantageous
to work in Nambu formalism. For this we introduce the Nambu spinor 
\begin{eqnarray}
\label{eq:nambulead}
\Psi_{s,k} = \left(  \begin{array}{c} c^{\phantom{\dag}}_{s,k,\uparrow} \\  c^{\dag}_{s,-k,\downarrow} 
\end{array} \right) .
\end{eqnarray} 
The Hamiltonian can then be written as
\begin{align}  
  H & =  \sum_{s=L,R}  H_{\rm lead}^{s}  + H_{\rm direct} \nonumber \\
 & =  \sum_{s=L,R} \sum_k \left( \epsilon_k \Psi_{s,k}^\dag \sigma_3 \Psi_{s,k}^{\phantom{\dag}} - 
 \Psi_{s,k}^\dag \bar{\Delta}_s  \Psi_{s,k}^{\phantom{\dag}} \right) - t_{\rm d} \left( \Psi_{L}^\dag \sigma_3 
 \Psi_{R}^{\phantom{\dag}}  + \mbox{\rm H.c.} \right),
\label{eq:Hjo}
\end{align}   
where $\sigma_i$ denotes the $i$-th Pauli matrix and 
\begin{eqnarray}
\label{eq:Deltamatrix}
\bar{\Delta}_s= \Delta \left(  \begin{array}{cc} 0 & e^{i \phi_s} \\   e^{-i \phi_s} & 0
\end{array} \right) .
\end{eqnarray} 

The left and right current operators $\hat J_s$ are defined as the time derivative of the 
left and right particle number operators $\hat N_s$  
\begin{eqnarray}
\label{eq:current}
\hat J_s = \partial_t \hat N_s = i [H,\hat N_s] .
\end{eqnarray} 
The expectation value of the commutator of $H_{\rm lead}^{s}$ with $\hat N_s$ vanishes if 
one takes into account the self-consistent definition of the superconducting 
s-wave order parameter $\Delta e^{i \phi_s} \sim \sum_k \left< c_{s,k,\uparrow}  c_{s,-k,\downarrow}  
\right>$
of BCS theory. The relevant part of the current operator is thus given by the 
commutator of $\hat N_s$ with the second addend of Eq.~(\ref{eq:Hjo}) and reads  
\begin{eqnarray}
\label{eq:currentdirect}
\hat J_s^{\rm direct} = i t_{\rm d} \Psi_{s}^\dag \Psi_{\bar s}^{\phantom{\dag}} ,
\end{eqnarray} 
with $\bar L = R$ and vice versa.

The thermal expectation value of the current operator can thus be written in terms of the Matsubara 
lead-lead Green function ${\mathcal G}_{\bar s,s}(i \omega)$ ($2 \times 2$-matrix in Nambu 
formalism)   
\begin{eqnarray}
\label{eq:expcur}
J_s^{\rm direct}  = - \frac{2 t_{\rm d}}{\beta} \sum_{i \omega} \mbox{Im} \, 
\mbox{Tr}  \, {\mathcal G}_{\bar s,s}(i \omega) ,  
\end{eqnarray} 
with $\beta = 1/T$.
Using the standard equation-of-motion technique the latter can be expressed in terms 
of the local Green function of an isolated (semi-infinite) superconducting lead 
evaluated at the open boundary 
\begin{eqnarray}
\label{eq:leadg}
g_{\rm s}(i \omega) = - \pi \rho_{\rm lead} \frac{1}{\sqrt{\omega^2+ \Delta^2}} 
\left(  \begin{array}{cc} i \omega & - \Delta e^{i \phi_s} \\   - \Delta e^{-i \phi_s} & i \omega
\end{array} \right) 
\end{eqnarray}  
as  
\begin{eqnarray}
\label{eq:GtoG}
{\mathcal G}_{s,\bar s}(i \omega) = - \frac{t_{\rm d} g_s (i \omega) \sigma_3 
g_{\bar s}(i \omega)}{1- t_{\rm d}^2 g_s(i \omega) \sigma_3 g_{\bar s}(i \omega) \sigma_3}. 
\end{eqnarray}
Here the local density of states $\rho_{\rm lead}(\omega)= \lim_{N \to \infty} \sum_k \delta(\omega-\epsilon_k)/N$ 
in the absence of superconductivity is assumed to be frequency independent (wide band limit). 
As we are not 
interested in effects of the details of the normal state band structure we focus on this limit  
throughout the review. Without loss of generality we fix the chemical potential to $\mu=0$.   
 
With these expressions the current Eq.~(\ref{eq:expcur}) can be computed for arbitrary 
tunnel coupling $t_{\rm d}$. As a simple limit we explicitely consider the case of a poor 
tunnel coupling and expand in $t_{\rm d}$. To lowest order this leads to the current 
\begin{eqnarray}
\label{eq:lowestorder}
J_{L}^{\rm direct} = -J_{R}^{\rm direct} = 
2 \pi^2  \rho_{\rm lead}^2 t_{\rm d}^2 
\Delta \tanh{\left(\frac{\beta \Delta}{2} \right)} \sin(\phi) + {\mathcal O}\left( t_{\rm d}^4 \right), 
\end{eqnarray}      
with the well-known sinusoidal dependence on the relative phase $\phi=\phi_L-\phi_R$. 
We emphasize that the supercurrent only depends on this relative phase $\phi$. This does not only 
hold to leading order in $t_{\rm d}$ but to all orders and can be verified straightforwardly using 
Eqs.~(\ref{eq:expcur}) to (\ref{eq:GtoG}). The absence of the average phase $\eta=(\phi_L+\phi_R)/2$ 
is a manifestation of gauge-invariance; we will further elaborate on this in Subsect.~\ref{subsec:model}. 
For $t_{\rm d} \ll \!\!\!\!\!\!\! / \;  1 $, the current has a more involved $\phi$-dependence. However, for 
$\phi \neq n \pi$, $n \in {\mathbb Z}$ one finds a nonvanishing equilibrium Josephson 
current which is a signature of the Josephson effect and was first described in 
1962.\cite{Josephson62} The current is periodic with period $2 \pi$ and an odd function 
of $\phi$. 

\subsection{Magnetic impurities in metals and superconductors}
\label{subsec:spinssup}

\subsubsection{Metallic leads}

The basic physics of a magnetic impurity in a metallic environment can, e.g., be studied within 
the SIAM. Left and right metallic leads are described by the first term in Eq.~(\ref{eq:Hleads}). 
The dot is modeled by 
\begin{eqnarray}
\label{eq:Hdot}
H_{\rm dot} = \sum_{\sigma} \epsilon_\sigma d_\sigma^\dag d_\sigma^{\phantom{\dag}} + 
U \left( n_\uparrow - 1/2 \right) \left( n_\downarrow - 1/2 \right) ,
\end{eqnarray}  
with $n_\sigma = d_\sigma^\dag d_\sigma^{\phantom{\dag}}$ and $d_\sigma^\dag$ being the creation operator 
of an electron of spin $\sigma$ on the dot level. Here $\epsilon_\sigma$ denotes the level 
energy which in the presence of a Zeeman field is spin-dependent and $U$ the local 
Coulomb repulsion. 
Note that we shifted the level occupancy $n_\sigma$ such that in the absence
of a Zeeman field $\epsilon_\sigma = \epsilon = 0$ corresponds to half filling of the dot. 
In experiments the level position $\epsilon_\sigma$ can be moved by applying a voltage to a properly designed gate.
The coupling 
between the dot and the leads is given by 
\begin{eqnarray}
\label{eq:Hcoup}
H_{\rm coup}^s = - t_s \sum_\sigma \left( c_{s,\sigma}^\dag d_\sigma^{\phantom{\dag}}  + \mbox{H.c.} 
\right) . 
\end{eqnarray}   
The relevant energy scale (rate) for tunneling (charge fluctuations) 
is $\Gamma = \Gamma_L + \Gamma_R$, 
with the frequency independent $\Gamma_s = \pi \rho_{\rm lead} t_s^2$ taken in the wide band limit.   
For vanishing Zeeman field, which we will assume from now on until stated differently,  
and $\epsilon = 0$ the dot is half filled and represents 
a localized spin. However, the spin can flip due to tunneling of a particle in and out of 
the dot (spin fluctuations). This leads to an effective exchange interaction between the dot and itinerant 
spins and for sufficiently large 
$U/\Gamma$ ultimately to the screening of the localized spin  by lead electrons and the 
formation of a nonlocal (Kondo) singlet. In this limit the occupancy is furthermore pinned to 
$1/2$ in a range of level positions of order $U$ around $\epsilon=0$ extending the screening 
to this parameter regime. Both these correlation effects are aspects of the 
Kondo effect. As textbooks\cite{Hewson97} and reviews\cite{Pustilnik04} on the Kondo effect are 
available we here will be rather brief about it. 

Projecting out the charge fluctuations by a Schrieffer-Wolff transformation\cite{Hewson97}
the SIAM can be reduced to the Kondo model with an explicit exchange 
interaction between the localized quantum spin and the lead spins. It 
can alternatively be used to describe the formation of the Kondo singlet.
We note, however, that when aiming at a comprehensive understanding of the spectral and
transport properties of a single-level quantum dot coupled to metallic leads in the
entire parameter space one cannot ignore charge fluctuations. The same holds
for a dot with superconducting leads. Starting with Sect.~\ref{sec:minimalmodel} we will thus
exclusively consider the SIAM with BCS leads.

A comprehensive understanding of the Kondo effect can be obtained using 
elaborate many-body methods such as, e.g., the NRG, the Bethe 
ansatz or QMC approaches. As the Bethe ansatz solution was not extended to the
case of superconducting leads it does not play a prominent role in the present review.
Certain aspects of the Kondo effect can also be understood 
employing more elementary approximate approaches as it was done before the development 
of these advanced tools. 

\begin{figure}[t]
\begin{center}
   \includegraphics[width=.48\linewidth,clip]{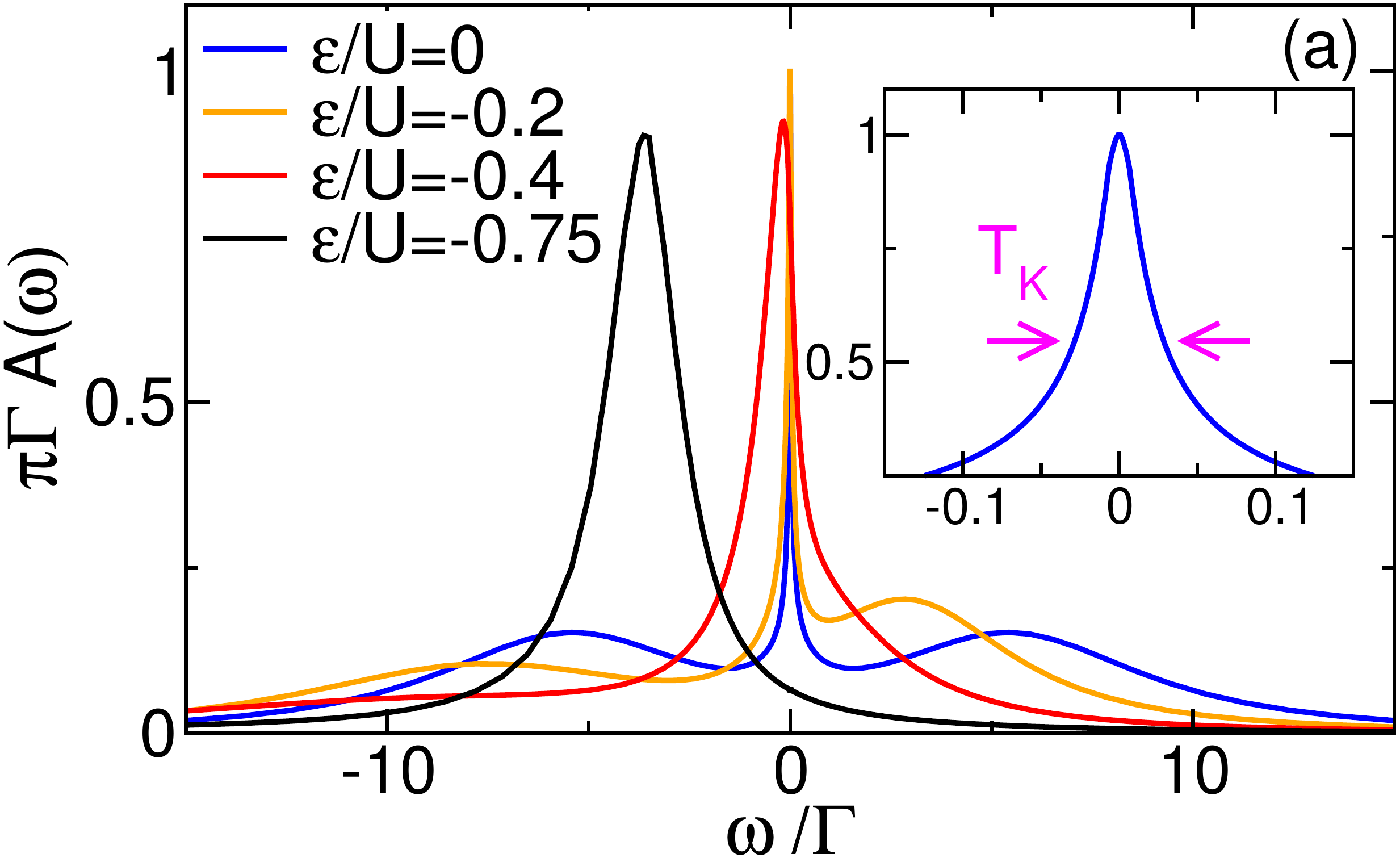}
 \includegraphics[width=.48\linewidth,clip]{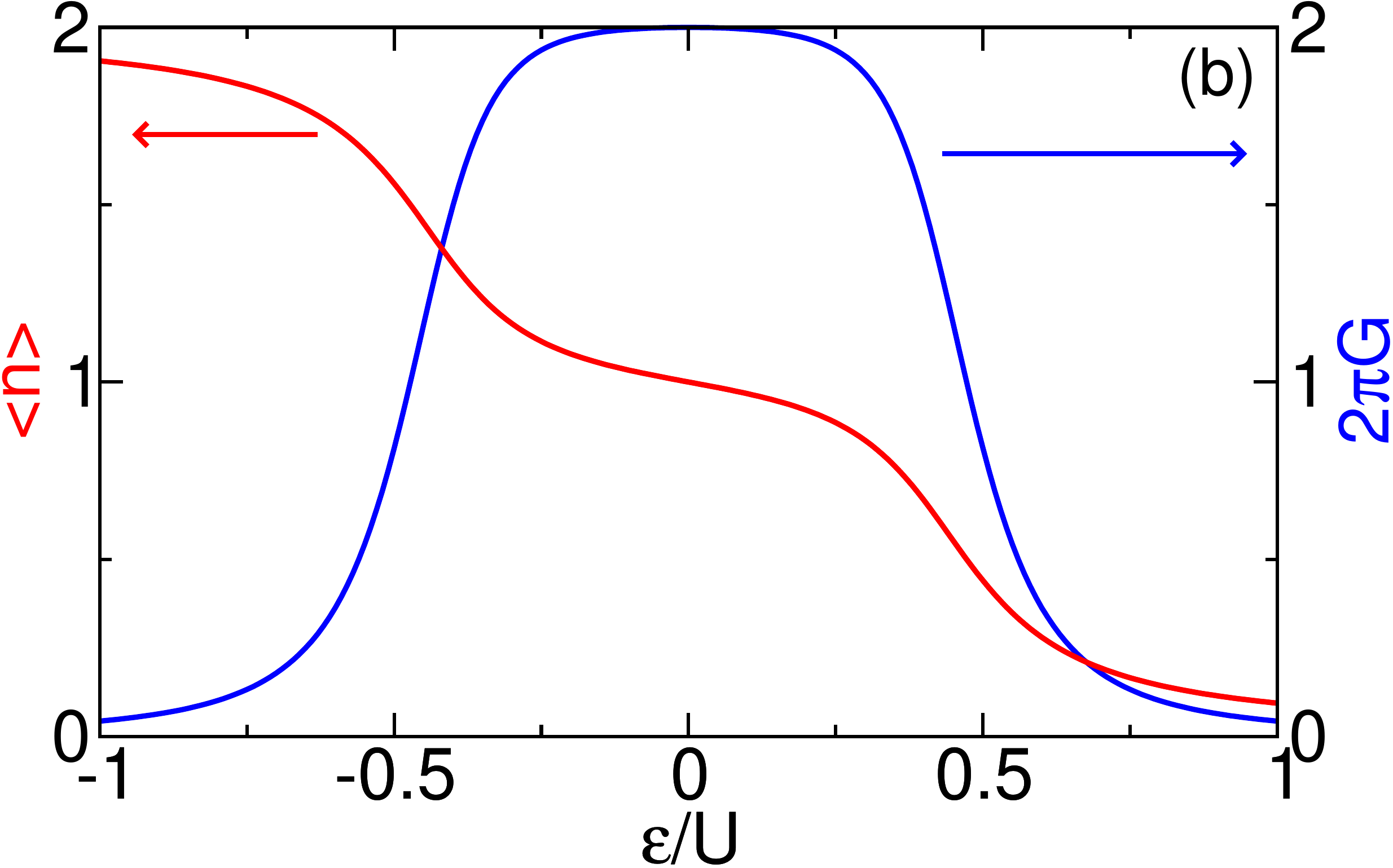}
   \caption{(a) Main plot: The single-particle dot spectral function for $U/\Gamma=4 \pi$ and different 
level positions. Inset: Zoom-in of the spectral function at $\epsilon=0$. (b) The level occupancy 
(left scale) and the linear conductance (right scale) as a function of $\epsilon$. The NRG
data were provided by T. Pruschke.}
   \label{fig1}
\end{center}
\end{figure}

For vanishing Zeeman field  
the (Matsubara) dot Green function ${\mathcal G}_\sigma(i \omega)$ is spin independent and we suppress the
spin index.
The spectral function can be computed from the Green function analytically continued
to real frequencies
\begin{eqnarray}
\label{eq:specfudef}
  A(\omega) = - \frac{1}{\pi} \mbox{Im} \, {\mathcal G}(\omega + i 0) .
\end{eqnarray}
In Fig. \ref{fig1} (a) we show the $T=0$ dot single-particle spectral function of the SIAM (with metallic 
leads) for $U/\Gamma=4 \pi$ and different $\epsilon$ obtained by 
NRG. 
We here do not give any details on the NRG approach as they can be found in 
the review \cite{Bulla08}. We merely note that it provides highly 
accurate numerical low-energy results for a variety of 
important observables of the SIAM (as well as of
other quantum impurity models) in equilibrium, provided the two numerical parameters--the logarithmic discretization 
parameter $\Lambda$ and the number of states $N_{\rm c}$ kept--are properly chosen.
The results at higher energies are not as accurate, which, however, does not play a significant
role for our purposes. Applying NRG it is also 
possible to gain direct physical insights, e.g., about the low-energy fixed point structure. 
This will turn out to be important when extracting the phase diagram for the SIAM with BCS leads.  
The Kondo effect leads to a sharp resonance of width $T_{\rm K} \ll \Gamma$--the latter being 
the width for $U=0$--which is pinned to the Fermi level (at $\omega=0$) when varying $\epsilon$;
see Fig.~\ref{fig1} (a). This Kondo temperature scales exponentially in $U/\Gamma$. 
One often refers to the analytic expression 
\begin{eqnarray}
\label{eq:T_Kana}
T_{\rm K}^{\rm ana}= \sqrt{U \Gamma/2} \exp{\left( - \frac{\pi}{8 U \Gamma} 
\left| U^2 - 4 \epsilon^2 \right|  \right)}
\end{eqnarray}   
extracted from the susceptibility of the local spin computed from the Bethe ansatz 
solution of the SIAM.\cite{Tsvelick83,Hewson97} 
However, a few words of caution are in order with respect to the use 
of this formula. (i) The prefactor of the exponential function depends 
on the observable studied. This concerns parameter dependencies as well as numerical factors. 
(ii) Even ignoring this prefactor issue Eq.~(\ref{eq:T_Kana}) should only be used deep in the
Kondo regime when the absolute value of the argument of the exponential function is much larger than
1. Importantly, the extreme
Kondo limit is not reached in most experiments; see Sect.~\ref{sec:experiments}.  
Besides the Kondo resonance 
the spectral function for $\epsilon \approx 0$ shows two Hubbard bands 
roughly located at $\pm U/2$.

At $T=0$ the linear conductance $G$ (infinitesimal bias voltage)
depends on the spectral function at the Fermi energy and is given 
by\cite{Meir92}\footnote{In units with $e=1=\hbar$ such that unitary conductance 
per spin becomes $1/(2 \pi)$.} 
\begin{equation}
\label{eq:GA} 
  G=4 \Gamma_L \Gamma_R  A(0) / \Gamma .
\end{equation}
The pinning of the spectral weight at $\omega=0$ thus leads to a plateau 
(Kondo ridge) in $G$ with width of the order of $U$ if $\epsilon$ is varied around $0$.
This is shown in Fig.~\ref{fig1} (b) (solid blue line) which in addition contains
data for the dot occupancy (solid red line). The latter follows by integrating $A(\omega)$
over frequency up to the Fermi energy at $\omega=0$. The pinning of the resonance
leads to a plateau of the dot occupancy at 1/2. 
We note that while the dot Green function and thus the dot spectral function 
(and the dot occupancy) only depends on $\Gamma= \Gamma_L + \Gamma_R$, due to the 
above prefactor $4 \Gamma_L \Gamma_R / \Gamma$ the conductance is in addition 
sensitive to the left-right asymmetry $a = \Gamma_L/\Gamma_R \neq 1$.  
The Kondo effect is destroyed by temperatures 
$T > T_{\rm K}$ or Zeeman splittings $\left|\epsilon_\uparrow-\epsilon_\downarrow\right| > T_{\rm K}$. 
Both degrade the conductance plateau and lead to a local minimum of $G(\epsilon) = 
G_\uparrow(\epsilon) + G_\downarrow(\epsilon)$ at $\epsilon=0$. We will return to this in 
Sect.~\ref{sec:experiments}.  
   
Standard techniques such as perturbation theory in $U$ or $\Gamma$, or
the mean-field approximation\cite{Bruus04} do not capture the Kondo effect in the SIAM 
(no Kondo peak of exponentially small width in $A(\omega)$, no conductance plateau in $G$). 
Within the unrestricted mean-field approach the spin-symmetry is spuriously broken in 
roughly the parameter 
regime in which the Kondo effect can be observed.\cite{Anderson61} This is a rather 
fundamental deficit of the unrestricted mean-field approximation which renders it an inappropriate tool 
to study correlated quantum dots. As discussed below the same holds if the metallic 
leads are replaced by superconducting ones.  

\subsubsection{Superconducting leads}

BCS superconductivity requires the formation of Cooper pairs of opposite spin (and momentum). 
Thus, scattering off magnetic impurities will affect superconductivity. This was 
investigated already at the end of the 50'ties and beginning of the 
60'ties in models describing bulk BCS superconducting electrons interacting with a 
single or a few localized spin-1/2 degrees of freedom ignoring charge fluctuations and 
Kondo physics (see, e.g., Ref.~\cite{Abrikosov61}). 
Shortly after Kondo's seminal work the interplay of the Kondo effect and superconductivity was 
investigated by several authors employing a Kondo model description 
of the localized spin (suppressed charge fluctuation).\cite{Liu65,Griffin65,Maki66} 
It was shown that the superconducting gap acquires the logarithmic temperature dependence 
as described by Kondo for the resistance of a metallic host.\cite{Kondo64} Note
that at that time neither NRG and QMC nor the Bethe ansatz solution were available which
limited the insights which could be gained. Furthermore, the Kondo model with a BCS reservoir
was considered in its own right but not derived from the SIAM. In fact, only later it became
apparent that a generalized Schrieffer-Wolff transformation leads to additional terms not present
for a metallic reservoir.\cite{Salomaa88}

In parallel the effect of a classical spin in a superconductor was studied.\cite{Yu65,Shiba68,Rusinov68} 
In such systems a pair of bound states with energies located in 
the superconducting gap forms which is nowadays 
referred to as Yu-Shiba-Rusinov states. The appearance of a pair of in-gap states 
at energies symmetrically located around the Fermi energy was also investigated for quantum spins, 
i.e., within the Kondo model with BCS leads.\cite{Maki66,Soda67,Mueller73} It was found 
that the energies of these bound states move when varying the ratio $T_{\rm K}/\Delta$ and that 
their nature changes from being a singlet for $T_{\rm K}/\Delta \ll 1$ to being 
a doublet for $T_{\rm K}/\Delta \gg 1$.\cite{Matsuura77}\footnote{The starting point
  of this paper is the SIAM with BCS leads. However, later an approximation is
introduced which reduces this model to the Kondo model with BCS leads.} This provided a first 
indication that a ground state level crossing (first order quantum phase transition) 
might occur for $T_{\rm K}/\Delta \approx 1$ between a doublet for  $T_{\rm K}/\Delta \ll 1$ 
and a singlet for $T_{\rm K}/\Delta \gg 1$. At the transition the first excited in-gap state becomes 
the ground state and vice versa, that is the bound states reach the Fermi energy at this point. 
However, due to restrictions of the applicability of the methods used in the two limits
of very large and very small $T_{\rm K}/\Delta$ no comprehensive understanding was achieved at 
that time. The results are in accordance 
with the physical picture that for $T_{\rm K}/\Delta \ll 1$ superconductivity prevails, 
the Kondo singlet cannot develop and the spin becomes free (doubly degenerate). In contrast for 
$T_{\rm K}/\Delta \gg 1$ the Kondo effect prevails and the spin is screened to form a 
(modified) Kondo singlet. This level crossing scenario and the associated change of the ground 
state degeneracy was unambiguously confirmed only 
two decades later when NRG was first applied to the Kondo model with BCS leads in 
1992.\cite{Satori92,Sakai93} In none of the above works a finite phase difference 
$\phi$ of the superconducting order parameter was considered. In bulk systems with 
embedded spin-1/2 degrees freedom the notion of a phase difference across the
impurity is meaningless.   

In the expressions of the last paragraph $T_{\rm K}$ refers to the Kondo temperature as it would 
emerge in the system after setting $\Delta=0$; we will refer to it as the normal state $T_{\rm K}$. 
For finite $\Delta$, and in particular for $\Delta$ larger than the normal state $T_{\rm K}$ one 
cannot expect that an inherent Kondo scale emerges; the Kondo effect does not (fully) develop. 
This indicates that in the interesting transition region from Kondo to superconductivity 
dominated behavior the ratio $T_{\rm K}/\Delta$ might not be the relevant parameter. The assumption
that this ratio is the only relevant variable to characterize observables (``Kondo universality'') 
becomes even more questionable if the SIAM (including charge fluctuations) instead of the Kondo 
model is used; see Sect.~\ref{sec:minimalmodel}.       

The tunneling of Cooper pairs leads to the Josephson current. Magnetic impurities 
(localized spin-1/2 degrees of freedom) in the barrier separating the two superconductors will 
thus alter this equilibrium current. A reduction of the current and potentially even a sign 
change  was predicted ignoring the Kondo effect by Kulik in 1966.\cite{Kulik66a,Kulik66b} 
The supercurrent in the presence of Kondo's logarithmic corrections to the scattering 
was computed shortly after.\cite{Shiba69}
Further indications of possible $\pi$-junction behavior were given in Ref.~\cite{Bulaevskii77}. 
The strong effect of the level-crossing transition 
discussed in the second to last paragraph on the Josephson current was only studied much later and will 
be one of the main topics of the rest of this review.
  
As mentioned above to give a full account of the physics of a correlated single-level 
quantum dot with two superconducting leads and to be in a position to 
compare to recent experiments we need to include charge fluctuations and thus to 
study the SIAM with BCS leads as a minimal model. In the next section we discuss 
the physics of this model and briefly account for the historic development (as we just did 
for the Kondo model with superconducting leads) in Subsects.~\ref{subsec:nearlyexact} 
and \ref{subsec:alternative}.  

\section{A minimal model and its physics}
\label{sec:minimalmodel}

\subsection{The model}
\label{subsec:model}

The Hamiltonian of a minimal model for transport through a single-level dot with two BCS leads 
including charge fluctuations can be obtained by adding the terms of 
Eqs.~(\ref{eq:Hleads}), (\ref{eq:Hdot}), and (\ref{eq:Hcoup}). The direct hopping 
between the superconducting leads Eq.~(\ref{eq:Hdirect}) is considered in addition 
in Subsect.~\ref{subsec:fano}.  

In a first step we also rewrite $H_{\rm dot}$ and $H_{\rm coup}^s$ in terms of the Nambu spinor
Eq.~(\ref{eq:nambulead}) and 
\begin{eqnarray}
\label{eq:nambudot}
\varphi = \left(  \begin{array}{c} d^{\phantom{\dag}}_{\uparrow} \\  d^{\dag}_{\downarrow} 
\end{array} \right) 
\end{eqnarray}   
as
\begin{eqnarray}
\label{eq:nambuHdot}
H_{\rm dot} = \epsilon \varphi^\dag \sigma_3 \varphi^{\phantom{\dag}} 
- U \left( \varphi_1^\dag \varphi_1^{\phantom{\dag}} -1/2 \right)
\left( \varphi_2^\dag \varphi_2^{\phantom{\dag}} -1/2 \right)
\end{eqnarray}
and 
\begin{eqnarray}
\label{eq:nambuHcoup}
H_{\rm coup}^s = 
-t_s \Psi^\dag_s \sigma_3 \varphi^{\phantom{\dag}} + \mbox{H.c.} , 
\end{eqnarray}
still assuming a spin-degenerate level $\epsilon_\uparrow=\epsilon=\epsilon_\downarrow$. 
Using the equation-of-motion technique we can compute the $U=0$ (Nambu space Matsubara) 
dot Green function including the lead self-energy
\begin{eqnarray}
\label{eq:G0}
{\mathcal G}_0(i \omega) =  \frac{-1}{D_0(i \omega)} \left(  
\begin{array}{cc} i \tilde \omega + \epsilon & - \tilde \Delta \\  - \tilde \Delta^\ast & i \tilde 
\omega - \epsilon
\end{array} \right) , \quad D_0(i \omega) = {\tilde \omega}^2 + \epsilon^2 + 
\left| \tilde \Delta \right|^2 ,
\end{eqnarray} 
with 
\begin{eqnarray}
\label{eq:defsforG0}
i \tilde \omega = i \omega \left(1 + \frac{\Gamma}{\sqrt{\omega^2 + \Delta^2}} \right), \quad 
\tilde \Delta = \Delta \sum_s \frac{\Gamma_s}{\sqrt{\omega^2 + \Delta^2}} e^{i \phi_s} .
\end{eqnarray}  
The anomalous off-diagonal terms of the dot Green function are signatures of the proximity 
effect.  
The full interacting dot Green function is obtained as
\begin{eqnarray}
\label{eq:Gfull}
{\mathcal G}(i \omega) =  \frac{-1}{D(i \omega)} \left(  
\begin{array}{cc} i \tilde \omega + \epsilon + \Sigma^\ast(i \omega)& - 
\tilde \Delta + \Sigma_\Delta(i \omega) \\  - \tilde \Delta^\ast+ \Sigma_\Delta^\ast(i \omega) 
& i \tilde \omega - \epsilon- \Sigma(i \omega)
\end{array} \right),  
\end{eqnarray} 
with
\begin{eqnarray}
\label{eq:Dfull}
D(i \omega) = \tilde \omega^2 + \left| \epsilon + \Sigma(i \omega) \right|^2 +
\left| \tilde \Delta - \Sigma_\Delta(i \omega) \right|^2 ,
\end{eqnarray} 
were a form of the self-energy matrix $\mathcal S$ (resulting from the local interaction) was used which 
obeys all symmetries (including those which follow from spin-symmetry)\cite{Karrasch10,Zonda16}
\begin{eqnarray}
\label{eq:sigmasym}
\mathcal{S}(i\omega) = \left( \begin{array}{cc} \Sigma(i \omega)& \Sigma_\Delta(i \omega) \\ 
\Sigma_\Delta^\ast(i \omega) 
& - \Sigma^\ast (i \omega)
\end{array} \right). 
\end{eqnarray}

The relevant part of the current operator is now obtained
from the commutator $\left[ H_{\rm coup}^s,\hat N_s\right]$ and will be denoted by 
$\hat J_s^{\rm imp}$. It reads
\begin{eqnarray}
\label{eq:Jsimp}
\hat J_s^{\rm imp} = - i t_s \varphi^\dag_{\phantom{s}} \Psi_s^{\phantom{\dag}} + \mbox{H.c.} . 
\end{eqnarray}
Its thermal expectation value can thus be expressed in terms of the lead-dot Green function 
${\mathcal G}_{s,{\rm d}}(i \omega)$ as
\begin{equation}
  \label{eq:Jsimpexp1}
J_s^{\rm imp} = \frac{2 t_s}{\beta} 
\sum_{i \omega}   \mbox{Im} \, 
\mbox{Tr}  \, {\mathcal G}_{s,{\rm d}}(i \omega) .
\end{equation}
The contribution to the current from the commutator 
of $H_{\rm lead}^s$ with $\hat N_s$ again vanishes due to the BCS self-consistency condition  
(see Subsect.~\ref{subsec:josephson}). 
Employing the equation-of-motion approach the lead-dot Green function can be written in terms 
of the full dot Green function as 
\begin{eqnarray}
\label{eq:GtoG1}
{\mathcal G}_{s,{\rm d}}(i \omega) = - t_s g_s(i \omega) \sigma_3 {\mathcal G}(i \omega) .
\end{eqnarray} 
In turn the current can be computed as  
\begin{eqnarray}
\label{eq:Jsimpexp}
J_s^{\rm imp}  = \frac{-2 t_s^2}{\beta} 
\sum_{i \omega}   \mbox{Im} \, 
\mbox{Tr}  \, \left[g_s(i \omega) \sigma_3 {\mathcal G}(i \omega)\right] .  
\end{eqnarray} 
Inserting Eqs.~(\ref{eq:Gfull}) and (\ref{eq:leadg}) this leads to an explicit expression 
for the Josephson current which involves the self-energy 
\begin{eqnarray}
\label{eq:Jsimpexpexp}
\hspace{-1.7cm} J_s^{\rm imp}  = 
\frac{4}{\beta} \sum_{i \omega} \left\{  
\frac{\Gamma_s \Gamma_{\bar s} \Delta^2}{\omega^2+\Delta^2} 
\frac{\sin\left(\phi_s - \phi_{\bar s}\right)}{D(i\omega)} 
- \frac{\Gamma_s \Delta \left[ e^{i \phi_s} \Sigma_\Delta^\ast(i \omega) -  e^{-i \phi_s} 
\Sigma_\Delta^{\phantom{\ast}}(i\omega) \right]}{2 i D(i \omega) \sqrt{\omega^2 + \Delta^2}}
  \right\} .
\end{eqnarray} 
The diagonal entry $\Sigma(i\omega)$ of the self-energy matrix enters only via
$D(i \omega)$ Eq.~(\ref{eq:Dfull}).   

We merely note that within an exact treatment of the model the supercurrent
is conserved: $ J_L^{\rm imp}   = -  J_R^{\rm imp} $. In case 
the self-energy is computed approximately it depends on the approximation scheme 
whether or not current conservation holds. For a discussion of this for the Anderson-Josephson dot, see, 
e.g., Ref.~\cite{Zonda16}. All results shown in this review were obtained by methods
which obey current conservation and we will only consider 
$J_L^{\rm imp} $ from now on.\footnote{In the truncated approximate FRG 
scheme of Subsect.~\ref{subsec:funrg} at least for the case $\Delta_L = 
\Delta_R$ considered here.\cite{Karrasch08}}     

The current can equivalently be computed as twice the derivative 
of the free energy with respect to the relative phase $\phi$. We show this 
for the Hamiltonian 
$H(\phi_L,\phi_R)=H_{\rm lead}^L(\phi_L) +H_{\rm lead}^R(\phi_R) + H_{\rm dot} + H_{\rm coup}^L + H_{\rm coup}^R + 
H_{\rm direct}$ Eqs.~(\ref{eq:Hleads}), (\ref{eq:Hdot}), (\ref{eq:Hcoup}), and (\ref{eq:Hdirect}) 
including a direct hopping between the two superconductors given by $H_{\rm direct}$; see Subsect.~\ref{subsec:fano}. 
To this end we perform a gauge transformation 
\begin{eqnarray}
\label{eq:gaugetrafo}
c_{s,k,\sigma} \to e^{-i \phi_s /2} c_{s,k,\sigma}, \quad d_\sigma \to e^{-i \phi_R /2} d_\sigma 
\end{eqnarray}
after which $H$ can be rewritten as 
\begin{align}
\label{eq:Htrafo}
\bar H(\phi) = & H_{\rm lead}^L(\phi_L=0) +H_{\rm lead}^R(\phi_R=0) + H_{\rm dot} \nonumber \\
& + H_{\rm coup}^L(t_L \to \bar t_L) 
+ H_{\rm coup}^R + 
H_{\rm direct} (t_{\rm d} \to \bar t_{\rm d}),
\end{align}
with $\bar t_L = e^{-i \phi/2} t_L$ and $\bar t_{\rm d} = e^{-i \phi/2} t_{\rm d}$ in self-explaining notation. 
The sum of the two current operators Eqs.~(\ref{eq:currentdirect}) and (\ref{eq:Jsimp}) (with $s=L$) can then 
be obtained by taking the derivative 
\begin{eqnarray}
\label{eq:curener}
\hat J_L = 2 \partial_\phi \left[  H_{\rm coup}^L(t_L \to \bar t_L) 
+ H_{\rm direct} (t_{\rm d} \to \bar t_{\rm d}) \right]
=  2 \partial_\phi \bar H(\phi) .
\end{eqnarray}
Exploiting the Hellmann-Feynman theorem this leads to 
\begin{eqnarray}
\label{eq:curenerexp}
J_L  =  2 \partial_\phi \Omega(\phi) ,
\end{eqnarray}
with the free energy $\Omega$. Obviously, this relation also holds if either the coupling via the dot or via
the direct link are set to zero; for now we will again focus on the latter case. 
These considerations also show explicitely that the supercurrent is only a function 
of the relative phase $\phi=\phi_L - \phi_R$; the average phase 
$\eta=(\phi_L+\phi_R)/2$ does not enter, a property which is rooted in the gauge invariance 
of the current.   

In the past it was generally believed that the case of left-right symmetric couplings with 
$a = \Gamma_L / \Gamma_R =1$ does not contain all informations to fully understand the more general 
asymmetric one with $a \neq 1$. However, in a recent important work Kadlecov\'a, 
\v{Z}onda, and Novotn\'y showed that typical observables such as, e.g., the free 
energy, the dot occupancy, and the current of a general asymmetric system can be computed from 
the corresponding expressions obtained for the symmetric case.\cite{Kadlecova17} For the current 
at given asymmetry $a$ and phase difference $\phi$ the exact transformation reads
\begin{align}
\label{eq:cursym}
J_L^{\rm imp} (\phi) = & \frac{\cos(\phi/2)}{\sqrt{\frac{(a+1)^2}{4a}-\sin^2(\phi/2)}} \nonumber \\
& \times
J_L^{\rm imp,sym} \left( 2 \arccos \sqrt{1- \frac{4a}{(a+1)^2} \sin^2[\phi/2]}  \right) ,
\end{align}   
with $J_L^{\rm imp,sym} (\psi)$ being the expression for the supercurrent at relative phase $\psi$ 
obtained in the symmetric case $a=1$. 

This has practical implications. In experimental 
setups it is nearly impossible to reach $a=1$. For a comparison still, only the current
for $a=1$ has 
to be computed which can then be transformed according to the experimental asymmetry; the parameter 
space is effectively reduced by one dimension. However, the above insight also has fundamental consequences. 
As discussed in Subsect.~\ref{subsec:spinssup} in the past it was often presumed that 
observables only depend on the ratio $T_{\rm K}/\Delta$ provided the 
system parameters for suppressed superconductivity are chosen such that the dot is in the Kondo 
regime. In systems with metallic leads a similar exclusive dependence on $T_{\rm K}/S$, with $S$ 
being an energy scale of the system (e.g.~the temperature), is referred to as  ``Kondo universality''. 
The supposed scaling in $T_{\rm K}/\Delta$ was used in theoretical as well as in experimental 
investigations of dots with superconducting leads. While according to Eq.~(\ref{eq:cursym}) the
supercurrent changes with varying asymmetry $a$ at fixed $\Gamma = \Gamma_L+ \Gamma_R$  
the ratio $T_{\rm K}/\Delta$ remains invariant, as $T_{\rm K}$ only depends on $\Gamma$; 
see Eq.~(\ref{eq:T_Kana}). The same holds for other observables. 
This provides a more clear-cut argument that this type of ``Kondo universality''
is of minor relevance for the Anderson-Josephson quantum dot
than the somewhat vague one given in Subsect.~\ref{subsec:spinssup}.

\subsection{The $U=0$ Josephson current}
\label{subsec:U0}

For $U=0$ the self-energy $\mathcal S$ vanishes. For $T=0$ and $\Gamma_L = \Gamma/2 = \Gamma_R$ 
(without loss of generality, see the second to last paragraph of Subsect.~\ref{subsec:model})  
the left current can then be written as
\begin{align}
\hspace{-1.7cm} J_L^{\rm imp}  & =  
\frac{\sin(\phi)}{2 \pi} \int_{-\infty}^\infty 
d \omega \frac{\Gamma^2 \Delta^2}{\Gamma^2 \Delta^2 \cos^2(\phi/2) + 
\omega^2 \left( \Gamma + \sqrt{\omega^2 + \Delta^2}   \right)^2 + \epsilon^2 \left( 
\omega^2 +\Delta^2 \right) } \nonumber \\
& =  \frac{\Delta \sin(\phi)}{2 \pi} \int_{-\infty}^\infty 
d x \frac{1}{\cos^2(\phi/2) + 
x^2 \left( 1 + \frac{\Delta}{\Gamma}\sqrt{x^2 + 1}   \right)^2 + 
\left(\frac{\epsilon}{\Gamma}\right)^2 \left( 
x^2 +1 \right) }.
 \label{eq:Jsimpexp0}
\end{align} 
The supercurrent is a $2 \pi$-periodic odd function of $\phi$. This also holds for $U>0$, 
which can, e.g., be inferred from Eq.~(\ref{eq:curenerexp}) 
and implies that later on we can restrict our attention to $\phi \in [0 , \pi]$.   

\begin{figure}[t]
\begin{center}
   \includegraphics[width=.6\linewidth,clip]{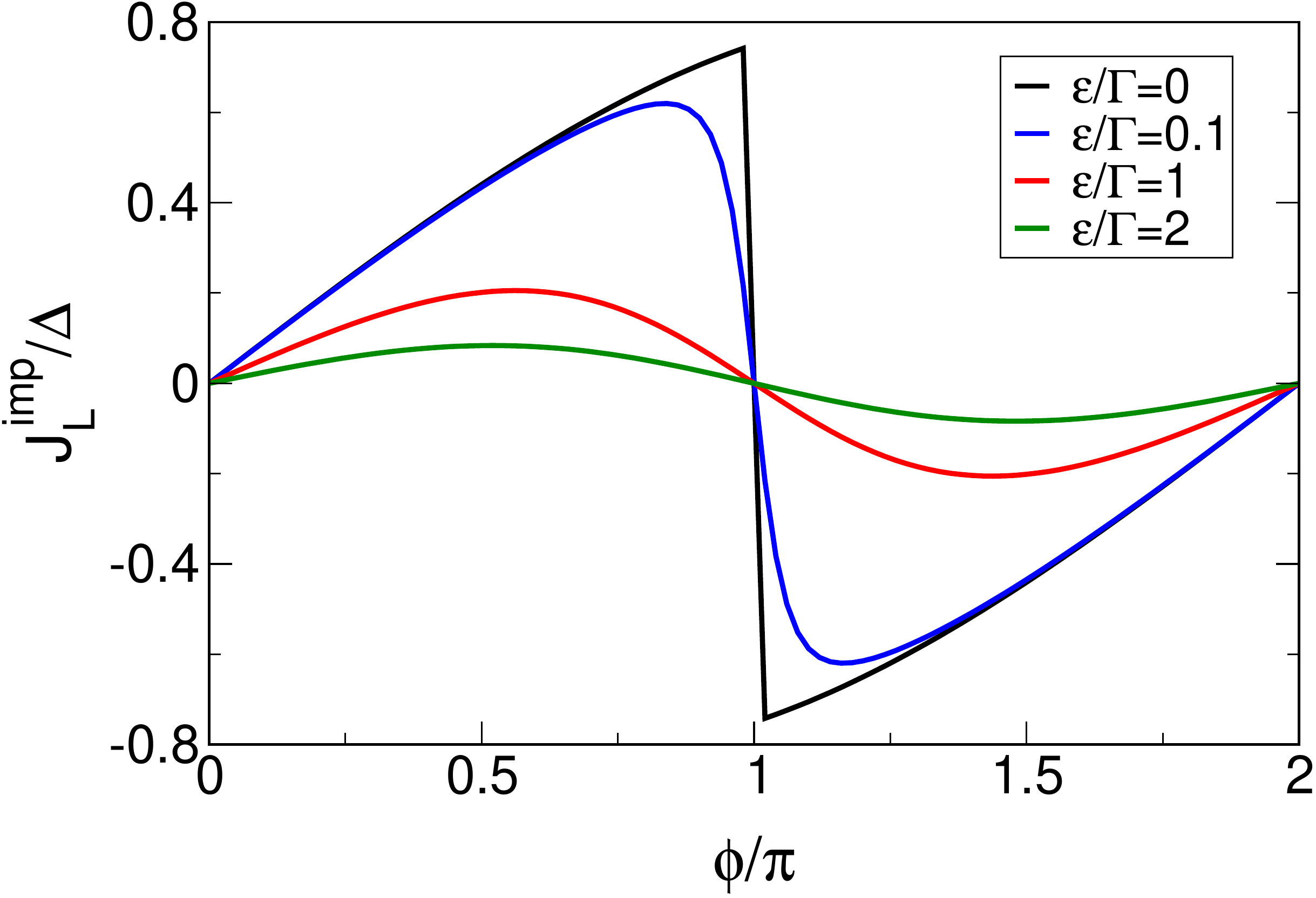}
   \caption{Noninteracting ($U=0$) zero-temperature Josephson current $J_L^{\rm imp}$ as a function 
of the phase difference $\phi$ Eq.~(\ref{eq:Jsimpexp0}) for $\Delta/\Gamma=1/3$ and 
different $\epsilon$.}
\label{fig2}
\end{center}
\end{figure}

For $|\epsilon|/\Gamma \gg 1$ all but the last term in the denominator of the integrand 
can be neglected. The integral can then be performed leading to 
 \begin{eqnarray}
\label{eq:J0largeeps}
 J_L^{\rm imp} = \Delta
\frac{\sin(\phi)}{2} \left(\frac{\Gamma}{\epsilon}\right)^2 , \quad \frac{|\epsilon|}{\Gamma} \gg 1. 
\end{eqnarray}   
As expected the CPR for a noninteracting dot with a large onsite energy becomes 
purely sinusoidal as it is the case for the current through a weak link connecting 
two superconductors; compare Eq.~(\ref{eq:lowestorder}).   
For small $|\epsilon|/\Gamma$, i.e., close to the transport resonance, the internal degree of 
freedom of the dot matters and higher harmonics affect the CPR. This can analytically be 
seen in the limit of either 
$\Delta/\Gamma \ll 1$ or $\Delta/\Gamma \gg 1$ in which
\begin{eqnarray}
\label{eq:J0limits}
J_L^{\rm imp}  = 
\frac{\sin(\phi)}{2 \sqrt{\left(\frac{\epsilon}{\Gamma} \right)^2 + \cos^2(\phi/2)}} 
\times \left\{ \begin{array}{l} \Delta \quad \mbox{for} \, \frac{\Delta}{\Gamma} \ll 1, 
\frac{|\epsilon|}{\Gamma} \ll 1 \\
\Gamma \quad \, \mbox{for} \, \frac{\Delta}{\Gamma} \gg 1, 
\frac{|\epsilon|}{\Gamma} \ll 1 \end{array}
\right. 
\end{eqnarray}   
For $\epsilon=0$ the Josephson current is proportional to $\pm \sin(\phi/2)$ with a sign change 
from $+$ to $-$ at $\phi=\pi$. For small $|\epsilon|/\Gamma$ the resulting jump at $\phi=\pi$ 
is smeared out as shown in Fig.~\ref{fig2} which displays $J_L^{\rm imp}/\Delta$ of Eq.~(\ref{eq:Jsimpexp0}) 
as a function of $\phi$ for various $\epsilon/\Gamma$. We took $\Delta/\Gamma=1/3$ which is 
roughly the value found in the experimental setups discussed in Sect.~\ref{sec:experiments}.
On the scale of the plot the data for $\epsilon/\Gamma=2$ (green curve) are already indistinguishable
from the large $\epsilon$ result Eq.~(\ref{eq:J0largeeps}).        

After analytic continuation from Matsubara to real frequencies the noninteracting Green 
function Eq.~(\ref{eq:G0}) has a pair of poles (zeros of the determinant $D_0$) 
located symmetrically around zero at energies inside the interval $[- \Delta, \Delta]$. 
They correspond to many-body states which are commonly referred to as Andreev bound states and  
indicated by in-gap $\delta$-peaks in the single-particle spectral function.
Employing the Lehmann representation of the lesser (photoemission) and greater (inverse photoemission)
spectral function it is obvious that the absolut value of the bound state energy corresponds
to the energy difference between the many-body ground state and first excited one. 
Although the 
spectral properties are not at the focus of our attention it 
is still interesting to investigate if these bound states survive for $U>0$ and if so, how they 
are related to the Yu-Shiba-Rusinov states obtained in spin models.\cite{Yoshioka00} We therefore 
comment on the bound states when appropriate. 

\subsection{The infinite gap limit: exact solution}
\label{subsec:infinitegap}

To gain insights into the physics for $U>0$ we first consider the limit 
$\Delta \to \infty$ (atomic limit). 
As indicated in the last subsection this is not the limit as realized in systems 
of experimental interest. However, it allows us to obtain the exact solution 
analytically.
We note that by first discussing the $\Delta \to \infty$ limit of the $U>0$ SIAM with
BCS leads we do not follow the historic development. In fact, this limit was considered
at a surprisingly late stage.\cite{Rozhkov00,Tanaka07} We here consider the general case with
asymmetry $a \neq 1$ as this does not cause 
any additional difficulties.
 
For $\Delta \to \infty$ the noninteracting dot Green function simplifies to 
[see Eqs.~(\ref{eq:G0}) and (\ref{eq:defsforG0})]   
\begin{eqnarray}
\label{eq:G0at}
{\mathcal G}_0^{-1}(i \omega) =  \left(  
\begin{array}{cc} i \omega - \epsilon & \Delta_{\rm d} \\  \Delta^\ast_{\rm d} & i  
\omega + \epsilon
\end{array} \right) , \quad \Delta_{\rm d} = \sum_s \Gamma_s e^{i \phi_s} .
\end{eqnarray}  
It is now obvious that all system properties which can be studied based on 
$\mathcal G$, that is ${\mathcal G}_0$ via a perturbative expansion, can
be computed considering the effective Hamiltonian
\begin{eqnarray}
\label{eq:Hat}
\hspace{-2.2cm} H_{\rm atom} = \epsilon \left(\varphi^\dag_1 \varphi^{\phantom{\dag}}_1 - 
\varphi^\dag_2 \varphi^{\phantom{\dag}}_2 \right)
- \Delta_{\rm d} \varphi^\dag_1 \varphi^{\phantom{\dag}}_2 
-\Delta_{\rm d}^\ast \varphi^\dag_2 \varphi^{\phantom{\dag}}_1 
- U \left( \varphi_1^\dag \varphi_1^{\phantom{\dag}} -1/2 \right)
\left( \varphi_2^\dag \varphi_2^{\phantom{\dag}} -1/2 \right) .
\end{eqnarray}
This includes the current and the eigenenergies as well as their degeneracy.
In the many-particle basis $\left\{ \left| 0,0 \right>,\left| 1,0 \right>, \left| 0,1 \right>, 
\left| 1,1 \right>\right\}$ (in self-explaining notation) it is represented by the matrix 
\begin{eqnarray}
\label{eq:Hatmat}
{\mathcal H}_{\rm atom} = 
\left(  
\begin{array}{cccc} 
-U/4 & 0 & 0 & 0 \\
0 & U/4+ \epsilon & - \Delta_{\rm d} & 0 \\
0 & - \Delta_{\rm d}^\ast & U/4 - \epsilon & 0 \\
0 & 0 & 0 & -U/4
\end{array} \right)
\end{eqnarray} 
The single-particle $2 \times 2$-block can easily be diagonalized leading to 
the eigenvalues $U/4 \pm \sqrt{\epsilon^2 + |\Delta_{\rm d}|^2}$. The ground state 
is thus nondegenerate (doubly-degenerate) if 
\begin{eqnarray}
\label{eq:condition}
U/4 - \sqrt{\epsilon^2 + |\Delta_{\rm d}|^2} \lessgtr -U/4 \quad \Leftrightarrow \quad U/2  \lessgtr   
\sqrt{\epsilon^2 + |\Delta_{\rm d}|^2} . 
\end{eqnarray}  
with 
\begin{eqnarray}
\label{eq:Deltadquad}
|\Delta_{\rm d}|^2 = 
\Gamma_L^2 + \Gamma_R^2 + 2 \Gamma_l \Gamma_R \cos \phi .
\end{eqnarray} 
The opposite holds for the first excited state. 
One can conclude that a ($T=0$) level-crossing quantum phase transition between a singlet and a doublet 
ground state can be driven by varying $U$, $\epsilon$, $\Gamma_s$ or $\phi$ if the other parameters
are taken from appropriate ranges. For properly chosen fixed $U$, $\epsilon$, 
$\Gamma= \Gamma_L + \Gamma_R$, and $a=\Gamma_L/\Gamma_R$ Eq.~(\ref{eq:condition}) defines a 
critical phase difference $\phi_{\rm c} \in [0, \pi]$ at which the transition from the singlet to 
the doublet state takes place. Alternatively, a critical interaction $U_{\rm c}$, a critical level position 
$\epsilon_{\rm c}$, a critical tunneling rate $\Gamma_{\rm c}$, or a critical 
asymmetry $a_{\rm c}$ can be defined.     

\begin{figure}[t]
\centering
\includegraphics[width=0.45\linewidth,clip]{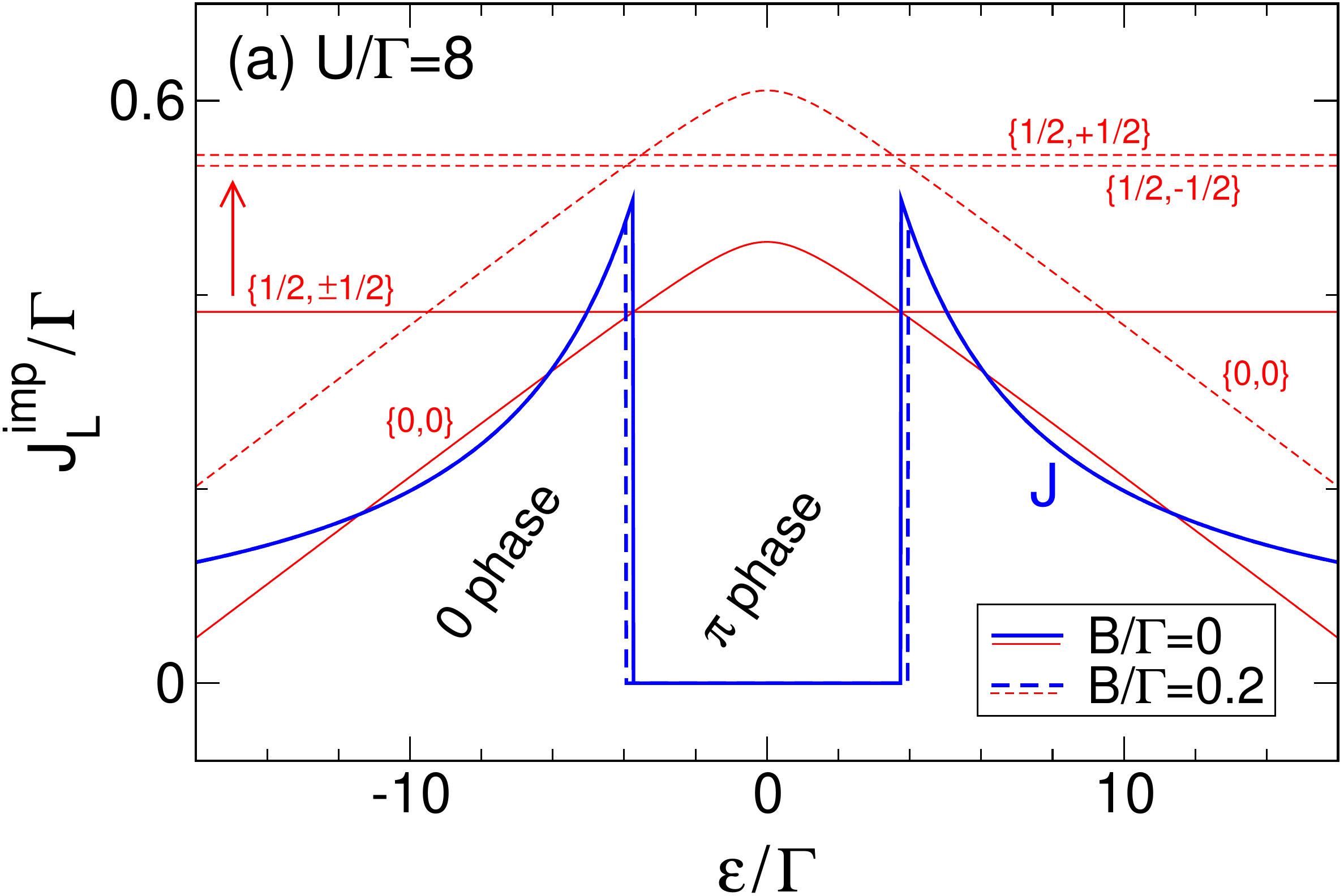}\hspace*{0.05\linewidth}
\includegraphics[width=0.45\linewidth,clip]{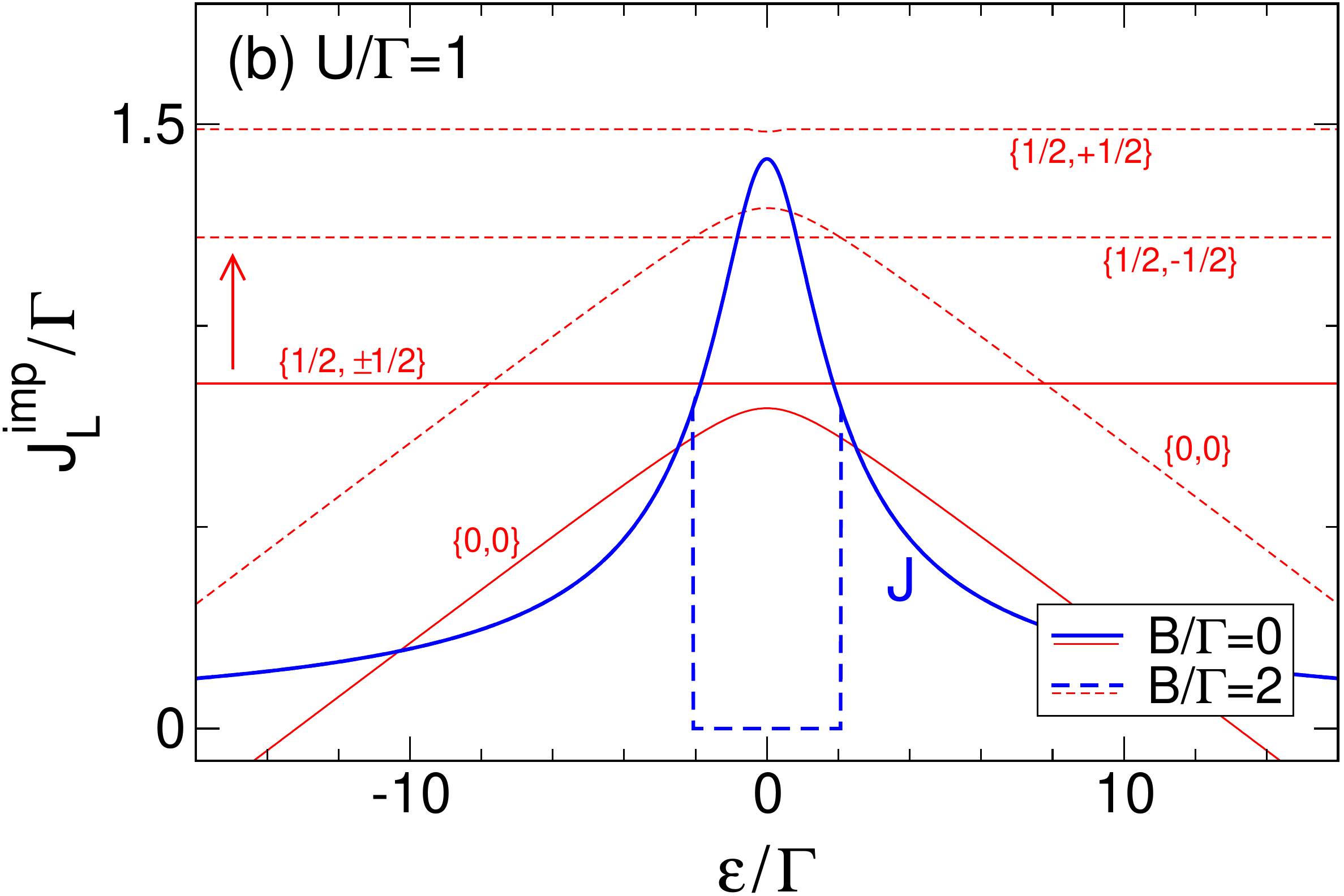}
\caption {Gate voltage $\epsilon$ dependence of the Josephson current (blue lines) as well as of the
  three lowest many-particle energies (red lines) in the large-gap limit $\Delta \to \infty$ at phase
  difference $\phi/\pi=1/2$, different Coulomb interactions $U$, and Zeeman fields $B$. The many-particle
  energies are shown in arbitrary units; those for finite $B$ (dashed lines) were shifted upwards for clarity as indicated by
  the arrow. The corresponding eigenstates are characterized by spin quantum numbers $\{s,m\}$.  Reprinted figure 
     with permission from C. Karrasch, S. Andergassen, and V. Meden, Phys. Rev. B
     {\bf 84}, 134512 (2011). Copyright (2011) by the American Physical Society.}
\label{fig2a}
\end{figure}

To further illustrate this it is useful to compute the spin quantum numbers $\{ s,m\}$ of the
states.
The spin operator is defined as $\hat{\vec s} = \sum_{\sigma,\sigma'} d_{\sigma}^\dag \vec \sigma_{\sigma,\sigma`}
d_{\sigma'}$, with $\vec \sigma$ being the vector of Pauli matrices. It is straightforward to show 
that $\hat{\vec s}^2$ and $\hat s_3$ commute with each other as well as with $H_{\rm atom}$.  
In Fig.~\ref{fig2a} the $\epsilon$ dependence of the energy of the three lowest lying
many-body states is shown for $a=1$, $\phi=\pi/2$ and two different $U/\Gamma$. The solid red
lines are labeled by their spin quantum numbers. In Fig.~\ref{fig2a} (a) $U/\Gamma$ is sufficiently
large such that a level crossing from the  $\{ s=0,m=0\}$ singlet to the
$\{ s=1/2,m=\pm 1/2\}$ doublet occurs around $\epsilon/\Gamma = \pm 4$ , while for
the smaller $U/\Gamma$ in (b), the singlet is the ground state for all $\epsilon$.  

One can straightforwardly extend the exact solution to the case with a Zeeman field of
amplitude $B$ and level energies $\epsilon_\uparrow = \epsilon+B$, $\epsilon_\downarrow =
\epsilon - B$ by replacing $\epsilon$ in the $(1,1)$-matrix element of Eq.~(\ref{eq:G0at})
by $\epsilon + B$ as well as $\epsilon \to \epsilon-B$ in the $(2,2)$-matrix element and
accordingly in the steps leading to the equation for the phase boundary. The levels
for this case are shown as dashed red lines in Fig.~\ref{fig2a}. This shows that
even for small $U/\Gamma$ a level crossing phase transition can be induced by a sufficiently
large Zeeman field; see Fig.~\ref{fig2a} (b).  However, the transition
is no longer one between a nondegenerate and a (almost) doubly-degenerate state.
For small enough $B$ at large $U/\Gamma$ [dashed lines in Fig.~\ref{fig2a} (a)] the
physics is still determined by the interplay of one nondegenerate $\{ s=0,m=0\}$
state and a pair of almost twofold-degenerate ones $\{ s=1/2,m=\pm 1/2\}$. Further down
we will return to these observations.

The first excited state of the present $\Delta \to \infty$ limit mimics the in-gap 
Andreev bound states discussed in the last subsection for $U=0$; the continuum was shifted to 
$\pm \infty$. Varying the parameters the bound 
state energy  moves and hits zero at the transition. Approaching the transition from the
singlet side at this point the singlet ground state and the doublet
first excited state become degenerate; the excitation energy given by the bound state energy
vanishes.
Beyond the transition (in the doublet phase) the doublet is the ground state and has a finite energy gap
to the first excited singlet. 
This is the same phenomenology as observed for the Yu-Shiba-Rusinov states in the Kondo model with BCS
leads (see Subsect.~\ref{subsec:spinssup}) and again raises the question about the relation of 
these and the Andreev bound states.\cite{Yoshioka00}        

As reviewed in Subsect.~\ref{subsec:spinssup} for the Kondo model with superconducting leads 
and discussed below for the SIAM the transition between a singlet and doublet ground state 
does not only occur for $\Delta \to \infty$  but also for generic $\Delta$.   
This type of physics is absent at $U=0$ and can thus be directly
linked to the two-particle interaction (the exchange interaction in the Kondo
model with BCS leads). However, for $\Delta \to \infty$ the Kondo effect 
is suppressed completely and it can thus not be assigned as the driving force for 
such a transition. For finite $\Delta$ remnants of the Kondo effect can, however, be 
expected to affect the transition. We investigate this in the next subsection. 

The simplest way to obtain the Josephson current in the atomic limit at $T=0$ is to take the 
derivative of the many-body ground state energy with respect to $\phi$ 
[see Eq.~(\ref{eq:curenerexp})].  This leads to 
\begin{eqnarray}
\label{eq:JDeltainf}
J_L^{\rm imp}  = 
\left\{ \begin{array}{ll}  \frac{2 \Gamma_L \Gamma_R \sin \phi}{\sqrt{\epsilon^2 + 
\Gamma_L^2 + \Gamma_R^2 + 2 \Gamma_L \Gamma_R \cos \phi}}, 
& \mbox{singlet phase} \\ 0, & \mbox{doublet phase}
\end{array}
\right. 
\end{eqnarray}
Note that the current in the singlet phase is equal to the $U=0$ current;
compare to the lower line of Eq.~(\ref{eq:J0limits}) for $\Gamma_L=\Gamma/2=\Gamma_R$. 
Equivalently, the 
interacting dot Green function and thus the self-energy can be 
computed using the Lehmann representation.\cite{Karrasch08} From this the current 
can be obtained performing the frequency integral of Eq.~(\ref{eq:Jsimpexpexp}).
Equation (\ref{eq:JDeltainf}) shows that the ($T=0$) quantum phase transition 
(resulting from the two-particle interaction) is indicated 
by a jump of the supercurrent from a finite value to zero.
The $\epsilon$ dependence of the current is shown as the
blue lines in Fig.~\ref{fig2a}.
The vanishing
of the current in the doublet phase is special to the atomic limit. However,  
the jump is also found for generic $\Delta$ as discussed next. 
In this case the current in the doublet phase is negative even for $\phi \in [0,\pi]$; 
compare to the $U=0$ current of Fig.~\ref{fig2} which is positive for $\phi \in [0,\pi]$.
For this reason one speaks interchangeably of the doublet- or $\pi$-phase and the
singlet- or $0$-phase.
The jump of the current as a function of $\phi$ at $\phi_{\rm c} < \pi$ (for properly chosen 
fixed $U$, $\epsilon$, $\Gamma$, and $a$) resulting from the quantum phase transition should not 
be confused with the jump at $\phi=\pi$ obtained for $U=0$ and $\epsilon=0$; see 
Subsect.~\ref{subsec:U0}, in particular the black curve in Fig.~\ref{fig2}.
Figure \ref{fig2a} shows that the line shape $J_L^{\rm imp}(\epsilon)$ resulting
from the Zeeman field induced level crossing transition at small $U/\Gamma$ [see Fig.~\ref{fig2a} (b)]
resembles the one resulting from the transition induced by the two-particle interaction
at $B=0$ [see Fig.~\ref{fig2a} (a)].  

\subsection{``Nearly'' exact results from elaborate quantum many-body methods}
\label{subsec:nearlyexact}

The expected equivalence of the SIAM with BCS leads to the Kondo model with such 
leads in the proper limit of large $U/\Gamma$ as well as $-U/2 < \epsilon < U/2$ 
and the results for the latter model reviewed in Subsect.~\ref{subsec:spinssup}  
suggest that the level crossing scenario of the $\Delta \to \infty$ limit
survives at finite $\Delta$. In this subsection we show this 
based on NRG and QMC results. 
We discuss the phase diagram as well as the
Josephson current and briefly the bound states. 

The SIAM with two BCS leads and $\Delta < \infty$ was first treated
by the NRG in Ref.~\cite{Yoshioka00}.        
Focusing on selected parameter sets, in particular $\phi=0$ and thus 
a vanishing Josephson current, the level crossing 
scenario was unambiguously confirmed.
The assumed left-right symmetry 
and vanishing of the phase difference allowed the authors to reduce the problem to one 
with a single lead (single-channel NRG) which renders the NRG computationally less demanding.     
Analyzing the first few low-energy many-body states it was shown
that increasing $U$ at fixed $\epsilon$ and $\Gamma$ leads
to a groundstate crossing from a singlet to a doublet.

\begin{figure}[t]
\begin{center}
   \includegraphics[width=.6\linewidth,clip]{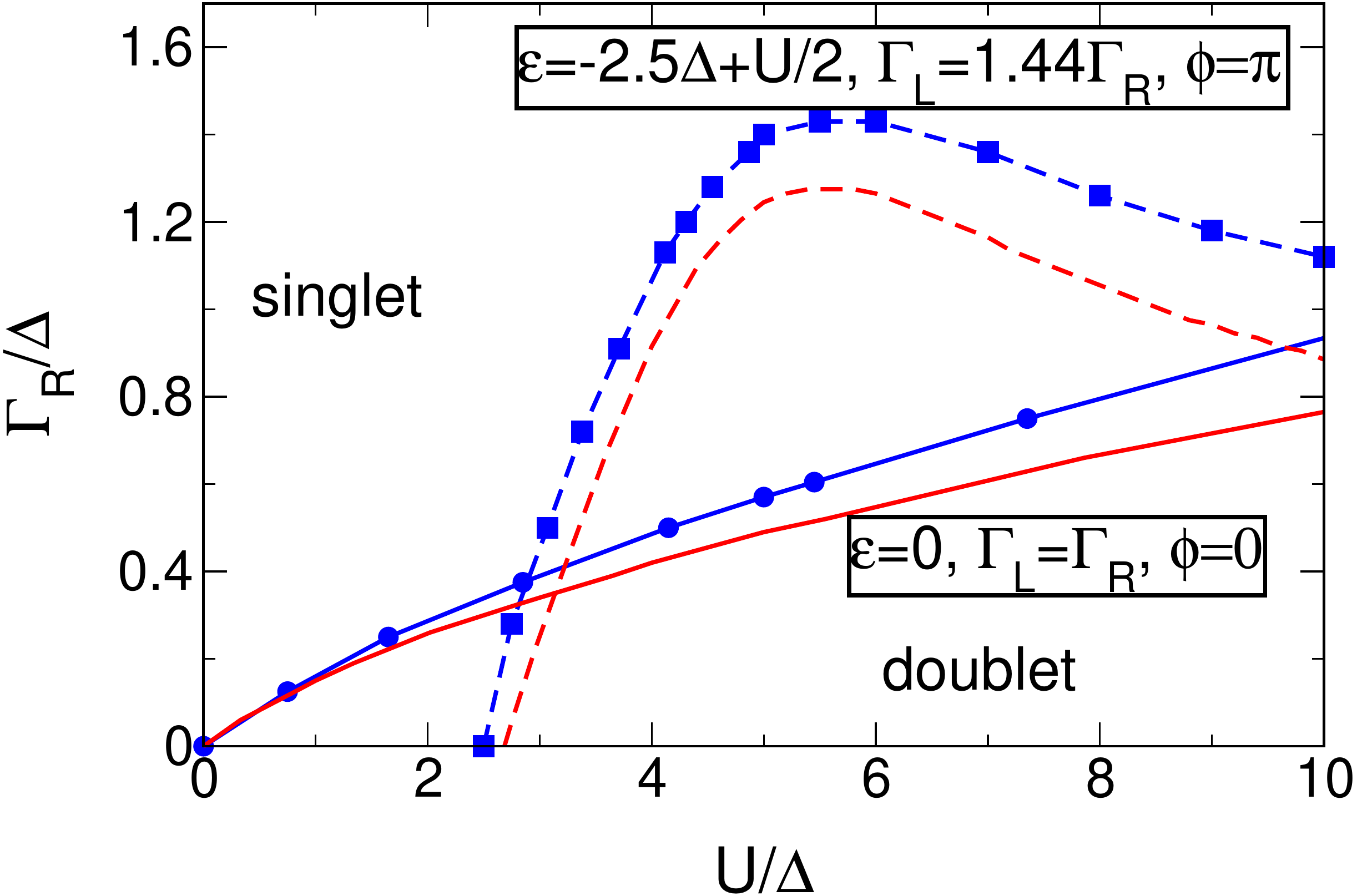}
   \caption{Phase diagram of the Anderson-Josephson dot obtained by NRG (blue symbols connected by lines)
     as a function of $U/\Delta$ and $\Gamma_R/\Delta$. Results for two different sets of the 
     other parameters, as indicated in the figure, are shown.
For efficiency reasons we included data obtained by an approximate FRG approach to be discussed in 
Subsect.~\ref{subsec:funrg} as red lines.
     The data were taken from Fig.~2 and Fig.~3 (b) 
of Ref.~\cite{Karrasch08}. 
}
\label{fig3}
\end{center}
\end{figure}

Figure \ref{fig3} shows the phase diagram extracted from the low-energy spectra obtained by 
NRG (blue symbols connected by lines) as a function of $U/\Delta$ and $\Gamma_R/\Delta$. Two 
parameter sets as given in the figure are considered--including one with $\Gamma_L \neq
\Gamma_R$ and $\phi \neq 0$. The data are taken from Ref.~\cite{Karrasch08}.
They fully confirm the $\Delta \to \infty$ scenario.
A systematic study shows that 
decreasing $U$, $\epsilon$, $|a-1|$ or $\phi$ favors the singlet phase.
To avoid a reproduction of the NRG data at a later stage of this review we
included results obtained by an approximate FRG approach to be discussed in 
Subsect.~\ref{subsec:funrg} as red lines in Fig.~\ref{fig3}. 

To compute the current at $T=0$ a two-lead NRG was first employed in 
\cite{Choi04}. 
At the quantum phase transition--the $0$-to-$\pi$-transition--the current jumps from 
a positive value 
to a negative one even for $\phi \in [0, \pi]$; the current shows $\pi$-junction behavior.    
It, however, turned out that the data for the supercurrent presented in this paper
are inaccurate; the amplitude is roughly a factor of two to small.\cite{Karrasch08} 
The most likely reason for this is an improper selection of the NRG numerical 
parameters $\Lambda$ and $N_{\rm c}$ (see Ref.~\cite{Karrasch08}).

To avoid the numerical
obstacles of two-channel NRG Oguri, Tanaka, and Hewson\cite{Oguri04} (see also Ref.~\cite{Tanaka07}) 
studied the model in the limit in which the gap of the left lead is sent to infinity, 
while the right one is kept finite. Again the level-crossing scenario was confirmed.
In contrast to the case in which both the left and the right gaps are sent to infinity 
the current in the doublet phase was found to be nonvanishing and negative even for 
$\phi \in [0,\pi]$.

\begin{figure}[t]
\begin{center}
   \includegraphics[width=.48\linewidth,clip]{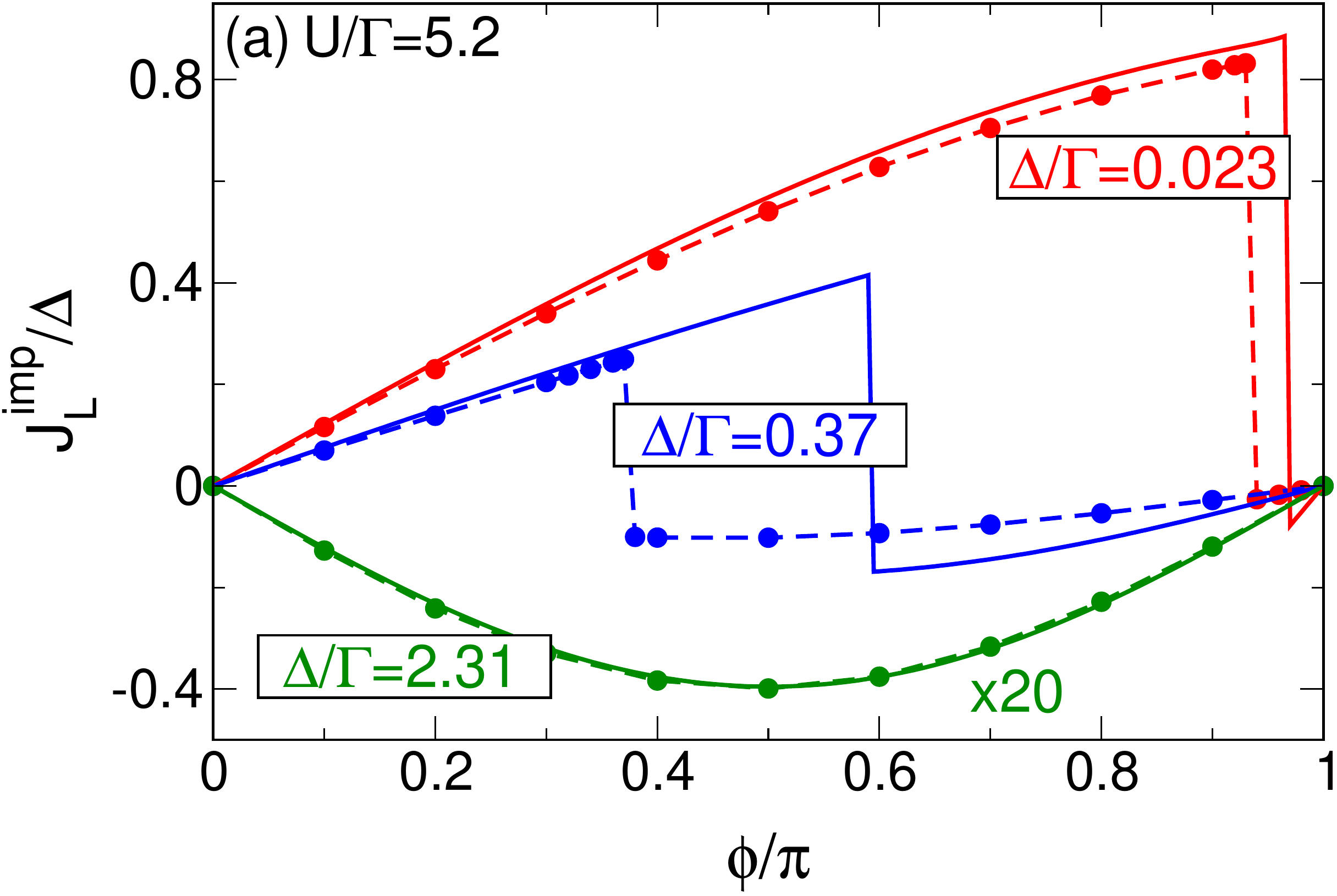}
 \includegraphics[width=.48\linewidth,clip]{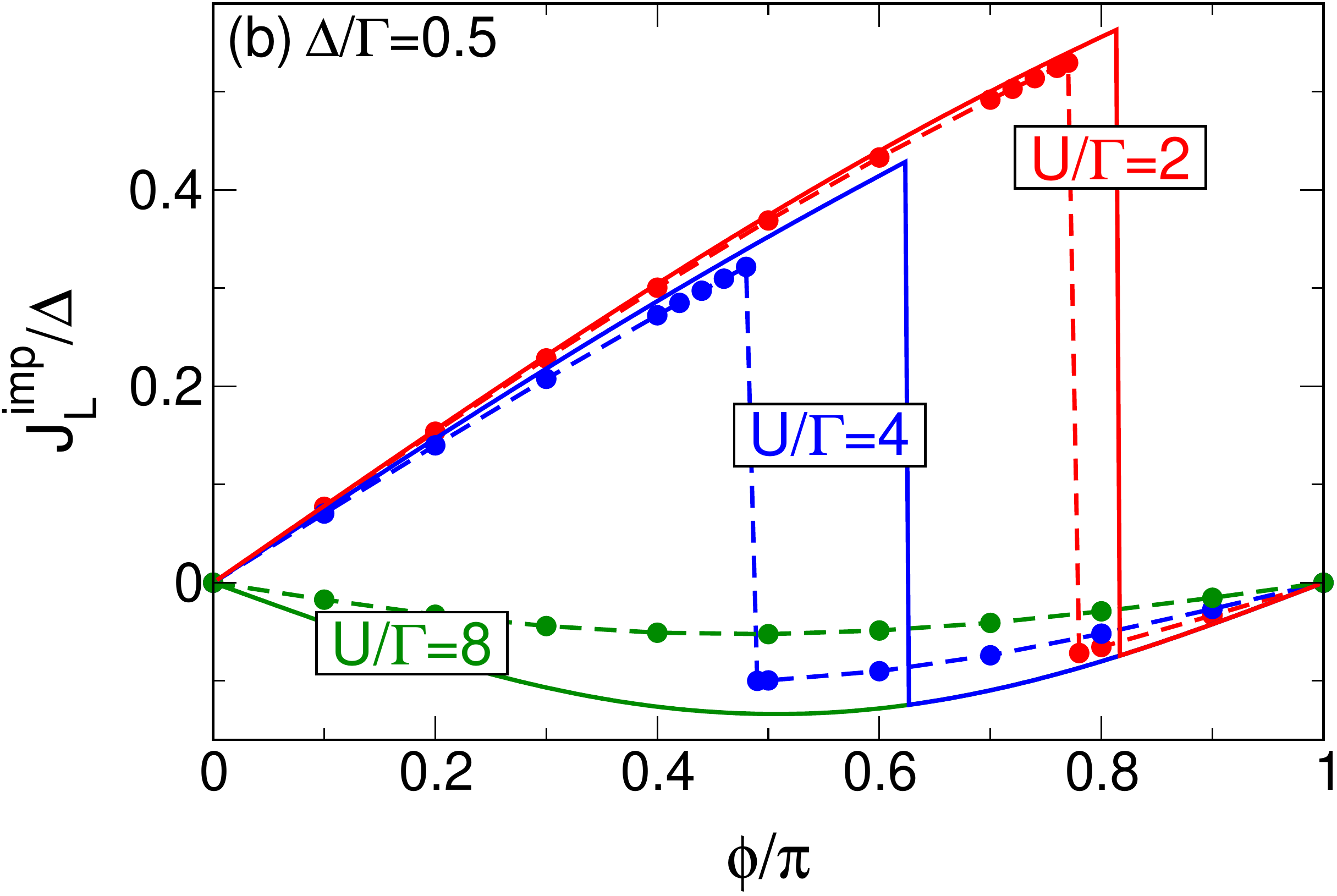}
   \caption{Zero-temperature Josephson current $J_L^{\rm imp}$ as a function of the phase difference 
$\phi$ computed with NRG (circles connected by dashed lines) and approximate FRG (solid lines) 
at $\epsilon=0$ and $\Gamma_L=\Gamma_R$. (a) $\Delta$ is varied at fixed $U/\Gamma=5.2$ 
($T_{\rm K}/\Gamma=0.209$). For clarity, the curves at $\Delta/\Gamma=2.31$ were scaled up by a 
factor of $20$. (b) $\Delta/\Gamma=0.5$ is fixed at different $U/\Gamma$ corresponding to 
$T_{\rm K}/\Gamma=0.45$, $T_{\rm K}/\Gamma=0.29$, and $T_{\rm K}/\Gamma=0.09$. Reprinted figures 
with permission from C. Karrasch, A. Oguri, and V. Meden, Phys. Rev. B {\bf 77}, 024517 (2008). 
Copyright (2008) by the American Physical Society.
}
\label{fig4}
\end{center}
\end{figure}

Figures \ref{fig4} (a) and (b) show the Josephson current as a function 
of $\phi \in [0, \pi]$ for various parameter sets with $\epsilon=0$, $\Gamma_L = \Gamma_R$, and 
finite gaps $\Delta$ computed using two-channel NRG with properly chosen numerical parameters
(circles connected by dashed lines).\cite{Karrasch08} In Fig.~\ref{fig4} (a) the interaction $U$
is fixed and $\Delta$ is varied; vice versa 
in Fig.~\ref{fig4} (b).  Remind that the current is a $2 \pi$-periodic and odd function of 
$\phi$ and can thus accordingly be extended to phases outside the shown range.  
The solid lines show FRG results; see Subsect.~\ref{subsec:funrg}.
In accordance with the $\Delta \to \infty$ limit the quantum phase transition 
is indicated by a jump of the supercurrent. For $\phi \in [0, \pi]$ the current is positive in the singlet 
phase and negative in the doublet one. The same behavior is found for $\epsilon \neq 0$. 
Results for $\Gamma_L \neq \Gamma_R$ can be obtained from the symmetric case using 
the transformation Eq.~(\ref{eq:cursym}).  
In the caption of Fig.~\ref{fig4} we give 
the values of $T_{\rm K}$ [computed from Eq.~(\ref{eq:T_Kana})] such that it is possible to estimate 
the strength of the correlations for suppressed superconductivity for a given parameter set. We, however, 
reemphasize that the ratio $\Delta/T_{\rm K}$ plays a by far less relevant role than assumed 
in large parts of the literature of the Anderson-Josephson quantum dot. For details on the NRG 
procedure in particular the NRG numerical parameters $\Lambda$ and $N_{\rm c}$, 
see Ref.~\cite{Karrasch08}.

\begin{figure}[t]
\begin{center}
   \includegraphics[width=.43\linewidth,clip]{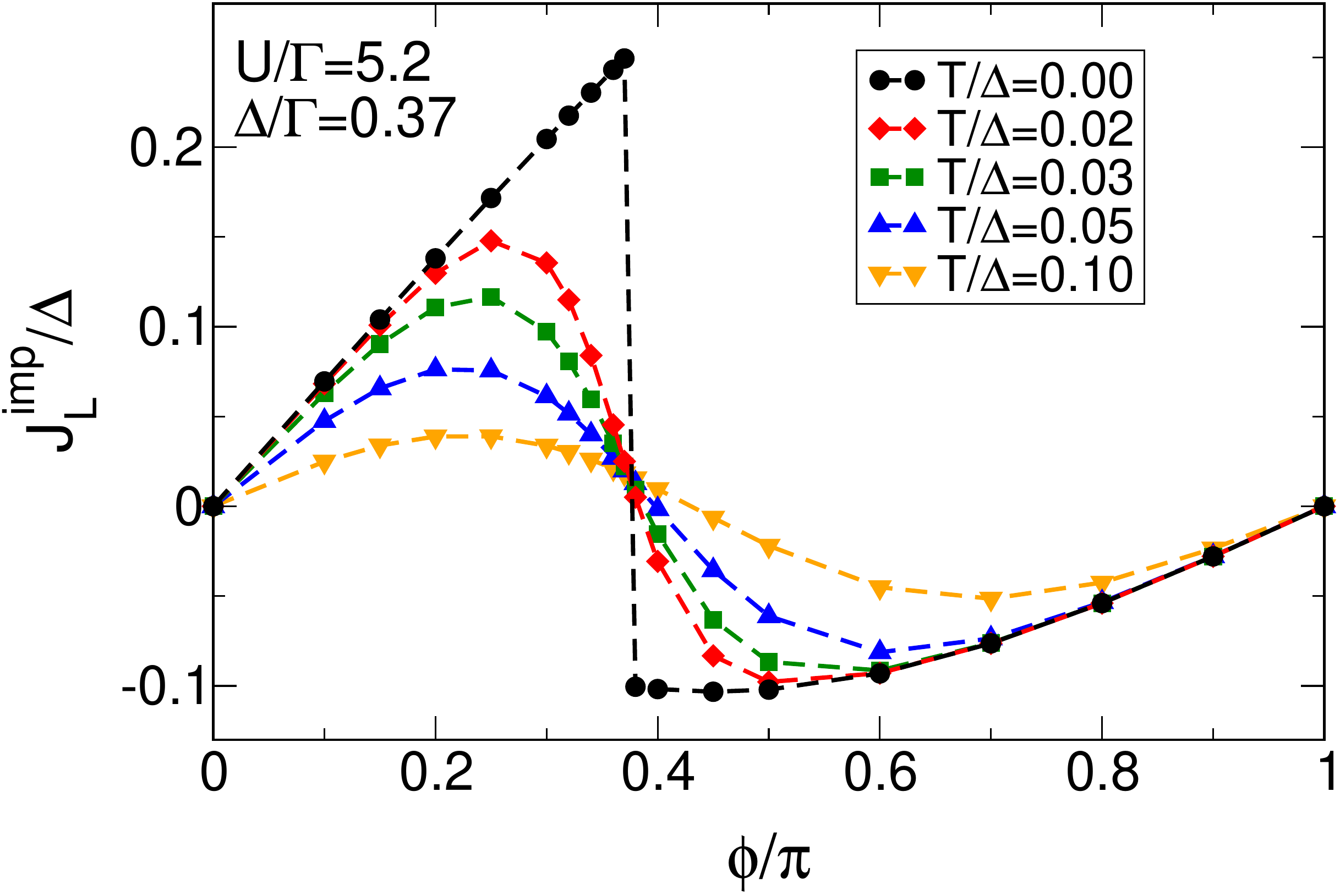}
 \includegraphics[width=.5\linewidth]{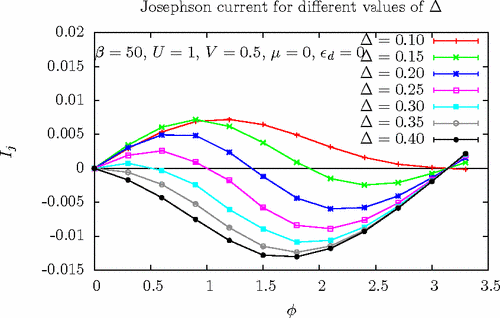}
   \caption{Finite-temperature Josephson current $J_L^{\rm imp}$ as a function of the phase difference 
$\phi$. Left panel: NRG data for the parameters of the blue curve of Fig.~\ref{fig4} (a).  Reprinted figure  
with permission from C. Karrasch, A. Oguri, and V. Meden, Phys. Rev. B {\bf 77}, 024517 (2008). 
Copyright (2008) by the American Physical Society. Right panel: CTINT QMC data for the left current 
$I_j$ ($=J_L^{\rm imp}$ in our notation). 
The parameters given in the figure must be translated to our model parameters as $V = t_L = t_R$ and 
$\epsilon_d=\epsilon$.
In the CTINT QMC computation a lead band with dispersion $\epsilon_k = -2 t \cos(k)$ is considered.
Band effects can be neglected and the model becomes equivalent to the one in the wide band limit
as long as all parameters of dimension energy, which here are given in units of $t$, are taken 
to be smaller than the hopping. Reprinted figure  
with permission from D.J. Luitz and F.F. Assaad, Phys. Rev. B {\bf 81}, 024509 (2010). 
Copyright (2010) by the American Physical Society.
}
\label{fig5}
\end{center}
\end{figure}

Figure \ref{fig5} (a) shows NRG data for the CPR at different temperatures.
The parameters of the blue curve 
in Fig.~\ref{fig4} (a) ($U/\Gamma=5.2$, $\Delta/\Gamma=0.37$) are considered.
The jump at $T=0$ results from a quantum phase transition and is thus smeared out 
out for $T>0$. Additionally, the amplitude of the current is strongly suppressed with increasing
$T$.\cite{Karrasch08} 

The supercurrent was also computed by the Hirsch-Fye QMC method\cite{Siano04} and the continuous-time 
interaction-expansion QMC (CTINT QMC) method,\cite{Luitz10} both inherently being $T>0$ techniques. While the results 
of the first approach were shown to be 
accurate in the $T=0$ doublet phase but imprecise in the singlet 
one\cite{Karrasch08} the CTINT QMC approach of Ref.~\cite{Luitz10} leads to highly accurate 
(``nearly'' exact) currents in both regimes. It was later used to directly compare to experimental 
data as reviewed in Sect.~\ref{sec:experiments}. The right panel of Fig.~\ref{fig5}
shows CTINT QMC data at fixed $T$ for varying $\Delta$.\cite{Luitz10}.  

\begin{figure}[t]
\begin{center}
   \includegraphics[width=.9\linewidth,clip]{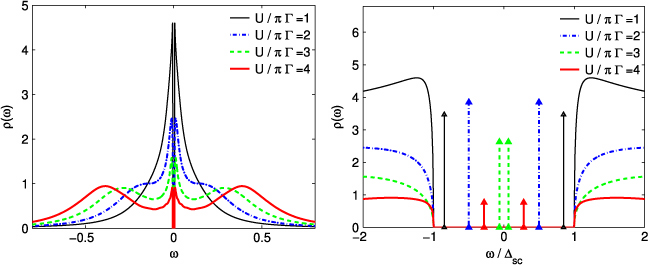}
   \caption{Dot spectral function at half dot filling $\epsilon=0$, vanishing phase $\phi=0$ 
   for a small superconducting gap $\Delta = 0.005 D$ and coupling $\Gamma =0.2D/\pi$, with
   $D$ being half the lead band width. The lead density of states is assumed to be constant, but no wide 
   band limit is taken. 
   This fully corresponds to the model in the wide band limit as long as all parameters are taken to be smaller than $D$. The spectral 
   function is computed by NRG and denoted as $\rho(\omega)$ (corresponding to  $D A(\omega)$ in our notation). 
   Left panel: Spectral weight on all relevant energy scales ($\omega$ is measured in units of $D$). Right panel: Zoom-in of 
   the weight on the scale $\Delta_{\rm sc}$ ($= \Delta$ in our notation). The vertical arrows indicate $\delta$-peaks 
   corresponding to in-gap bound states. Reprinted figure  
with permission from J. Bauer, A. Oguri, and A.C. Hewson, J. Phys.: Condens. Matter {\bf 19}, 486211 (2007). 
Copyright (2007) by the Institute of Physics.
}
\label{fig6}
\end{center}
\end{figure}

A comprehensive NRG study of the $T=0$ spectral properties of the Anderson-Josephson 
quantum dot was presented in \cite{Bauer07} (see 
also \cite{Hecht08}). Figure \ref{fig6} shows the (dot) spectral function at half filling of the 
dot ($\epsilon=0$) and for $\phi=0$ in the limit of a small gap on all 
relevant energy scales (left panel) and in a zoom-in on the scale of the gap (right panel). 
As the gap is small a Kondo resonance around $\omega=0$ starts to form with increasing $U$ and 
Hubbard bands develop; compare to the $\epsilon=0$-curve (blue) of Fig.~\ref{fig1}.
However, due to the superconductivity the spectral weight at low energies $|\omega| < \Delta$ 
is suppressed and the resonance does not fully develop. A symmetrically located pair of in-gap bound 
states appears which in Fig.~\ref{fig6} are shown as vertical arrows ($\delta$-peaks),
the heights indicating the weight. 
Their position $E_{\rm b}$ 
and weight $w_{\rm b}$ depends on the parameters; in Fig.~\ref{fig6} on $U$. 
For increasing $U/\Gamma$ one crosses over from noninteracting Andreev bound states to the 
Yu-Shiba-Rusinov states both discussed above.

\begin{figure}[t]
\begin{center}
   \includegraphics[width=.9\linewidth,clip]{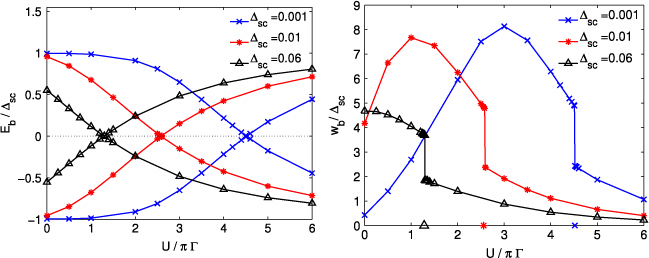}
   \caption{Parameter dependence of the position $E_{\rm b}$ (left panel) and the weight $w_{\rm b}$ (right panel) 
   of the in-gap bound states. The symbols on the $x$-axis of the right panel indicate the position of the 
   quantum phase transition (zeros of the left panel). The gap $\Delta_{\rm sc}$ ($=\Delta$ in our notation) 
   is measured in units of $D$. The other parameters are $\epsilon=0$, $\phi=0$ and $\Gamma =0.2D/\pi$. Reprinted figure  
with permission from J. Bauer, A. Oguri, and A.C. Hewson, J. Phys.: Condens. Matter {\bf 19}, 486211 (2007). 
Copyright (2007) by the Institute of Physics.
}
\label{fig7}
\end{center}
\end{figure}

The parameter dependence of the bound
states is further illustrated in Fig.~\ref{fig7} which shows $E_{\rm b}$ and $w_{\rm b}$ 
as a function of $U$ for different $\Delta$. In full accordance with the results for the 
Kondo model with BCS leads the energy of the pair of in-gap states hits zero at the quantum phase
transition. The gap between the many-body ground state and the first excited one which is given
by $\left|E_{\rm b}\right| $ vanishes. The first excited state becomes the ground state and
vice versa.
Note that the appearance of a pair of $\delta$-peaks is unrelated to the observation that one of the two
involved states is a doublet but rather follows from the fact that the total spectral function
shown contains the photoemission as well as the inverse photoemmission part. 
In addition, the weight of the $\delta$-peaks depends on the parameters and jumps at the transition. 
Additional NRG data of $A(\omega)$ for other parameter sets can be found in 
Ref.~\cite{Bauer07}. In particular, this includes data for larger $\Delta$ in which the Kondo peak is not 
formed even for large $U$ (superconductivity prevails) and for $\epsilon \neq 0$. In both 
cases the parameter dependence of $E_{\rm b}$ and $w_{\rm b}$ shows the same behavior as just 
described.  

Already in 1990 Jarrell, Sivia, and Patton\cite{Jarrell90} used Hirsch-Fye QMC to compute the dot 
single-particle spectral function of the SIAM with superconducting leads for generic 
$\Delta$ but $\epsilon=0$, i.e. half filling of the dot.\footnote{We note that the 
authors went beyond a BCS treatment of the leads and instead considered phonons leading 
to the attractive interaction in the leads and thus superconductivity.} The analytic 
continuation was performed by the maximum entropy method. The authors found in-gap states at 
a rapidly moving energy when varying the parameters $U/\Gamma$ and $\Delta/\Gamma$. 
However, due to inherent restrictions of the method, in particular, of the maximum entropy approach, 
the energy resolution was insufficient to demonstrate the level crossing. 
This was achieved with the CTINT QMC of Ref.~\cite{Luitz10}.
We note that it was recently shown that in addition a continuous-time QMC approach alternative to 
CTINT QMC, namely the hybridization-expansion (CTHYB) QMC can be used to obtain $T>0$ results for the 
dot spectral function.\cite{Pokorny17} 

Very recently CTHYP QMC was also used to compute the finite temperature supercurrent. The results
showed a very good agreement to $T>0$ NRG data obtained for the same parameters.\cite{Kadlecova18} 

As the discussion of the last three subsections shows a comprehensive understanding of the 
physics of the Anderson-Josephson quantum dot can be obtained using the analytical insights
of the $U=0$ and the $\Delta \to \infty$ limits as well as the results of highly accurate 
numerical NRG and CTINT QMC approaches for arbitrary parameters. We reemphasize that those two methods 
are accurate even if the parameters are chosen such that the dot is in the Kondo regime for 
suppressed superconductivity. As we will discuss in Sect.~\ref{sec:experiments} the NRG and 
CTINT QMC results for the single-level Anderson-Josephson dot can even be used for a direct 
comparison to experimental data. A satisfying quantitative agreement of the model 
calculations and measurements can be achieved. Before reviewing this, we will give a brief account
of alternative theoretical approaches used to investigate the Anderson-Josephson quantum dot
and comment on their reliability.     

\subsection{Alternative approaches: an overview}
\label{subsec:alternative}

Even before the two highly accurate numerical approaches 
were applied to the SIAM with superconducting leads the use 
of other approximate analytical methods provided indications of $\pi$-junction 
behavior of the current for generic parameters. In particular, a combined expansion 
in $\Gamma$ and effective model treatment of Glazman and Matveev in 1989\cite{Glazman89} 
indicated the sign change of the current when varying the other 
parameters.\footnote{For another effective model treatment showing a similar sign change of the Josephson 
current, see \cite{Spivak91}.}
Expansions in $\Gamma$ were later also used in Refs.~\cite{Novotny05} and \cite{Governale08}.
Strictly speaking this approximate method is limited to the regime in which $T \gg \Gamma$
and cannot be used to approach the quantum phase transition underlying the $\pi$-junction behavior. 
It also does not capture Kondo physics and cannot be used for a quantitative 
comparison to experiments in the most interesting parameter regime of competing
superconductivity and Kondo correlations.

Over the last decades several types of mean-field-like approaches were employed to 
investigate the Anderson model with BCS leads. In the early attempts
of the 60s and 70s to understand the  physics of dilute magnetic impurities in 
bulk superconductors including charge fluctuations the self-consistency problem was 
not fully considered.\cite{Zuckermann65,Takanaka66,Shiba73} 
In particular, the generation of an anomalous impurity self-energy was ignored. 
In the first mean-field treatment in the context of mesoscopic transport, 
Ref.~\cite{Rozhkov99}, the anomalous term was neglected as well. In this work it 
was pointed out that the spin symmetry of 
the model is spontaneously broken leading to a nondegenerate ground state in roughly 
the parameter regime in which the exact ground state is a doublet. The Josephson current 
in the symmetry broken state is negative. However, the breaking of the spin symmetry is an 
artifact of the mean-field approximation familiar from the SIAM with metallic 
leads.\cite{Anderson61} The same spurious symmetry breaking is found in the full unrestricted 
mean-field treatment of Ref.~\cite{Yoshioka00} which includes the solution of a self-consistency 
equation for the off-diagonal self-energy.
The spin-symmetry breaking corresponds to the spontaneous generation of a Zeeman field.
It is thus the Zeeman field induced level crossing transition discussed in connection
with Fig.~\ref{fig2a} (b) (in the limit $\Delta \to \infty$) which leads to the
negative supercurrent. 
We note in passing that Ref.~\cite{Karrasch08} reported 
on difficulties to reproduce the mean-field results of Ref.~\cite{Yoshioka00} but confirmed
the spurious spin-symmetry breaking as the reason for $\pi$-junction behavior. As the unrestricted 
mean-field approach breaks a fundamental symmetry, induces the level-crossing transition
for the wrong reason, namely the effective Zeeman field, and does not produce the correct
degeneracies we believe that it should not be used  to study the Anderson-Josephson quantum dot,
not even for qualitative estimates of the Josephson
current or the dot spectral function.

In Refs.~\cite{Zonda15} and \cite{Zonda16} it was suggested to use the spin-symmetric 
restricted mean-field solution to determine the phase boundary. In practice this means 
that the same self-consistency equations as in the unrestricted mean-field approach 
are solved as long as they lead to a nonmagnetic solution; the phase boundary 
is determined by the point at which this ceases to exist and shows 
a qualitatively correct dependence on the parameters as compared to NRG results.
It was shown that the restricted mean-field approach can serve as a simple starting 
point for a thermodynamically consistent perturbative treatment in $U$ which 
includes dynamical corrections to the self-energy. This leads to very good results 
for the phase boundary as well as the Josephson current and single-particle dot 
spectral function in the $0$-(singlet-)phase. However, within the restricted 
mean-field approach and any technique build on it   
the $\pi$-(doublet-)phase is inaccessible. Already earlier it was suggested 
to use fully\cite{Alastalo98} or partly\cite{Vecino03} self-consistent second
order perturbation theory in $U$ to study the SIAM with BCS leads. 

Also the noncrossing approximation which was successfully used for the SIAM with 
metallic leads\cite{Bickers87} and captures aspects of the Kondo effect was extended 
to the SIAM with BCS leads. Formally it corresponds to an expansion in  the inverse 
degeneracy of the dot level. It was shown to give reasonable results for the dot spectral 
function and the Josephson current including $\pi$-junction 
behavior.\cite{Ishizaka95,Clerk00,Sellier05} However, the results of this approximate approach 
were never directly compared to the ``nearly'' exact ones obtained by NRG or CTINT QMC. 
It is thus difficult to judge if the agreement goes beyond a qualitative one, in 
particular in the most interesting parameter regime of strongest competition between 
Kondo correlations and superconductivity that is if the scales $T_{\rm K}$ and $\Delta$
are comparable.   

As reviewed in Subsect.~\ref{subsec:infinitegap} studying 
the exactly solvable atomic limit $\Delta \to \infty$ 
is very instructive to gain a detailed understanding of the physics. 
Based on this insight Meng, Florens, and Simon\cite{Meng09} set up a systematic self-consistent 
expansion around this point. It was mainly used to determine the phase diagram and the 
in-gap bound state energy.\cite{Meng09,Wentzell16} The results for these observables agree 
very well with NRG data and also the current shows the characteristic behavior at the 
$0$-to-$\pi$-transition. By construction the expansion does not capture Kondo correlations 
but still constitutes a promising easy to handle approximation which was even used to 
directly compare to the experimental phase diagram.\cite{Maurand12} It would be
interesting to see how it performs in more complex models of localized levels coupled 
to superconducting leads, as e.g. investigated using FRG in Sect.~\ref{sec:complexdots}.      

Surprisingly, even a very simple model in which the two leads are replaced by two lattice
sites carrying an effective pairing potential, commonly referred to as the narrow-band limit,
captures the basic phenomenology of the Josephson current if the parameters are
varied.\cite{Vecino03,Bergeret06,Bergeret07,Allub15,Kirsanskas15} Needeless to say, this model
is by construction unable to capture any aspects of the Kondo effect. The Kondo singlet is
nonlocal and its formation requires extended leads.

Finally, purely phenomenological approaches were used to model the Anderson-Josephson quantum 
dot.\cite{Vecino03,Jorgensen07} In these the main characteristics of the observables as known from 
microscopic models (see above) is to a large extend already build in ``by hand''. In one type
of approach the spin-symmetry was broken ``by hand'' which for sufficiently large ``artificial''
Zeeman field leads to the field induced transition discussed in connection with
Fig.~\ref{fig2a} (b) even for $U=0$.  

\subsection{The functional renormalization group approach}
\label{subsec:funrg}

We next give a more detailed account of the approximate FRG approach.\cite{Metzner12} The FRG is a
flexible tool which was not only used to study the SIAM with two BCS leads\cite{Karrasch08,Wentzell16} but
also for more complex dot setups with superconducting reservoirs showing interesting many-body physics
as reviewed in Sect.~\ref{sec:complexdots}.\cite{Karrasch09,Karrasch11}
The basic steps of the application of FRG to interacting mesoscopic systems coupled to
noninteracting leads\footnote{The FRG approach
  to quantum many-body systems was originally developed for two-dimensional bulk systems in the context of high-temperature
  superconductivity.\cite{Metzner12,Kopietz10}} are the
following:
\begin{enumerate}
\item Write the partition function as a coherent state functional integral.
  \item Integrate out the noninteracting leads by
    projection. They are incorporated exactly as lead self-energies to the propagator of the interacting part.
    \item Replace the
reservoir-dressed noninteracting propagator of the system (the dot) by one decorated by a cutoff $\Lambda$ (not to be
confused with the NRG numerical parameter $\Lambda$). For the initial value $\Lambda_{\rm i}$ the free propagation
must vanish, for the final one $\Lambda_{\rm f}$ the original propagation must be restored.
One often uses ${\mathcal G}_{0}^\Lambda(i\omega) =  \Theta(|\omega| - \Lambda) {\mathcal G}_{0}(i\omega)$,
$\Lambda_{\rm i}= \infty$, and $\Lambda_{\rm f}=0$.  When $\Lambda$ is sent from $\infty$ to $0$ (see below)
this incorporates the
RG idea of a successive treatment of energy scales. This cutoff function was also used for quantum dots with
BCS leads.\cite{Karrasch08,Karrasch09,Karrasch11,Wentzell16}
\item Differentiate the generating functional of one-particle irreducible vertex functions with respect to
  $\Lambda$.
\item Expand both sides of the functional differential equation with respect to the vertex functions. This leads
  to an infinite hierarchy of coupled differential equations for the vertex functions. 
\end{enumerate} 
The hierarchy of coupled flow equations presents an exact reformulation of the quantum many-body problem
and integrating it from $\Lambda_{\rm i}$ to $\Lambda_{\rm f}$ leads to exact expressions for the vertex functions.  
From those observables such as the system spectral function, the current, etc. can be computed.
In practice truncations of the hierarchy are required resulting in a closed finite set of equations.
The integration of this leads to approximate expressions for the vertices and thus for observables.
Different truncation schemes and the application of FRG to (nonrelativistic) quantum many-body systems are
reviewed in Refs.~\cite{Metzner12} and \cite{Kopietz10}.

\begin{figure}[t]
\begin{center}
   \includegraphics[width=.6\linewidth,clip]{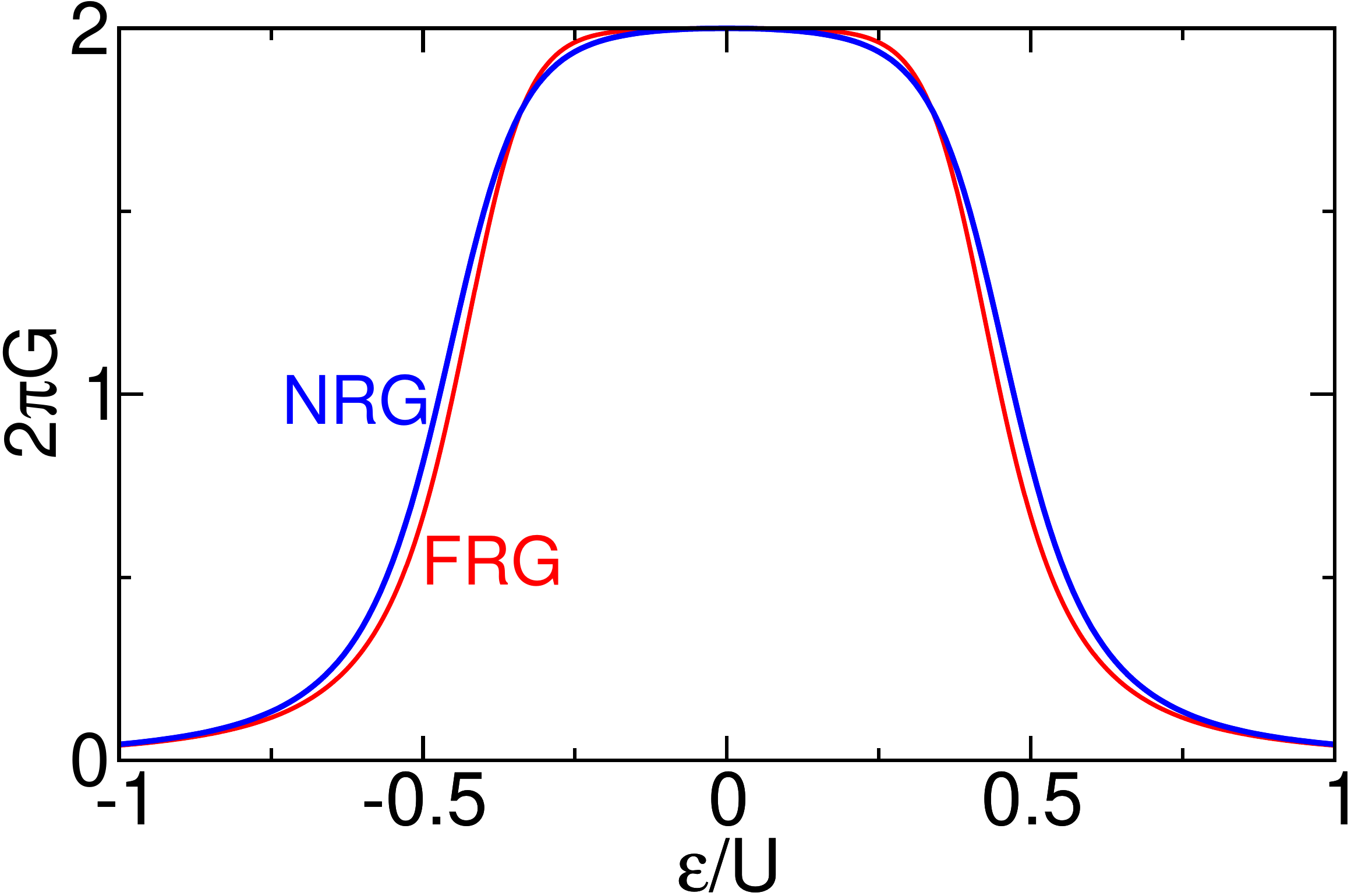}
   \caption{Comaprison of the linear conductance of the SIAM with two metallic leads
     as a function of $\epsilon$ obtained by approximate FRG and
     NRG for $U/\Gamma=4 \pi$.}
\label{fig7a}
\end{center}
\end{figure}

We here restrict ourselves to a truncation scheme in which the
flowing two-particle vertex is replaced by its static part, that is a flowing $U^\Lambda$, and higher
order vertices (generated during the flow) are neglected. The flowing self-energy then becomes static.
This approximation is controlled for small two-particle interactions. The self-energy contains all diagrams
to order $U$ but higher order ones are partly resummed in addition. Crucially, the resummation is not identical
to the one achieved in the mean-field approach and the approximate FRG does not suffer from the artificial
spin-symmetry breaking discussed above. For the SIAM with metallic leads it captures certain aspects of Kondo
physics, such as the (exponential) pinning of the spectral weight at the Fermi energy when varying the level
position.\cite{Karrasch06} This can be inferred from the red curve of the linear conductance
as a function of $\epsilon$ in Fig.~\ref{fig7a} computed within this truncation scheme taking into
account the relation between the spectral weight at the Fermi energy and the conductance Eq.~(\ref{eq:GA}).
The FRG data for $G(\epsilon)$ show an excellent agreement with the highly accurate NRG curve (blue line) even
for strong interactions ($U/\Gamma=4 \pi$ in the figure). Despite this success the truncated FRG approach has
obvious limitations. E.g.~within the above static approximation the dot spectral function is a Lorentzian of (noninteracting)
width $\Gamma$ and does not develop a sharp Kondo resonance [compare Fig.~\ref{fig1} (a)].\cite{Karrasch06} This
can partly be improved in a higher order truncation. Keeping the frequency dependence of the two-particle vertex leads
to a frequency dependent self-energy and a sharp resonance in the spectral function. While its width agrees well
with the one obtained by NRG at small to intermediate $U/\Gamma$ it does
not scale exponentially in $U/\Gamma$ as in Eq.~(\ref{eq:T_Kana}).\cite{Karrasch08a,Karrasch10}  

For the spin-degenerate SIAM with BCS leads the
FRG flow equations for the self-energy and the effective interaction within the low order static approximation
read
\begin{align}
  \partial_\Lambda \Sigma^\Lambda & = - \frac{U^\Lambda}{\pi} \frac{\epsilon+ \Sigma^\Lambda}{D^\Lambda(i \Lambda)} ,
  \label{eq:flowSigma}\\
  \partial_\Lambda \Sigma_\Delta^\Lambda & = - \frac{U^\Lambda}{\pi}
                                           \frac{\Sigma_\Delta^\Lambda- \tilde \Delta}{D^\Lambda(i \Lambda)} ,
                                           \label{eq:flowSigmaDelta}\\
  \partial_\Lambda U^\Lambda & = - \frac{2 \left( U^\Lambda \right)^2}{\pi \left[ D^\Lambda( i \Lambda) \right]^2}
                               \left[ \left( \epsilon + \Sigma^\Lambda \right)^2
                               + \left| \tilde \Delta - \Sigma_\Delta^\Lambda \right|  \right] ,
  \label{eq:flowU}
\end{align}
with the initial conditions
\begin{align}
\Sigma^{\Lambda = \infty}=0, \quad \Sigma_\Delta^{\Lambda = \infty}=0, \quad U^{\Lambda = \infty}= U.
\label{eq:flowini}
\end{align}
Here $D^\Lambda$ corresponds to Eq.~(\ref{eq:Dfull}) with $\Sigma$ and $\Sigma_\Delta$ replaced by
$\Sigma^\Lambda$ and $\Sigma_\Delta^\Lambda$, respectively and $\tilde \Delta$ is defined by
Eq.~(\ref{eq:defsforG0}). The three coupled equations can easily be solved numerically. The
Josephson current can be computed at the end of the RG flow inserting $\Sigma^{\Lambda = 0} $ and
$\Sigma_\Delta^{\Lambda = 0}$ in Eq.~(\ref{eq:Jsimpexpexp}).  Without loss of generality
we now focus on the left-right symmetric case
$\Gamma_L = \Gamma_R = \Gamma$; see Subsect.~\ref{subsec:model}.
FRG results for the current are compared to highly accurate NRG ones in Fig.~\ref{fig4}. It turns out
that the truncated FRG captures the quantum phase transition including the jump from $0$- to $\pi$-junction
behavior of the current. The exact position of the jump (in the figures as a function of
$\phi$) is sensitive to the approximation, however, the overall picture is reproduced quite well by
the FRG. It systematically overestimates the current in the doublet phase. 

Alternatively the two phases can be identified as follows. In the doublet phase one finds that at a certain
$\Lambda_{\rm c}$, $\epsilon + \Sigma^{\Lambda_{\rm c}} =0$. Equation
(\ref{eq:flowSigma}) implies that $\Sigma^\Lambda = - \epsilon$ for all $\Lambda < \Lambda_{\rm c}$.
The off-diagonal component of the self-energy continues to flow and reaches $\Sigma_{\Delta}^{\Lambda=0} = \Gamma
\cos(\phi/2)$.
Varying the parameters one can identify this behavior and thus determine the phase boundary.
The red lines of Fig.~\ref{fig3} were obtained this way. They show excellent agreement with the NRG data even
if the interaction $U$ is not small. Inserting $\Sigma^{\Lambda=0} = - \epsilon$ and
$\Sigma_{\Delta}^{\Lambda=0} = \Gamma \cos(\phi/2)$ in Eq.~(\ref{eq:Jsimpexpexp}) yields a result
which is independent of $U$ and $\epsilon$. This is an artifact of the approximation and
explains why the FRG currents of Fig.~\ref{fig4} do not depend on $U$ in the $\pi$-phase.

The discussion shows that the truncated static FRG provides a easy to use approximate approach which captures
the main characteristics of the Anderson-Josephson quantum dot at $\Delta < \infty$
even for fairly large two-particle
interactions without suffering from fundamental artifacts such as spurious spin-symmetry breaking. In addition
it is flexible and can straightforwardly be extended to more complex dot setups (multi-level, different
geometry) to reveal interesting physics. This is exemplified in Sect.~\ref{sec:complexdots}. However,
as for metallic leads a word of warning must be added. It turned out that a systematic extension of the
truncation keeping the frequency dependence of the two-particle vertex is not straight forward as briefly
discussed in Ref.~\cite{Karrasch10}.

\section{Comparison to experiments}
\label{sec:experiments}

In this section we compare results of calculations within our minimal model with experimental
data. We emphasize that it is not the aim of the present review to give a comprehensive
account of the experimental status of the Anderson-Josephson quantum dot; this would require a
separate effort. In fact, the presentation of the experiments is reduced to its minimum.

The $T=0$ level crossing quantum phase transition cannot be accessed directly in $T>0$ experiments.
Indications of the transition can, however, be found in observables at finite temperatures;
see Fig.~\ref{fig5} for the Josephson current. The two most prominent
ways of experimentally investigating the physics of the transition induced by the local
two-particle interaction are the measurement of the Josephson current and the spectroscopy of the in gap bound
states. Here we will focus on the equilibrium current through quantum dots coupled to two conventional
s-wave superconducting reservoirs and do not dwell on bound-state spectroscopy; for experiments
on the latter, see, e.g., Refs.~\cite{Pillet10,Deacon10,Franke11,Pillet13,Chang13,Kim13,Grove18}.

The Josephson current was measured in several experiments; recent ones showed indications
of the $0$-to-$\pi$ transition.\cite{vanDam06,Cleuziou06,Jorgensen07,Eichler09,Maurand12,Delagrange15,Delagrange16}
Indium arsenide nanowires\cite{vanDam06} and carbon
nanotubes\cite{Cleuziou06,Jorgensen07,Eichler09,Maurand12,Delagrange15,Delagrange16}
were used as quantum dots. The experimental challenge in measuring the true magnitude of the current
in SQUID geometries consists in suppressing uncontrolled fluctuations of the
superconducting phase difference. This can be achieved using particularly designed on-chip
circuits produced by highly advanced nanostructuring techniques.
For a short review on the experimental status until 2010, see Ref.~\cite{DeFranceschi10}.

In Ref.~\cite{vanDam06,Cleuziou06,Jorgensen07,Eichler09} the current-voltage [$I(V)$]
characteristics of the quantum dot Josephson junction was measured
at different gate voltages, i.e., level positions. 
Applying the extended resistively and capacitively shunted-junction (RCSJ)
model\cite{Ambegaokar69} from the theory of conventional (structureless) Josephson junctions
in an electro-magnetical environment allows one to extract the so-called critical current
$J_{\rm c}$ as a function of the gate voltage (but not the full CPR) from the $I(V)$ curves.
In this model the CPR is assumed to be purely sinusoidal [compare Eq.~(\ref{eq:lowestorder})]
with $J_{\rm c}$ being the amplitude, i.e., the current at $\phi=\pi/2$.
A positive $J_{\rm c}$ is thus indicative of the $0$-phase (singlet ground state) and a
negative one of the $\pi$-phase (doublet ground state); the gate voltage is the parameter
to be used to switch from one to the other. 
However, due to the assumed form $J(\phi) = J_{\rm c} \sin{\phi}$  this type of analysis is unjustified
and leads to meaningless results for $J_{\rm c}$ if higher harmonics contribute to the CPR.
Higher harmonics were found to be large in CPRs in which the dot parameters--in particular the
gate voltage--are fixed such that the $0$-to-$\pi$-transition is driven by varying the phase
difference $\phi$; compare to Figs.~\ref{fig4} and \ref{fig5}.\footnote{Even for $U=0$ and a level
  position close to resonance at $\epsilon=0$ this analysis is inapplicable; see Fig.~\ref{fig2}.}
One might thus wonder how this analysis can at all be useful to investigate the level-crossing phase
transition. We will comment on this in Subsect.~\ref{subsec:current}.

In Refs.~\cite{Maurand12,Delagrange15,Delagrange16} the full CPR was measured for different
gate voltages. See Ref.~\cite{Basset14} for a detailed description of the setup and measurement
protocol of how to achieve this.

We already now note that the dot levels investigated in 
Refs.~\cite{Jorgensen07,Eichler09,Maurand12,Delagrange15,Delagrange16}
are in the Kondo regime, however, not very deeply. As the gap $\Delta$ will turn out to be
larger than the normal state $T_{\rm K}$ by a factor between 3 and 10 the systems are in the most
interesting parameter regime of the competition of Kondo correlations
and superconductivity.

\subsection{How to extract the parameters}
\label{subsec:parameters}

\begin{figure}[t]
\begin{center}
   \includegraphics[width=.6\linewidth,clip]{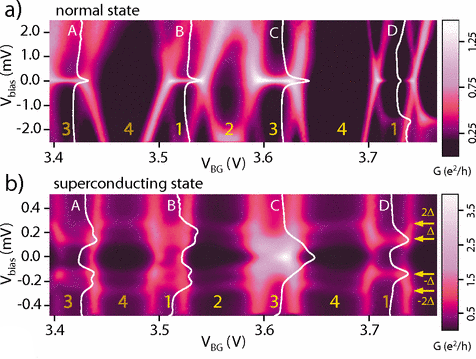}
   \caption{Experimental differential conductance as a function of gate voltage $V_{\rm BG}$ and bias voltage
     $V_{\rm bias}$ for normal state leads a) and superconducting ones b).   
     Reprinted figures 
     with permission from A. Eichler, R. Deblock, M. Weiss, C. Karrasch, V. Meden, C. Sch\"onenberger, and H. Bouchiat,
     Phys. Rev. B {\bf 79}, 161401 (2009). 
Copyright (2008) by the American Physical Society.}
\label{fig8}
\end{center}
\end{figure}

In a first step of the comparison of theory and experiment one has to extract the (model)
parameters from the experiment as accurately as possible.
The temperature of the system can be estimated with an error of the order of 10\% from the nominal
temperature of the cryostat and the experience of how to translate this into the electron temperature. 
Besides $J_{\rm c}$ or the CPR in all experiments considered here the differential conductance $dI/dV$ as a function
of bias and gate voltage was measured for superconducting as well as for normal state leads.
Superconductivity is destroyed by applying a magnetic (Zeeman) field of strength $h$ known within tight
bounds.\footnote{We intentionally give the experimental magnetic field (unit Tesla) a different symbol as compared
to the Zeeman field $B$ (unit of an energy) of the model first introduced in Subsect.~\ref{subsec:infinitegap}.}
This also splits the spin-degenerate dot levels and we have to consider the
case $\epsilon_\uparrow \neq \epsilon_\downarrow$ when discussing the situation
with normal state leads.
Typical $dI/dV$-data taken from Ref.~\cite{Eichler09} are shown in Fig.~\ref{fig8} a) for the normal
state and in b) for the superconducting one. Varying the gate voltage the differential
conductance shows partly overlapping repeating structures of comparable shape indicated
by the letters A to D. A single one of these structures can be described by the single-level SIAM;
for varying gate voltage individual levels move through the transport window.
For suppressed superconductivity Fig.~\ref{fig8} a) the bright region at zero bias
corresponds to the Kondo ridge of the linear conductance as discussed in
Sect.~\ref{subsec:spinssup}; see Fig.~\ref{fig1} (b). For the corresponding
gate voltages the occupancy is odd [see the yellow numbers in Fig.~\ref{fig8} a)].
In the surrounding dark regions transport is blocked by Coulomb blockade. 
From this we can conclude that the levels associated to A, B, and C are in the Kondo regime
while this does not hold for D. This is supported by the zero bias peaks of the
(vertical) white lines which show $dI/dV$ for a fixed gate voltage in the center of the corresponding
ridge. The peaks are related to the Kondo peak of the spectral function (see Ref.~\cite{Pustilnik04} and
references therein). From the edges of each Kondo ridge four lines originate when changing the bias
voltage which merge above or below the Kondo ridge center. They are best seen for ridge B and negative bias voltages
and are commonly denoted as Coulomb diamonds. From the height of the diamonds the local Coulomb
interaction $U$ of the corresponding level A to D can be estimated.\cite{Jorgensen07,Eichler09} For
superconducting leads the $dI/dV$-curves show peaks if the bias voltage equals the gap $\Delta$ as can
be seen in Fig.~\ref{fig8} b); see the yellow arrows. They originate from the leads BCS density of states and
allow to read off $\Delta$. At this stage of the analysis one thus obtained estimates of $U$, $\Delta$, $T$, and
$h$ all up to approximately 10\% error.

\begin{figure}[t]
\begin{center}
   \includegraphics[width=.49\linewidth,clip]{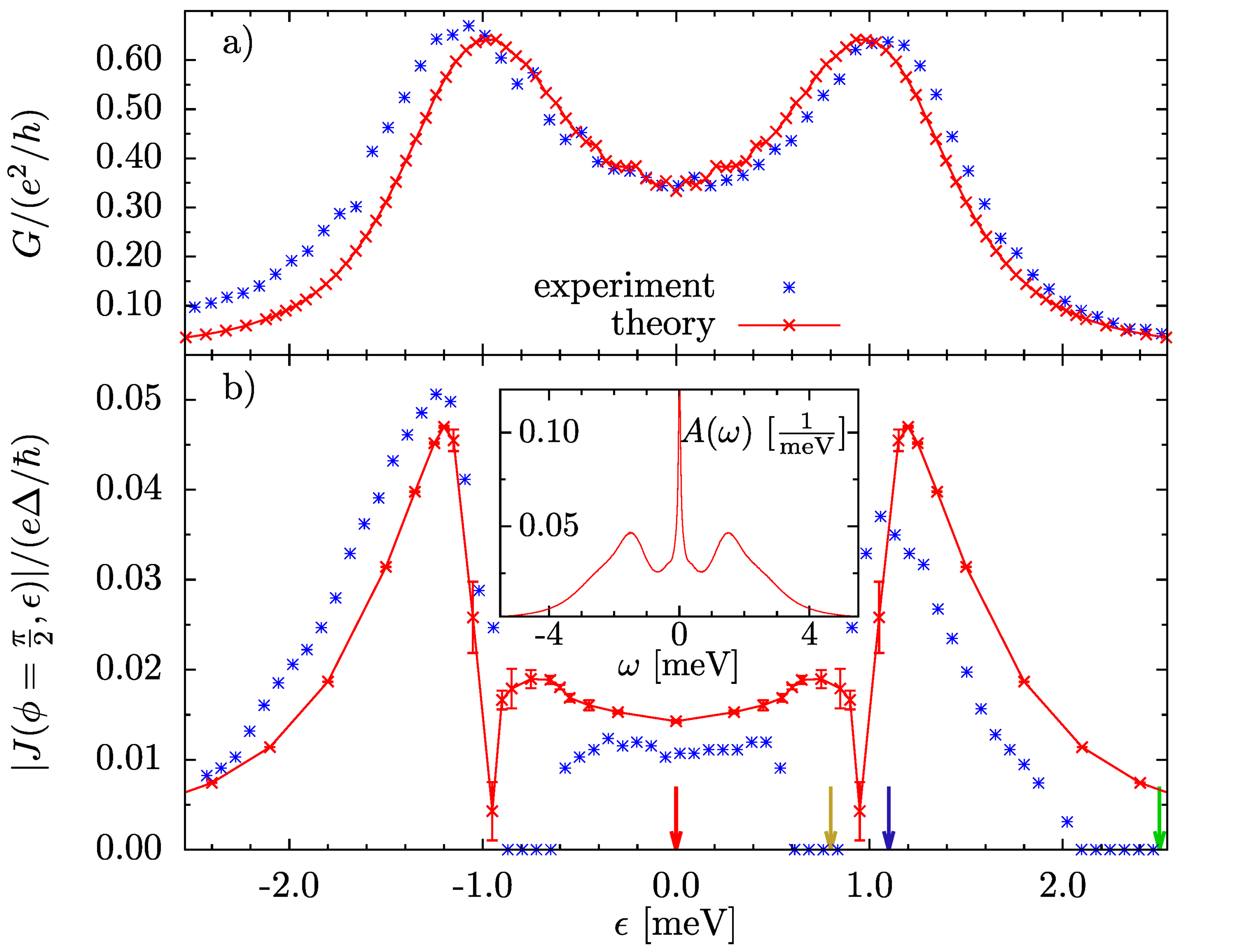}
 \includegraphics[width=.49\linewidth,clip]{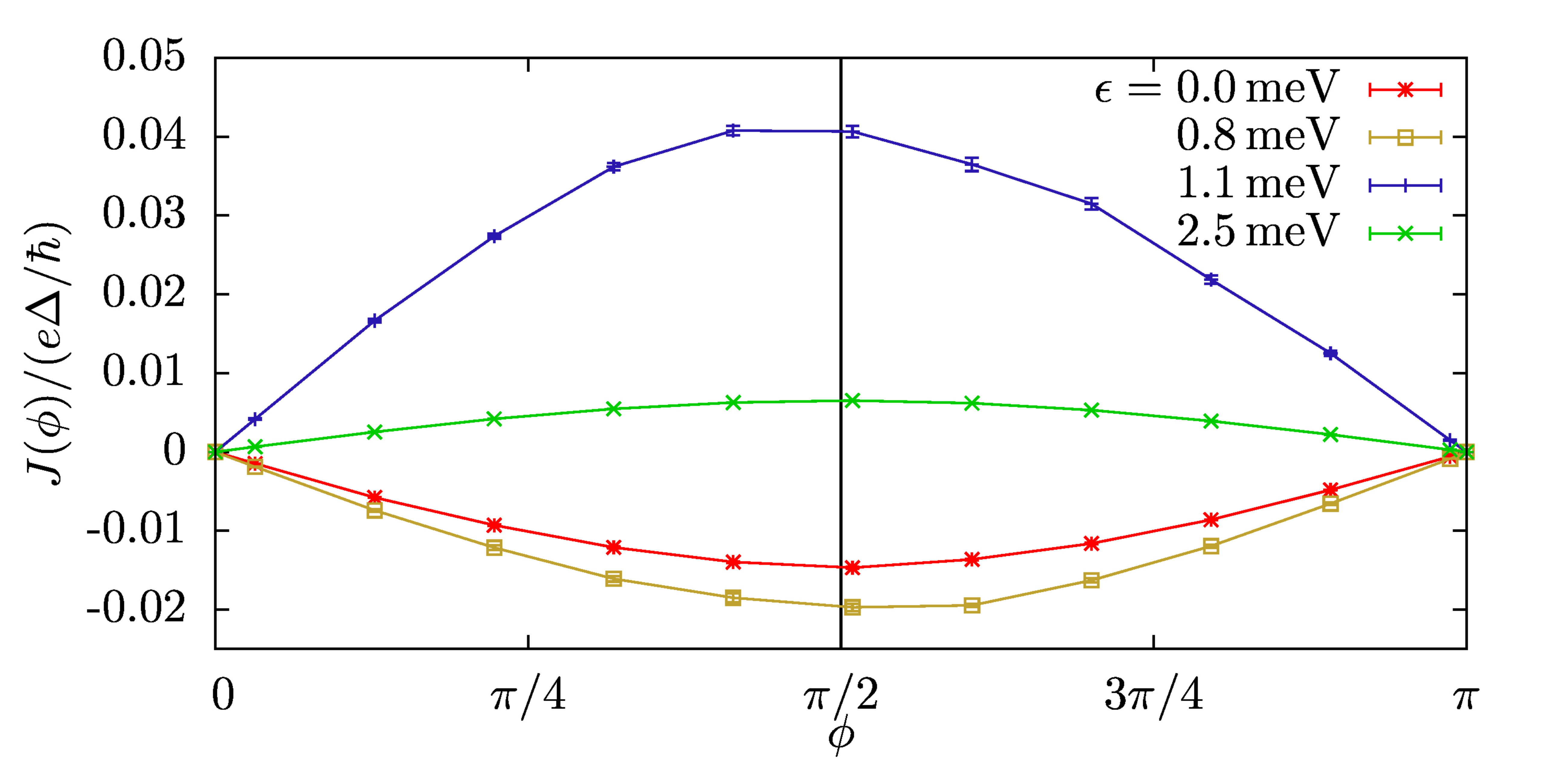}
 \caption{Left panel a): Linear conductance as a function of the level position which is tuned by an
   applied gate voltage. The blue symbols are experimental data and the red ones theoretical results
   obtained by fitting $\Gamma$, $a$, and the gate conversion factor $\alpha$ employing CTINT QMC. For
   more, see the text.
   Left panel b) main plot: Comparison of experimental and theoretical data for the gate-voltage dependence of
   the critical current $J_{\rm c}$.   Left panel b) inset: The spectral function with normal state leads
   computed by CTINT QMC. Right panel: Theoretical CPR data for the gate voltages indicated by the arrows in
   b) (color coded). Reprinted figures 
   with permission from D. Luitz, F. F. Assaad, T. Novotn\'y, C. Karrasch, and V. Meden,
   Phys. Rev. Lett.~{\bf 108}, 227001 (2012). 
Copyright (2012) by the American Physical Society.
 }
\label{fig9}
\end{center}
\end{figure}

The parameters which are most delicate to determine but strongly affect the Josephson current are
$\Gamma$ and $a$. In particular, if the level is in the Kondo regime estimating those from
$dI/dV$-curves requires input based on serious many-body theory; simpler theories, e.g., capturing
Coulomb blockade only cannot be used. As no fully reliable method to compute finite bias (nonequilibrium)
$dI/dV$-curves in the Kondo regime is available it was suggested  in Ref.~\cite{Luitz12} to fit theoretical
linear ($V_{\rm bias}=0$)  conductance curves as a function of $\epsilon$  [see Fig.~\ref{fig1} (b)]
obtained by CTINT QMC (or NRG) to the experimental data to determine $\Gamma$ and $a$. Here
the values for 
$U$, $T$, and $h$
as determined from the procedure described in the last paragraph are used.
As discussed in Subsect.~\ref{subsec:spinssup}
the width of the Kondo ridge is set by $U/\Gamma$ and allows to determine $\Gamma$, while the 
height is given by the asymmetry; see Eq.~(\ref{eq:GA}). We already now emphasize that both finite temperatures
and finite Zeeman fields have to be considered in the calculation as the associated energy scales will turn
out to often be comparable in size. Both have the effect to suppress the Kondo
ridge in its center (split the ridge), eventually leading to a two peak structure\cite{Hewson97};
see Fig.~\ref{fig9} a), left panel.   The fitting also allows to determine the so-called gate conversion
factor $\alpha$ which relates the change of the level position (unit of energy) to the change
of the applied gate voltage (unit of voltage).

Figure~\ref{fig9} a), left panel, exemplifies that rather accurate fits can be achieved. The blue symbols
show the linear conductance as measured for one of the dot levels (gate voltage regimes) studied in
Ref.~\cite{Jorgensen07}. For this the procedure described in the second to last paragraph gave
$U \approx 3$ meV, $\Delta \approx 0.1$ meV, $T \approx 75$ mK and $h \approx 150$ mT. Given these values
the best fit shown as red symbols was obtained for $\Gamma=0.27$ meV, $a=9.3$, and $\alpha=0.011$ V/meV.
This implies $U/\Gamma  \approx 11.15$ which is sufficiently large for the dot to be in the Kondo regime
and to show Kondo physics provided $T$ and $h$ are not too large. To estimate the Kondo scale in the center
of the ridge (at exactly odd filling of the dot) we use Eq.~(\ref{eq:T_Kana}) and obtain
$k_{\rm B} T_{\rm K} \approx 8$ $\mu$eV. This is roughly a factor of $0.03$ smaller  
than $\Gamma$ (Kondo regime) and of the same order as the energies associated to temperature
$k_{\rm B} T \approx 8$ $\mu$eV and the Zeeman field $\mu_{\rm B} h \approx 8.7$ $\mu$eV. Therefore,
as already mentioned above, neither the finite temperature nor the field can be neglected when computing
observables (e.g. the linear conductance) for suppressed superconductivity. The linear conductance of
Fig.~\ref{fig9} a), left panel, shows a Kondo ridge split by temperature and Zeeman field which should not
be mistaken with Coulomb blockade peaks which would be located at larger energies
$\epsilon \approx \pm U/2 \approx \pm 1.5$ meV. That the dot level is
in the Kondo regime is also confirmed by the spectral function in the center of the
Kondo ridge computed using CTINT QMC for the extracted parameters; see the inset
of Fig.~\ref{fig9} b), left panel. It clearly shows a sharp
Kondo resonance and Hubbard bands at the expected energies $\omega \approx \pm U/2$. The splitting of the Kondo
resonance by the Zeeman field\cite{Hewson97} is invisible within the energy resolution of the plot.
The spectral function is broadened by the analytic continuation required to obtain real frequency data
from CTINT QMC Matsubara ones.\cite{Luitz12} While for the dot level considered
$k_{\rm B} T_{\rm K} \approx k_{\rm B} T \approx \mu_{\rm B} h$ the gap $\Delta$ is roughly 10 times larger.
Kondo correlations can still not be neglected fully and affect the Josephson current.  

\subsection{The Josephson current}
\label{subsec:current}

After having determined all parameters it is now possible to compute the CPR by either CTINT QMC or NRG
in a parameter-free way and compare to the critical current extracted from the measured current-voltage
characteristics of the quantum dot Josephson junction or to the measured full CPR. We start out with the former.

\subsubsection{The critical current}

The blue symbols in the main plot of Fig.~\ref{fig9} b), left panel, show $\left| J_{\rm c} \right|$ as a function
of the level position for the same gate voltage regime as shown in Fig.~\ref{fig9} a), left panel. The data
are extracted from the experiment Ref.~\cite{Jorgensen07} employing the RCSJ model.
Similar results were presented in Ref.~\cite{Eichler09}. 
The extraction of the critical current provides only access to its absolute value and not its
sign. This is a major drawback (we did not
mention above) when it comes to the study of the $0$-to-$\pi$-transition indicated by a sign change of the current.
However, solely based on the experimental data one is tempted to conclude that the system is in the $\pi$-(doublet-)\-phase
for $-1$ meV $<\epsilon < 1$ meV and in the $0$-(sing\-let-)\-phase outside; see Fig.~\ref{fig9} b), left column. As all
parameters are known this can be confirmed by calculations within our minimal model which provide access to the
sign. The right panel of
Fig.~\ref{fig9} shows the computed CPR (by CTINT QMC) for the level positions indicated by the vertical arrows in
Fig.~\ref{fig9} b), left column (note the color coding).\cite{Luitz12} As expected the current is negative for $-1$ meV
$<\epsilon < 1$ meV and positive outside this regime. It furthermore shows that even very close to the level-crossing
transition at $\epsilon_{\rm c} \approx \pm 1 $ meV the CPR is rather sinusoidal (see the light green and blue curves of
the right panel of Fig.~\ref{fig9}). This is a consequence of $T>0$ and the fairly large left-right asymmetry of the
experimental level-lead coupling. Thus the RCSJ model can be employed even close to
$\epsilon_{\rm c} \approx \pm 1 $ meV and the extracted $\left| J_{\rm c}\right|$
can be used as an indicator of the transition. We emphasize, that these computations 
are mandatory for the a posteriori justification of the analysis in terms of the RCSJ model. This insight
also allows one to use $J(\phi=\pi/2)$ as a measure for the critical current even close to the transition. The red
symbols of Fig.~\ref{fig9} b), left column, show the computed $\left|J(\phi=\pi/2)\right|$ in comparison to
the experimental $\left| J_{\rm c} \right|$. The error bars indicate the statistical error of the CTINT QMC
results. A very satisfying agreement
is reached. We reemphasize that the Josephson current is computed without any fitting after the parameters have
been extracted in the normal state as described above. It is crucial to use highly accurate methods such as CTINT QMC or
NRG to achieve this type of quantitative agreement for dot levels in the (normal-state) Kondo regime as well as
$\Delta$ and $T_{\rm K}$ being comparable.   

\subsubsection{The current-phase relation}

\begin{figure}[t]
\begin{center}
   \includegraphics[width=.9\linewidth,clip]{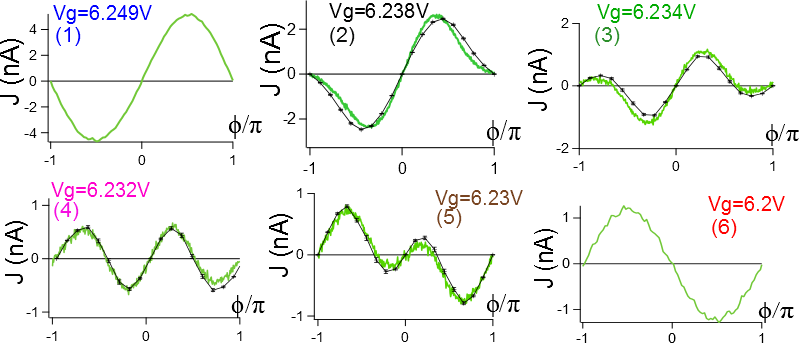}
   \caption{The measured full CPR (green lines) for gate voltages across the phase transition from the $0$- (upper left) to the
     $\pi$-phase (lower right). For comparison data computed with CTINT QMC are shown as black symbols (the black lines
     are guides to the eyes). Note that the regime $\varphi \in [-\pi,\pi]$ is shown. For more see the text. Reprinted figure 
     with permission from R. Delagrange, D.J. Luitz, R. Weil, A. Kasumov, V. Meden, H. Bouchiat and R. Deblock, Phys. Rev. B
     {\bf 91}, 241401 (2015). Copyright (2015) by the American Physical Society.}
\label{fig10}
\end{center}
\end{figure}

Reference~\cite{Delagrange15} reports on the successful measurement of the CPR of a carbon nanotube based
quantum dot junction across the entire $0$-to-$\pi$-transition. Additional data were presented in Ref.~\cite{Delagrange16}
and details on the setup  and measurement protocol are given in Ref.~\cite{Basset14}. These works constitute 
convincing experimental demonstrations of the level-crossing transition controlled by the phase
difference $\phi$. The CPR was recorded
for different gate voltages, i.e., different positions of the dot level. Figure~\ref{fig10} shows the measured
CPRs (green lines) with the amplitude being scaled-up by a unique gate-voltage independent factor (see below).
Note that the regime $\phi \in [-\pi,\pi]$ is shown. For the gate voltage of panel (1) the
dot is in the $0$-phase with a sinusoidal current of positive amplitude for $\phi \in [0,\pi]$. In (2) the
CPR is deformed for $\phi$ close to $\pi$ and no longer a simple harmonic. In (3) the current is negative
in this $\phi$ range. For the corresponding level position, varying $\phi$ drives the system through the
transition (compare the green curve in Fig.~\ref{fig5}, right panel). This trend continues with further
decreasing the gate voltage until in panel (6) the dot is in the $\pi$-phase for all $\phi$.

Following exactly the same steps as described above the parameters for the dot level considered
in Fig.~\ref{fig10} were determined. The values are $U \approx 3.2$ meV, $\Delta \approx 0.17$ meV,
$T \approx 150$ mK and $h \approx 1$ T, $\Gamma=0.44$ meV, $a=4$, and $\alpha=0.026$ V/meV all
up to an error of the order of 10\%.
This implies $U/\Gamma=7.27$ and the level is at the boundary of the Kondo regime. To estimate
$T_{\rm K}$ we still use Eq.~(\ref{eq:T_Kana}) (see the discussion in connection with this equation)
which gives $k_{\rm B} T_{\rm K} = 48$ $\mu$eV. We thus obtain $\Delta/(k_{\rm B} T_{\rm K}) \approx 3.5$.

With the estimated parameters the CPR was computed using CTINT QMC. After
shifting the level position by $\delta \epsilon=0.28$ meV and properly adjusting the scaling factor
of the amplitude mentioned in the last paragraph the theoretical and experimental data match almost
perfectly; compare the green lines and black data points of the central four panels of
Fig.~\ref{fig10}. It is not surprising that the amplitude of the experimental current has to be scaled
up. By construction the measured current is necessarily smaller than the true Josephson current. 
The QMC calculation predicts a transition region centered around a smaller $\epsilon$
than measured experimentally requiring the shift by $\delta \epsilon=0.28$ meV to superimpose
the experimental and theoretical data sets. This shift is currently not understood. Even with this
caveat in mind the excellent agreement of the entire line shape of the experimental and theoretical
curves for varying level position across the transition region provides a strong indication
that (i) the single-level SIAM with BCS leads can be used to describe the experiment and that (ii) the
latter shows the finite temperature signatures of the $T=0$ quantum phase transition. Computing the
gate voltage dependence of the Fourier coefficients of the experimental and theoretical CPRs  by a
numerical Fourier transform the comparison was brought to an even higher level in Ref.~\cite{Delagrange15};
see Fig.~4 of this publication.  

The quantitative agreement between the measured critical current and CPRs and the ones computed
in our model provide a convincing example that the model calculations can directly be applied to
measurements. In the minimal model many details (e.g., the details of the leads band
structure) are ignored. Focusing on the minimal model, however, allows one to investigate and
understand the quantum many-body physics lying at the heart of the level-crossing phase transition
in due detail using a combination of well controlled analytical and numerical approaches. 

\section{More complex dot geometries}
\label{sec:complexdots}

We finally exemplify the rich many-body physics which can be found in more complex setups
of quantum dots coupled to two superconducting leads. In the first, in addition to being
coupled via a single spin-degenerate level, the two BCS leads are coupled directly
by a hopping term of the form Eq.~(\ref{eq:Hdirect}). In other words, the quantum dot
Josephson junction is embedded in a Aharonov-Bohm-like geometry. For small tunnel-coupling via
the dot we expect that the Fano effect will play a role. It results from interference of a
structureless transport path (via the direct link) and one characterized by a narrow resonance
(via the dot). The physics of our second example is driven by the interplay of superconductivity
and almost degenerate singlet and triplet two-body states in multi-level dots. 

\subsection{Reentrance behavior and the Fano effect}
\label{subsec:fano}

The Hamiltonian of the Aharonov-Bohm-like setup is given by the one of our minimal model
of Subsect.~\ref{subsec:model} supplemented by the term Eq.~(\ref{eq:Hdirect}). For simplicity
we here focus on a left-right symmetric setup with $t_L = t_R$. Following the same steps as in
Subsects.~\ref{subsec:josephson} and \ref{subsec:model} it is straightforward to show that the
expectation value of the current operator is given by the sum of Eqs.~(\ref{eq:expcur}) and (\ref{eq:Jsimpexp1}) and
can thus be written in terms of the lead-dot and lead-lead Green functions
\begin{equation}
  \label{eq:directplusdot}
  J_L = \frac{2}{\beta}\sum_{i\omega}\mbox{Im}\,\mbox{Tr}\,\left[t_L{\mathcal G}_{L,{\rm d}}(i\omega)-t_{\rm d}
      {\mathcal G}_{L,R}(i\omega)\right] .
\end{equation}
The two terms have a natural interpretation as the dot and direct contribution to the Josephson current.
Note however, that the two Green functions are computed in the presence of both the direct link as well as the
one across the dot and are thus not identical to the ones derived in Subsects.~\ref{subsec:josephson} and
\ref{subsec:model}, respectively. Employing the equation-of-motion technique the two Green functions can be
expressed in terms of the interacting dot Green function and $g_s(i \omega)$ Eq.~(\ref{eq:leadg}). We refrain
from reproducing the corresponding expressions which can be found in Ref.~\cite{Karrasch09}.

\begin{figure}[t]
\includegraphics[width=0.9\linewidth,clip]{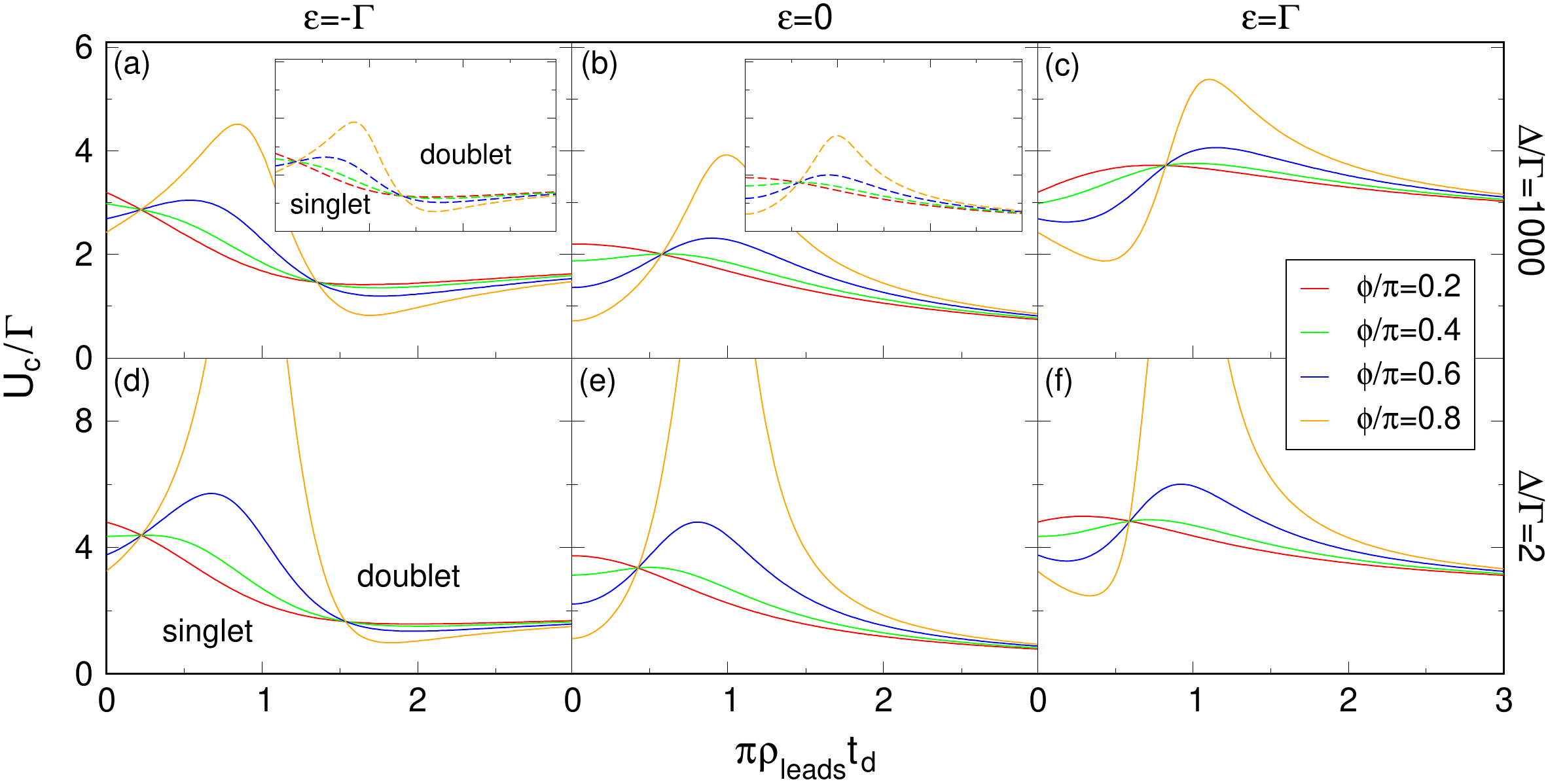}
\caption{The critical interaction strength $U_{\rm c}$ as a (nonmonotonic) function of the direct coupling $t_{\rm d}$
  for different BCS gaps $\Delta$ and impurity energies $\epsilon$, altogether characterizing the singlet-doublet
  level-crossing phase transition of the Aharonov-Bohm quantum dot Josephson junction. Solid lines where obtained
  from the FRG approach, dashed lines display the analytic result derived in the limit $\Delta \to \infty$
  [see Eq.~(\ref{eq:boundaryfano})].
  The axis of the insets are scaled the same as the axis of the corresponding main part.
Reprinted figure 
     with permission from C. Karrasch and V. Meden, Phys. Rev. B
     {\bf 79}, 045110 (2009). Copyright (2009) by the American Physical Society.
}
\label{fig11}
\end{figure}

To investigate the physics of the Anderson-Josephson-Aharonov-Bohm quantum dot we copy the successful
strategy employed above and first study the limit $\Delta \to \infty$. In this limit the noninteracting dot
Green function reduces to Eq.~(\ref{eq:G0at}) with the replacements
\begin{align}
 & \epsilon \to \tilde \epsilon = \epsilon+\Gamma\frac{\tilde t_{\rm d}\cos(\phi)+\tilde{t}_{\rm d}^3}{1+2\tilde t_{\rm d}^2
  \cos(\phi)+\tilde t_{\rm d}^4}, \\
& \Delta_{\rm d} \to   \tilde \Delta_{\rm d} = \Gamma\cos(\phi/2)\frac{1+{\tilde t}_{\rm d}^2}{1+2\tilde t_{\rm d}^2\cos(\phi)+\tilde t_{\rm d}^4} ,
  \label{eq:replacements}
\end{align}  
with $\tilde t_{\rm d} = \pi \rho_{\rm lead} t_{\rm d}$. Fundamental properties of the ground state can again be
obtained considering the effective (atomic limit) Hamiltonian Eq.~(\ref{eq:Hat}) with the same
replacements. The condition for a nondegenerate (doubly-degenerate) ground state is thus the same as
Eq.~(\ref{eq:condition}) with the proper replacements. The equation for the phase boundary separating the two
is
\begin{equation}
U^2 = 4\tilde\epsilon^2+4\tilde\Delta_{\rm d}^2 .
  \label{eq:boundaryfano}
\end{equation}  
The transition is again a first order level crossing one. However, there is one crucial difference to the
$t_{\rm d}=0$ case. Since $\cos(\phi)$ can become negative for $\phi \in [0,\pi]$, the right-hand side of
Eq.~(\ref{eq:boundaryfano}) is not necessarily a monotonic function of the bare parameters
$\epsilon/\Gamma$ and $t_{\rm d}/\Gamma$, immediately indicating reentrance behavior and multiple singlet-doublet
phase transitions. This is illustrated in the two insets to Fig.~\ref{fig11} which show the critical $U_{\rm c}$ obtained
from Eq.~(\ref{eq:boundaryfano}) for a variety of $\epsilon$ and $\phi$ as a function of $t_{\rm d}$.\cite{Karrasch09}     

This picture of reentrance quantum phase transitions was confirmed for $\Delta < \infty$ by the approximate
FRG approach reviewed in Subsect.~\ref{subsec:funrg}. We here refrain from giving the three coupled
RG flow equations for the components of the (static) self-energy and the effective two-particle interaction
and merely emphasize that they have a structure similar to the ones of Eqs.~(\ref{eq:flowSigma})-(\ref{eq:flowU}).
They can be found in Ref.~\cite{Karrasch09}. The phase diagram resulting from the numerical solution of these
is shown in the main panels of Fig.~\ref{fig11} for a variety
of parameter sets. The upper row was computed for $\Delta/\Gamma \gg 1$ and quantitatively confirms the
exact $\Delta \to \infty$ results shown in the insets even for sizeable $U/\Gamma$; remind that $U/\Gamma$ is
considered as the small parameter in the approximate FRG approach. The behavior remains qualitatively the same
for $\Delta/\Gamma$ of order one as shown in the lower row.

\begin{figure}[t]
\includegraphics[width=0.9\linewidth,clip]{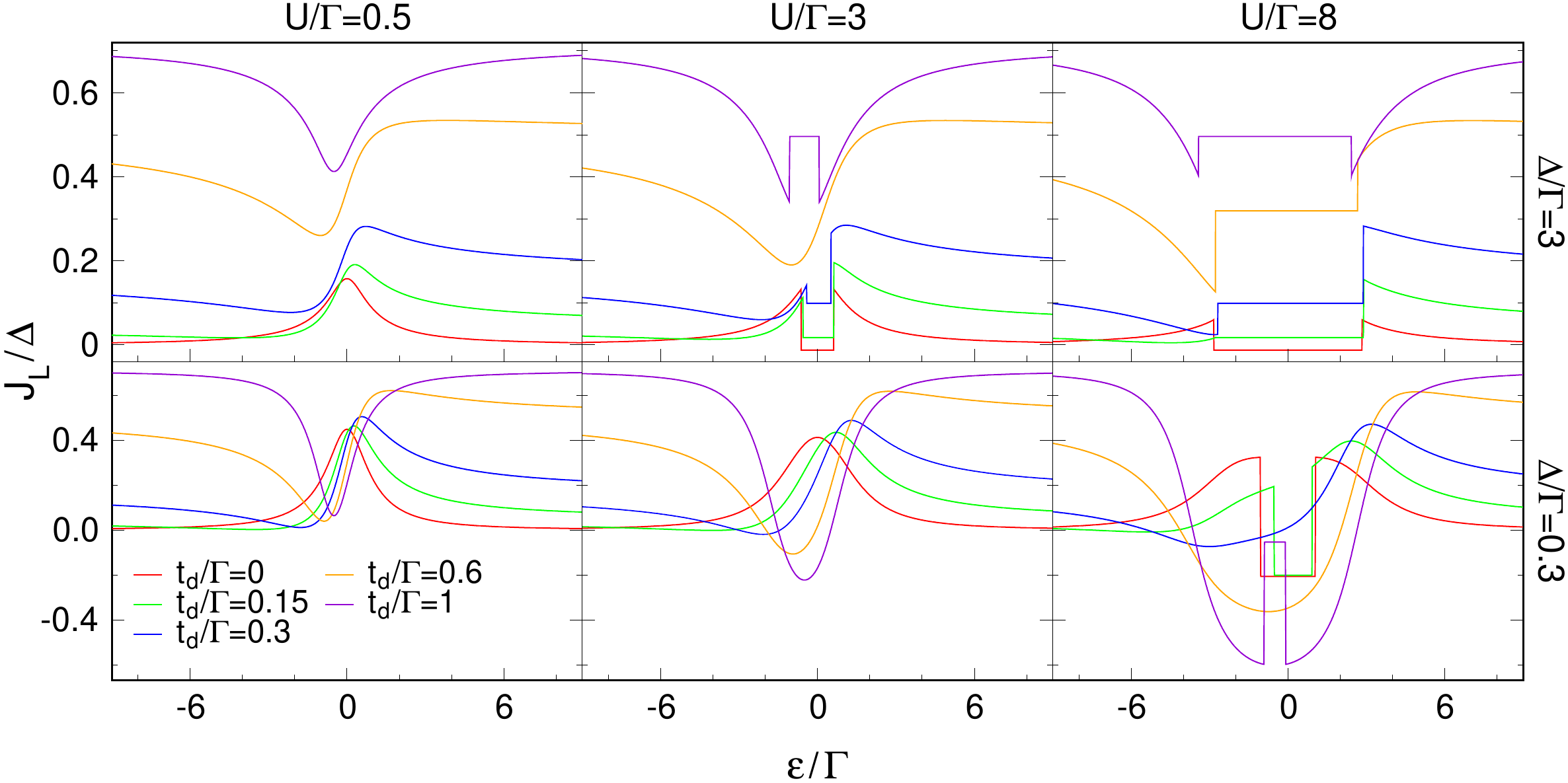}
\caption{Josephson current $J_L$ as a function of the impurity energy $\epsilon$ for
  constant $\phi=0.5\pi$ and different $t_d/\Gamma$ at $T=0$. The results were obtained from
  the approximate
  FRG approach. Reprinted figure 
     with permission from C. Karrasch and V. Meden, Phys. Rev. B
     {\bf 79}, 045110 (2009). Copyright (2009) by the American Physical Society.}
\label{fig12}
\end{figure}

Figure \ref{fig12} shows the Josephson current as a function of level position for $\phi=\pi/2$ and
several sets of the other parameters obtained by FRG. For small interactions (left colum), for which the system is in the
singlet state for the considered $t_{\rm d}$ and all $\Delta$ as well as $\epsilon$ the line shape changes from
that of a resonance (red line)
to the typical S-like Fano line shape (yellow line) to an inverted resonance (violet line) for increasing
$t_{\rm d}$. This can similarly be found in the linear-response conductance of the SIAM with metallic leads
(see, e.g., Ref.~\cite{Karrasch06} and references therein).
For small $\Delta/\Gamma$ this characteristics persists even up to intermediate $U/\Gamma$ (see
the central panel of the lower row of Fig.~\ref{fig12}). 
The doublet phase is characterized by the flat parts of $J_L(\epsilon)$ [compare to Fig.~\ref{fig9} b), left panel]. 
The nonmonotonic dependence of the phase boundary on the coupling strength $t_{\rm d}$, see Fig.~\ref{fig11}, results in
multiple singlet-doublet phase transitions manifesting as the appearance and disappearance of discontinuities
of $J_L(\epsilon)$ (see in particular the central panel in the upper row of Fig.~\ref{fig12}).
The behavior of Fig.~\ref{fig12} is generic for arbitrary phase differences $\phi$.
For the quantum dot Josephson junction with $t_{\rm d}=0$, the current is
positive (negative) in the singlet (doublet) phase. Both facts no longer necessarily hold for $t_{\rm d} > 0$ and it
would thus be misleading to refer to $0$- and $\pi$-junction behavior. It is
intuitive that $J_L$ can become positive in the doublet regime due to the additional contribution via the direct
link [see Eq.~(\ref{eq:directplusdot})]. Surprisingly one can also observe a negative current in the
singlet phase, particularly at small BCS energy gaps $\Delta$ (see the lower row of Fig.~\ref{fig12}). We, however,
emphasize that this is solely caused by the Coulomb interaction, and the supercurrent at $U=0$ always remains
positive. In contrast, Zhang\cite{Zhang05} obtained a negative current in the noninteracting
limit (for a $U=0$ study, see also Ref.~\cite{Cheng09}),
rendering his results on the current of the  Anderson-Josephson-Aharonov-Bohm quantum dot a priori highly
questionable.
The current of the present systems was also investigated in Ref.~\cite{Osawa08}, however, employing
the inappropriate unrestricted mean-field approach; see above. 

It would be very interesting to confirm the reentrance phase transition and the Fano line shape
of the current for the quantum dot Josephson junction in Aharonov-Bohm geometry in an experiment.
For a quantitative comparison between theory and experiment it might then be necessary to apply
CTINT QMC or NRG.

\subsection{Multi-level dots and singlet-triplet transitions}
\label{subsec:multilevel}

As a second example of a more complex setup we investigate a geometry with two single-level
dots aligned in series and coupled by a hopping $\tau$. The left dot is in addition coupled to
a left BCS lead and the right dot to a right BCS lead each by a tunnel coupling as considered
before. For definiteness we investigate a fully left-right symmetric setting with
$\Gamma_L = \Gamma_R$, left ($i=1$) and right ($i=2$) dot levels of equal energy
as well as equal on-dot and nearest-neighbor two-particle interactions. We emphasize,
however, that within the methods used to investigate the system 
all these restrictions can be relaxed. In the present subsection the Zeeman field splitting
the level energies of up and down spins will play an important role. The dot Hamiltonian 
replacing Eq.~(\ref{eq:Hdot}) reads
\begin{align}
  H_{\rm dot}  & =
  \sum_{i=1,2}\left[\left(\bar \epsilon+B\right) n_{i,\uparrow}+\left(\bar\epsilon-B\right)n_{i,\downarrow}\right]
  \nonumber \\
   & + U \sum_{i,\sigma\neq i',\sigma'} n_{i,\sigma} n_{i',\sigma'}
 - \tau \sum_\sigma\left( d_{1,\sigma}^\dagger d_{2,\sigma}^{\phantom{\dagger}} + \mbox{H.c.}\right),
\label{eq:ddham}
\end{align}
in self-explaining notation. The gate voltage $\bar \epsilon=\epsilon-3U/2$ is shifted such that $\epsilon=0$
corresponds to the point of half filling of the two dots at zero Zeeman field $B=0$. Here we will
only discuss the limit $\tau \gg \Gamma$ of two strongly coupled dots. In this case it is sometimes useful
to rotate to a bonding and antibonding single-particle basis. For $B=0$ the molecular spin-degenerate levels are
energetically well separated. For a discussion of the physics
in the opposite limit $\tau/\Gamma \lessapprox 1$, see Ref.~\cite{Zitko10}.

Here we are interested in the following two situations.
(i) For vanishing (or small) Zeeman field we expect to find the level-crossing scenario of the
single-level Anderson-Josephson dot for both the bonding as well as the antibonding molecular levels
if the gate voltage is properly tuned. We will indeed confirm this, which provides further justification
for the use of the single-level model when it comes to the comparison to experiments with multi-level
dots of sufficiently large level spacings; see Fig.~\ref{fig9}.  
(ii) For a Zeeman field $B \approx \tau$ the single-particle
(interaction $U=0$) energies of the bonding spin-up and the anti-bonding spin-down states are almost equal; or
rephrased, the smallest two-particle eigenenergies of an isolated dot (with $\Gamma_L = 0 = \Gamma_R$ but $U>0$),
which are a spin singlet as well as one out of a triplet, are almost degenerate. One might expect that for such
Zeeman fields the level crossing scenario is realized as well, with according signatures in the Josephson
current. However, this turns out to be only partially true: whereas various characteristics, e.g., the very
idea of a level crossing phase transition as well as the corresponding line shapes and parameter
dependencies of the current, are just as they are in a single-level case, the magnitude of the supercurrent
changes discontinuously at $B=\tau$ in one of the phases. This indicates an additional first order
singlet-triplet level-crossing quantum phase transition. 

\begin{figure}[t]
\centering
\includegraphics[width=0.45\linewidth,clip]{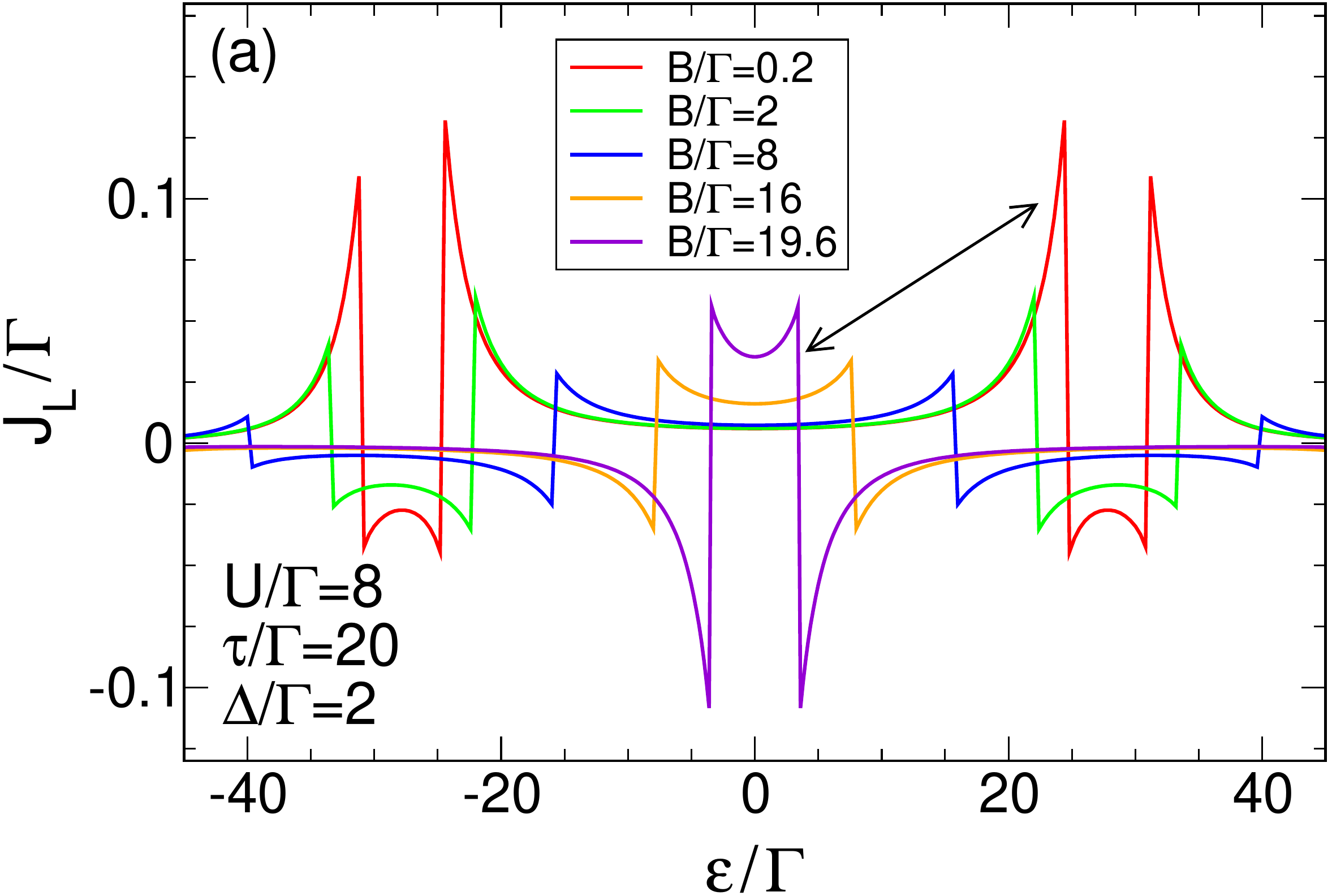}\hspace*{0.05\linewidth}
\includegraphics[width=0.45\linewidth,clip]{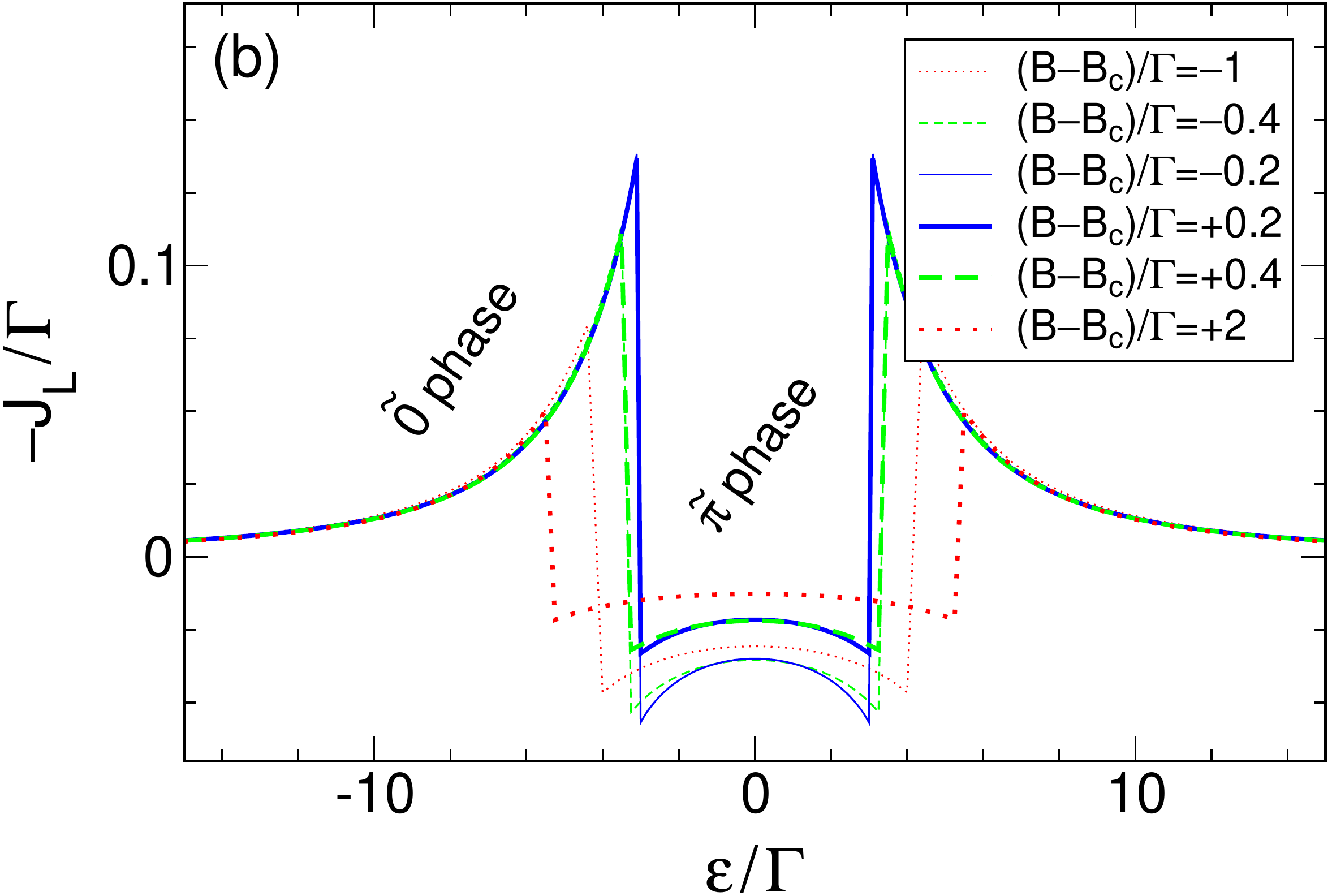}
\caption {Functional RG results for the $\epsilon$-dependence of the Josephson
  current $J_L$ flowing through the serial quantum dot for generic system parameters
  $U/\Gamma=8$, $\Delta/\Gamma=2$, $\phi/\pi=0.5$, $\tau/\Gamma=20$, zero temperature, and
  various Zeeman fields: (a) $B/\Gamma=0.2,2,8,16,19.6$.
  (b) $B/\Gamma=19,19.6,19.8,20.2,20.4,22$ or $(B-B_{\rm c})/\Gamma=-1,-0.4-0.2,0.2,0.4,2$ with $B_{\rm c}=\tau$.
Reprinted figure 
     with permission from C. Karrasch, S. Andergassen, and V. Meden, Phys. Rev. B
     {\bf 84}, 134512 (2011). Copyright (2011) by the American Physical Society.}
\label{fig13}
\end{figure}

As in the last subsection we will investigate the problem considering $\Delta \to \infty$ (exact solution)
and by approximate FRG (for $\Delta < \infty$). As nothing fundamentally new has to be added when setting
up the equations of the two approaches we refrain from giving any technical details and focus on the
results. The corresponding equations and derivations can be found in Ref.~\cite{Karrasch11}.

\begin{figure}[t]
\centering
\includegraphics[width=0.45\linewidth,clip]{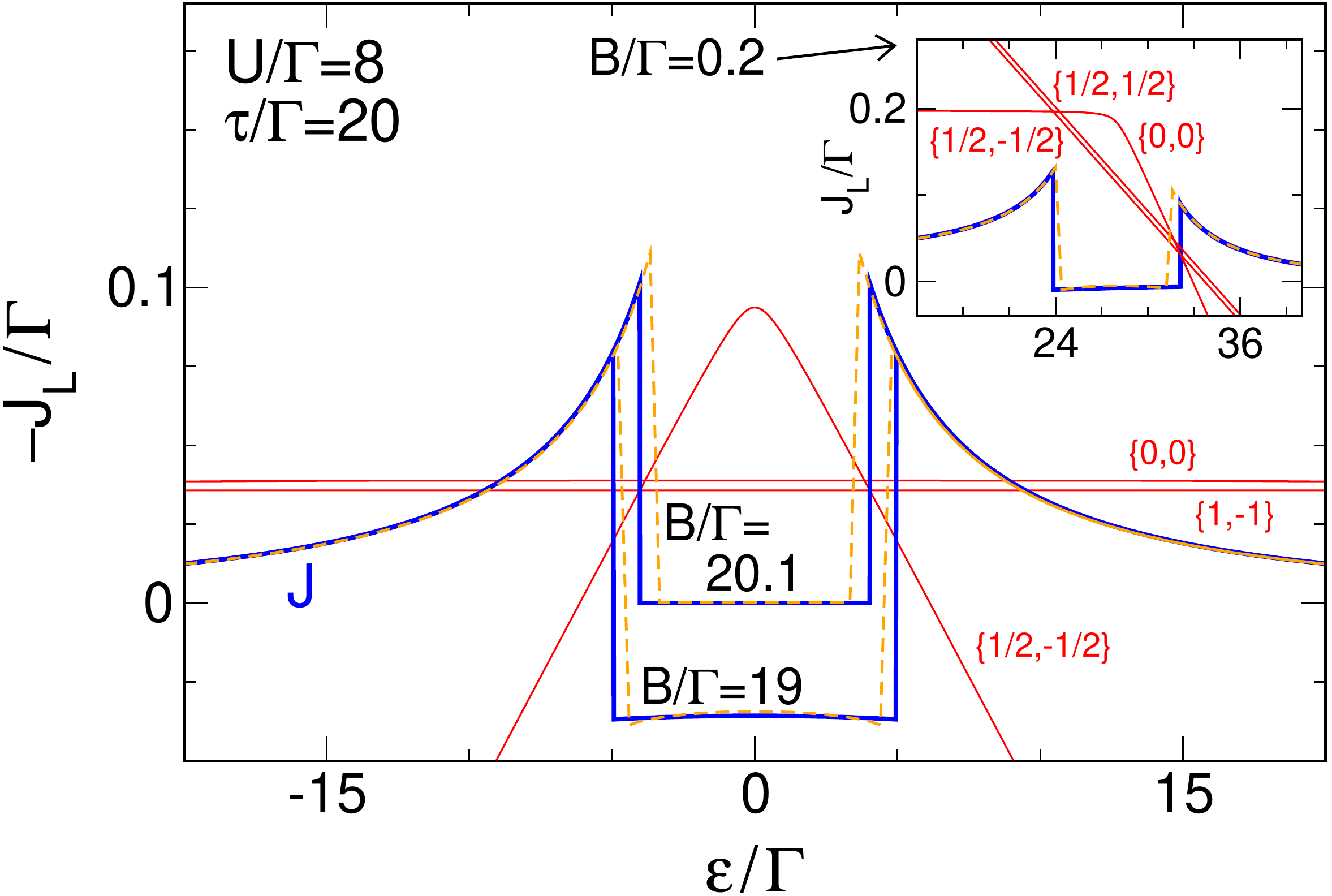}\hspace*{0.05\linewidth}
\includegraphics[width=0.45\linewidth,clip]{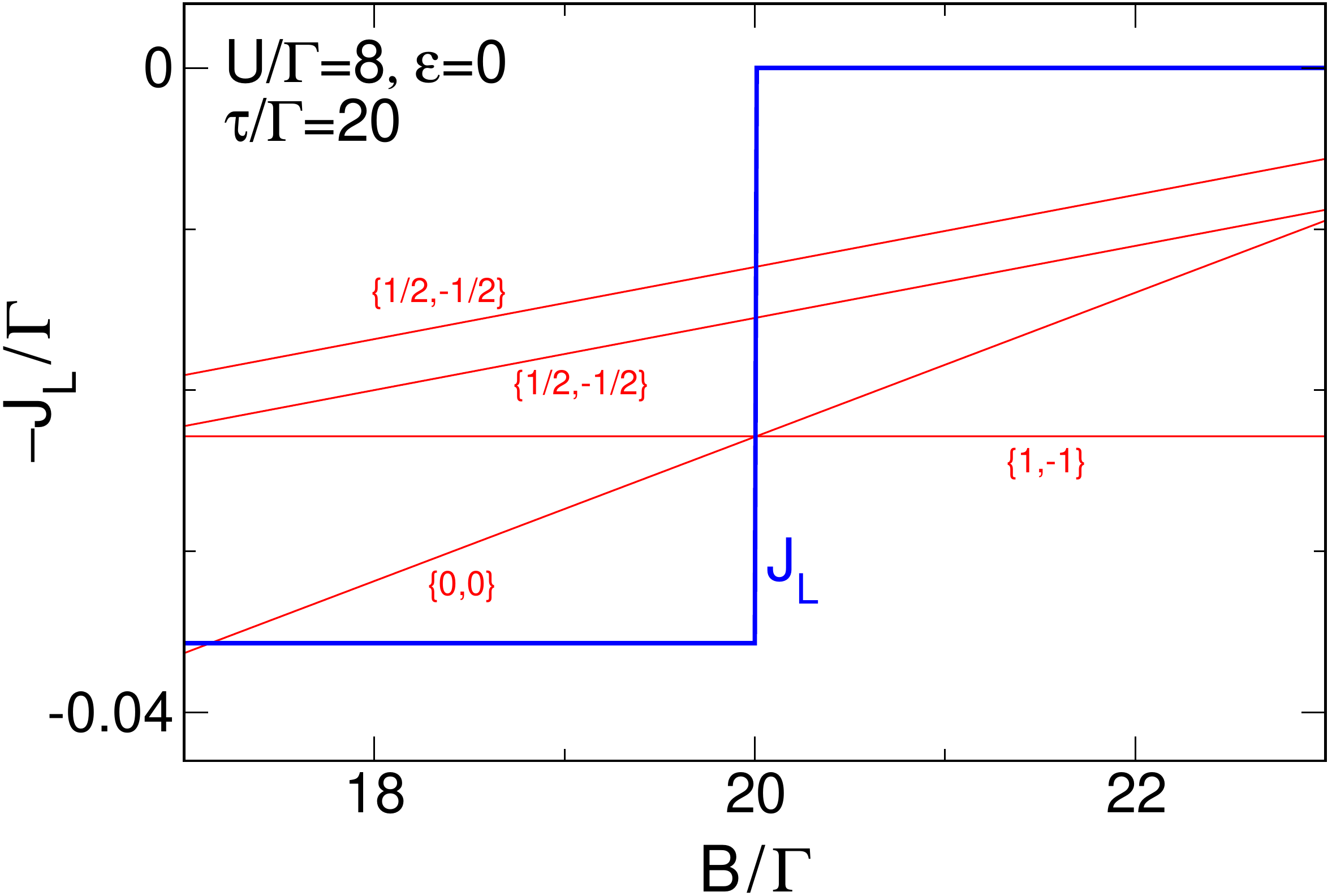}
\caption {Left panel: Exact results for the zero-temperature Josephson current in the infinite-gap limit
  (solid blue lines) compared to approximate FRG data obtained at $\Delta/\Gamma=2000$ (dashed orange lines)
  for three different Zeeman fields. The other parameters are the same as in Fig.~\ref{fig13}. Red lines
  show the three lowest many-particle eigenenergies for $B/\Gamma=20.1$ (main panel) and $B/\Gamma=0.2$ (inset)
  in arbitrary units. They are labeled by the corresponding spin quantum numbers $\{s,m\}$.
  Right panel: Atomic-limit ($\Delta \to \infty$) Josephson current (blue line) and the four lowest
  eigenenergies (red lines; arbitrary units) as a function of the Zeeman field $B$ at $\epsilon=0$. The
  other parameters are as in Fig.~\ref{fig13}. Reprinted figures 
     with permission from C. Karrasch, S. Andergassen, and V. Meden, Phys. Rev. B
     {\bf 84}, 134512 (2011). Copyright (2011) by the American Physical Society.}
\label{fig14}
\end{figure}

Figure \ref{fig13} shows FRG results for the Josephson
current as a function of the level position $\epsilon$ obtained for generic system parameters
(in particular $\Delta < \infty$) at different
$B$. For vanishing or small $B$ [red curve in Fig.~\ref{fig13} (a)] one indeed finds two copies of the
single-level behavior located at $\epsilon \approx \pm \epsilon_0$ (up to a slight asymmetry
around the center of each copy); compare to the red symbols in Fig.~\ref{fig10} for
$J_L(\epsilon)$. The jump in the current indicates the $0$-to-$\pi$-transition with a negative current in
the $\pi$-phase. The inset of Fig.~\ref{fig14}, left panel, shows the $\epsilon$-dependence of the current
(blue line) and the energy of the lowest lying levels (red lines) for $\Delta \to \infty$,
$\epsilon \approx \epsilon_0$, and a small Zeeman field. As for the single-dot case we characterize
the states by computing the total spin.
The ground state is a $\{s=0,m=0\}$ singlet in the $0$-phase or an almost ($B>0$) two-fold generate state
$\{s=1/2,m=\pm 1/2\}$ in the $\pi$-phase in strict analogy to the single-level case; compare
to Fig.~\ref{fig2a} (a).

As $B$ increases, the size of the regions with $J_L<0$ increases; see Fig.~\ref{fig13} (a). The transition
leading to the jump is a mixture of the transition induced by the two particle interaction,
as in Fig.~\ref{fig2a} (a), and that induced by the Zeeman field, as in Fig.~\ref{fig2a} (b).
At $B \approx B_{\rm c}=\tau$, however, a particular behavior can be observed. For this field
the single-particle ($U=0$) energies of the anti-bonding
spin-up and bonding spin-down states become equal; they cross the Fermi level at $\epsilon=0$.
Alternatively this point in parameter space can be characterized as follows: the smallest 
eigenvalue of $H_{\rm dot}$ (at $U>0$; no leads attached) with two particles,
which is the smallest overall one close to $\epsilon=0$, becomes twofold degenerate. The associated spin
configuration is either a singlet or (one out of) a triplet. We thus investigate closer whether
for $B\approx B_{\rm c}$, $\epsilon\approx0$ the physics can again be described in a pure (and simple)
single-impurity fashion. This is the case for normal leads where finite-$B$ Kondo ridges appear.\cite{Grap11} 
The Josephson current of the multi-level quantum dot with $B\approx B_{\rm c}$, $\epsilon\approx0$ indeed
resembles the one for $B\approx0$, $\epsilon\approx\pm\epsilon_0$ up to an irrelevant overall sign
[see Fig.~\ref{fig13} and compare the curves in (a) which are connected by an arrow].
Characteristically, there are discontinuities associated with a sign change.
The regimes of negative and positive current, denoted by $\tilde0$ and $\tilde\pi$, respectively, show
the same dependencies on system parameters as the $0$- and $\pi$-phases of a single impurity. Namely,
decreasing $U$, $\epsilon$, $|B-B_{\rm c}|$, or $\pi-\phi$ favors the $\tilde0$ regime. Analogously to
the line shape of $J_L(\epsilon)$, the CPR around $B\approx B_c$, $\epsilon\approx0$ is similar to the
single-level case: It is half-sinusoidal (sinusoidal) in the $\tilde0$ ($\tilde\pi$) regime; compare to
Fig.~\ref{fig5}.

There is, however, one obvious and
interesting difference to the single-level case. For sufficiently large $U/\Gamma$ the current
displays another discontinuity at $B=B_{\rm c}$ in addition to the $\tilde 0$-$\tilde \pi$ transition. It
is associated with a change in magnitude of $J_L$ but not a sign flip as shown in Fig.~\ref{fig13} (b).
On both sides of this additional supposedly first-order quantum phase transition,
one observes $\tilde\pi$ phase behavior as described above.
In order to obtain a more thorough understanding of this, we again consider the exactly solvable atomic limit
$\Delta=\infty$. The main plot of the left panel of Fig.~\ref{fig14} shows $J_L(\epsilon)$ for
$B< B_{\rm c}$ and $B>B_{\rm c}$ (blue lines). As for generic $\Delta$ the current is smaller (in fact
vanishing for $\Delta \to \infty$) in the latter case, compare to Fig.~\ref{fig13} (b).
The main plot of the left panel of Fig.~\ref{fig14}
in addition shows the $\epsilon$-dependence of the three lowest lying levels for the case $B>B_{\rm c}$.
For such a Zeeman field the ground state in the $\tilde \pi$-phase has $s=1$, i.e., is (one out of)
a triplet. As shown in the right panel of Fig.~\ref{fig14} (red lines) this changes if $B$ is decreased beyond $B_{\rm c}$
and the ground state becomes a singlet state with $s=0$ in the $\tilde \pi$-phase with a discontinuously
increased current (blue line). Within the  $\tilde \pi$-phase one thus finds an additional first order
level-crossing phase transition of a singlet and (one out of a) triplet. The same happens for
$\Delta < \infty$. It is intuitive that the current is larger in the singlet state as a triplet should
prevent Cooper pair tunneling. Using perturbation theory in $\Gamma_{L/R}$ this intuitive picture
can be made quantitative as discussed in Ref.~\cite{Karrasch11}. For a similar perturbative
analysis which however turned out to be incomplete,\cite{Karrasch11} see Ref.~\cite{Lee10}. 
We finally note that the above scenario is generic in a broader sense: it is not altered
qualitatively if different local- and nearest-neighbor interactions, level detunings or source-drain
coupling asymmetries are introduced.

The main plot as well as the inset of the left panel of Fig.~\ref{fig14}
show in addition to the exact $\Delta \to \infty$ current (solid blue lines) the one computed by approximate
FRG at $\Delta/\Gamma=2000$ (dashed orange lines). The excellent agreement between the two reconfirms
that the approximate FRG leads to reliable results even for fairly large $U/\Gamma$ ($=8$ in the figure).

It would be very interesting to experimentally observe the two transitions in a double-dot geometry;
the one from the $\tilde 0$- to the  $\tilde \pi$-phase and the singlet-triplet transition within
the $\tilde \pi$-phase. For a quantitative comparison between theory and experiment it might again be
necessary to use either NRG or CTINT QMC for the model calculations. In fact, a first step was 
taken recently. Reference \cite{Estrada18} reports on the successful measurement of the critical
current in a double-dot structure based on an indium arsenid quantum wire. However, it was only
possible to observe the series of $0$-to-$\pi$-transitions at $B=0$; see the red curve in Fig.~\ref{fig13} (a).
In this experiment increasing $B$ led to effects which do not seem to be explainable within the
simple model and obscure the physics described above. The observed $B$ field dependence appears
to originate in the details of the realization of the double-dot.\cite{Estrada18}
Thus more experimental work is needed.  

An extension of the above double-dot model is one in which the coupling $\tau$ is modified
such that it includes Rashba spin-orbit coupling. This has interesting effects on the
Josephson current as discussed in Refs.~\cite{Droste12} and \cite{Brunetti13}. Note that
the spin-orbit coupling might be of relevance in double-dots based on indium arsenid
quantum wires (see, e.g., Ref.~\cite{Estrada18}). 

Other aspects of the Josephson current through multi-dot or multi-level systems were studied
in Refs.~\cite{Choi00,Zhu02,Bergeret06,Bergeret07,Lopez07,Zazunov10,Yi16}.

\section{Summary}
\label{sec:summary}  

The aim of the present review is threefold.
First, a comprehensive account on the theoretical understanding
of the level-crossing phase transition physics of the single-level Anderson-Josephson quantum dot
was given. The first order quantum phase transition results from local on-dot correlations. The same are
responsible for the Kondo effect. The Kondo effect is not the driving force behind the level-crossing physics
as can, e.g., be seen from considering the large gap limit. It is, however, interesting to study the interplay
of BCS superconductivity and the Kondo effect. Combining analytical insights from the $U=0$ and the
$\Delta \to \infty$ limit as well as highly accurate results obtained by the two numerical approaches
NRG and CTINT QMC a deep understanding of the basic physics underlying the transition
from a nondegenerate singlet to a doubly-degenerate ground state is gained.
This includes an understanding of the finite temperature signatures of the $T=0$ transition. 
We briefly reviewed the historical development leading to this understanding and evaluated the methods
which can be used to obtain reliable results. Our main focus was on the Josephson current as the observable
showing the transition. It is characterized by a discontinuous sign change of the current.
In the singlet phase the current is positive (for $\phi \in [0,\pi]$) while it is negative
in the doublet one, i.e., shows $\pi$-junction behavior.   
In addition, we reviewed results for the dot spectral function which shows
in-gap bound states. As corroborated by our discussion we believe that the physics of the
single-level Anderson-Josephson dot is fully understood. 

Secondly, we showed that a quantitative agreement between experimental data for the supercurrent and the ones
computed with accurate methods applied to the SIAM with BCS leads can be reached.
This holds for the gate-voltage dependence of the critical current as well as for the full CPR at varying
gate voltage. Details ignored in
the model thus do not affect the basics physics captured by the model. For dots which show the Kondo effect
at suppressed superconductivity one has to employ methods capturing this to even properly extract the
parameters, in particular the hybridizations $\Gamma_{L/R}$. As the experimental current displays the
finite temperature signatures of the transition known from the model calculations, which can be
carried out at $T=0$ and $T>0$, one can be confident that the experiments indeed show the transition.  

Thirdly, we reviewed two examples of how the interaction induced level-crossing physics manifests in more
complex dot setups. Experiments on both should be realizable with todays nanostructuring and measurement
technology. For an Anderson-Josephson dot embedded in an Aharonov-Bohm-like geometry we discussed
reentrance phase transitions showing in the supercurrent. The linear double-dot geometry
showed two ground state transitions for a properly tuned Zeeman field. The one known from the
single-level case [nondegenerate singlet to (almost) doubly degenerate state] and additionally
one in which the ground state turns from a state with spin quantum numbers $\{s=0,m=0\}$
into one with $\{s=1,m=-1\}$. In accordance with the intuition that the former supports Cooper
pair transport while the latter inhibits it, the current is larger in the singlet and jumps
to a smaller value across this second transition, however, without a sign change. As a general
strategy to tackle more complex systems we suggest to use method combination.  The parameter space
can efficiently be scanned using the computationally cheap FRG. Analytical
insights can be obtained in the $\Delta \to \infty$ limit.  
The parts of the parameter space which show interesting many-body effects can be
further investigated with the computationally more demanding NRG or CTINT QMC, in particular,
when aiming at a quantitative comparison to experimental data. 

\section*{Acknowledgement}

Over the last years I had the pleasure to work with a large group of theorists and experimentalists
on the topics presented in this review. Many results were obtained in a fruitful
collaboration with Christoph Karrasch and David Luitz. I would like to thank both of them and 
Christoph (in addition) as well as Kurt Sch\"onhammer for a critical reading of an earlier version
of this manuscript.
I enjoyed additional collaboration on the theory of the Anderson-Josephson quantum
dot with Akira Oguri, Tom\'a\v{s} Novotn\'y, Sabine Andergassen, Fakher Assaad, Nils Wentzell, Serge Florens,
and Tobias Meng. I am delighted that Richard Deblock, Rapha\"elle Delagrange, H\'el\`{e}ne Bouchiat,
Christian Sch\"onenberger, Alexander Eichler, Markus Weiss, Raphael Weil, and Alik Kasumov were willing to
share their experimental insights with me and to embark on an endeavor to better understand the physics which led
to several common experiment-theory publications. My understanding of the problem reviewed was promoted in
many discussions with Jens Paaske and Kasper Grove-Rasmussen which I gratefully acknowledge. I would like to
thank the Deutsche Forschungsgemeinschaft for funding within the framework of FOR 723. 

\vspace{1.0cm}

\bibliography{ref}

\end{document}